\documentclass[smallextended]{svjour3}       % onecolumn (second format)
\smartqed  % flush right qed marks, e.g. at end of proof
\usepackage{graphicx}
\usepackage{natbib}
\begin{document}

\title{Molecular gas in distant galaxies from ALMA studies%\thanks{Grants or other notes
}

\titlerunning{Molecular gas in high-z galaxies with ALMA}        % if too long for running head

\author{Fran\c{c}oise Combes  
}


\institute{F. Combes \at
 Observatoire de Paris, LERMA, Coll\`ege de France, CNRS, PSL Univ., Sorbonne University, UPMC, Paris, France \\
              Tel.: +33-1-40512077\\
              Fax: +33-1-40512002\\
              \email{francoise.combes@obspm.fr}           %  \\
}

\date{Received: date / Accepted: date}

\maketitle

\begin{abstract}
ALMA is now fully operational, and has been observing 
in early science mode since 2011.
 The millimetric (mm) and sub-mm domain is ideal to tackle
galaxies at high redshift, since the emission peak of the dust at 100$\mu$m is
shifted in the ALMA bands (0.3mm to 1mm) for z=2 to 9, and the CO lines,
stronger at the high-J levels of the ladder, are found all over the 0.3-3mm range.
 Pointed surveys and blind deep fields have been observed, and the wealth of data collected reveal a
drop at high redshifts (z $>$ 6) of dusty massive objects, although surprisingly active
and gas-rich objects have been unveiled through gravitational lensing.
The window of the reionization epoch is now wide open, and ALMA has detected galaxies
at z=8-9 mainly in continuum, [CII] and [OIII] lines.
Galaxies have a gas fraction increasing steeply with redshift, 
as (1+z)$^2$, while their star formation efficiency increases also but more slightly,
as (1+z)$^{0.6}$ to (1+z)$^1$.
Individual object studies have revealed luminous quasars, with black hole
masses much higher than expected, clumpy galaxies with resolved star formation rate
compatible with the Kennicutt-Schmidt relation, extended cold and dense gas
in a circumgalactic medium, corresponding to Lyman-$\alpha$
blobs, and proto-clusters, traced by their brightest central galaxies. 
\end{abstract}

\section{Introduction}
\label{sec:intro}
The recent years have seen a breakthrough in the domain of high redshift
galaxies, and in particular in the knowledge of their molecular gas and
their star formation with ALMA. While the uv/optical/infrared domains give information
on the star formation density, and its evolution in the Universe
 \citep[e.g.][]{Madau2014}, the efficiency of star formation requires knowledge
of the gas content. Molecular gas is the fuel of star formation, and its observation
is necessary to understand galaxy formation. 

ALMA with its 66 dishes (54 antennae of 12m and 12 antennae of 7m, located in 
a unique high and dry site, has increased the power of previous (before 2010) 
millimetric arrays by an order of magnitude. Baselines from 20m to 16km,
at wavelengths between 3mm and 0.3mm, provide spatial resolutions up to 15mas.
The large bandwidth of 7.5 GHz/polar ensures a high sensitivity for
continuum observations, and allows to search and determine redshifts.
ALMA is well adapted for deep fields, but not for big surveys, the field of 
view is from 1 arcmin (at 3mm) to 6 arcsec (at 0.3mm). Mapping small regions with
 mosaics is very efficient.

The main advantage for high redshift galaxies is that the peak of dust
emission usually around 100 microns, for star forming objects,
 is redshifted to the submm and mm domain. This produces a negative K-correction,
i.e. continuum emission from dust is as easy to detect at z=10 than at z=1
 \citep[e.g.][]{Blain2002}.
Already in the pre-ALMA era, it was possible to detect hundreds 
 of high-z galaxies with L(IR) $>$ 10$^{12}$ L$_\odot$, up to z=6
\citep[e.g.][]{Omont2007}.  The large derived dust masses of 
$\sim$10$^8$ M$_\odot$, mean that dust forms early in the universe.
These sub-millimeter galaxies (SMG) contribute significantly to the sub
millimetre background, their redshift distribution peaks at z=2-3
\citep{Chapman2005}.

For the CO lines, there is in general no negative K-correction; there is only a possible 
increase as the frequency square of the CO flux in the first transitions of the ladder (low J), 
when the gas is dense enough to be thermalized \citep[e.g.][]{Combes1999}.
Distant galaxies have started to be explored in molecular lines in 1992,
with lensed objects \citep{Brown1992, Downes1995}, and line detections 
followed at a high rate, about 50 objects up to z=6.4 with the quasar
J1148+5251 \citep{Walter2003, Cox2005, Maiolino2005}.

With ALMA, it is now possible to detect CO lines in a large amount of high-z
galaxies, even not amplified by gravitational lensing.
It is possible to discover obscured objects in deep fields, from their dust emission, 
and search for their redshift, when it is not possible in the optical domain.
For z$>$6 galaxies, the high-J CO lines (J$>$7) are observed at
low frequencies (3mm) with a field of view of 1 arcmin, and a bandwidth of 2x 8GHz
$\sim$ 16\%, or 50 000km/s. At z=6, the spacing between the various CO lines of the
rotational ladder is of 16 GHz, so that the redshift may be obtained
with 2 tunings only.

This review highlights the main results of ALMA
observations of distant galaxies, from the dust emission to molecular lines.
The CO are the most usual gas tracers, but at very high z, the [CII] and CI
lines bring important information, in a domain where the CO lines are
not or little excited. Dense gas tracers (HCN, HCO$^+$, CS, etc) and isotopes,
 bring complementary knowledge on gas properties, and are frequently
observed simultaneously, thanks to the wide bandwidth of ALMA.

The review emphasizes only the recent results
since the previous reviews in the domain, pre- or post-starting of ALMA 
 \cite{Solomon2005, Carilli2013}.

\section{The CO lines as tracer of the molecular gas}
\label{sec:SLED}

The main H$_2$ molecule is symmetric and has no dipolar transition. At the low
temperature of the interstellar medium ($\sim$ 20K), the quadrupolar transitions
are not excited, and moreover they have a very weak Einstein coefficient. The main
tracer of the molecular gas is then the CO molecule (with solar abundance of CO/H$_2\sim$ 10$^{-4}$),
tracing the bulk of the gas, with a critical density of the order of 10$^3$-10$^4$ cm$^{-3}$ for the lowest
levels. Other molecules with a higher critical density (more than 2 orders of magnitude higher),
like HCN, HCO$^+$ or CS are used as high density tracers. Their intensity is usually 10-30 times lower.

\subsection{Emission lines}
\label{sec:em}

At high redshift, the large advantage of the CO tracer is the high probability to find
any line J from the rotational ladder, since the ladder spacing decreases as (1+z)$^{-1}$;
and when the gas is excited, the line strength of the J levels increases almost as the
square of the frequency. There are good reasons to expect higher density molecular gas
in high-z galaxies (they are more gas rich, and their volumic density is higher), such that
higher excitation is the norm. This situation favors the detection of molecular gas at high-z,
while the atomic gas has large difficulties with the unique 21cm line. 

The distribution of radiating energy among the various J-lines of the CO ladder
called the SLED (Spectral Line Energy Distribution), is a very useful diagnostic of the
physics of the emitting interstellar medium (ISM), in particular its density and temperature.
 It has been shown that the CO SLED can distinguish between
quiescent Milky Way-like galaxies, where the emission is peaking at J=3,
and dense and warm starbursts, where the peak is up to J=8 \citep{Weiss2007}.
These excitations come from star formation processes (PDR, or photo-dissociation regions),
but near AGN, higher excitation is possible, in particular through hard X-rays
(XDR, or X-ray dominated regions) \citep{vanderWerf2010}. In some cases of
very concentrated starbursts, the dust opacity could also perturb the SLED
\citep{Papadopoulos2010}.

At high redshift, galaxies have a higher gas fraction \citep[e.g.][]{Geach2011,Tacconi2010}.
The gas fraction can be defined by the gas to stellar mass ratio, or the gas to baryonic mass ratio.
Not only the gas is more abundant, but it is also denser, and star formation
rates are higher in average. It is therefore expected that the CO lines are
more excited, favoring the detection of the molecular gas. There is also the 
possibility of radiative excitation from the cosmic background, which temperature
varies in (1+z), reaching $\sim$ 30K at z=9. It is not obvious that this
excitation helps the detection, since the detected signal is only the excess above the
background. This has been simulated, and indeed, there is no negative K-correction
for the CO lines, contrary to the dust continuum emission,
\citep[e.g.][]{Combes1999,daCunha2013}.

\begin{figure}
\begin{center}
  \includegraphics[width=0.8\textwidth]{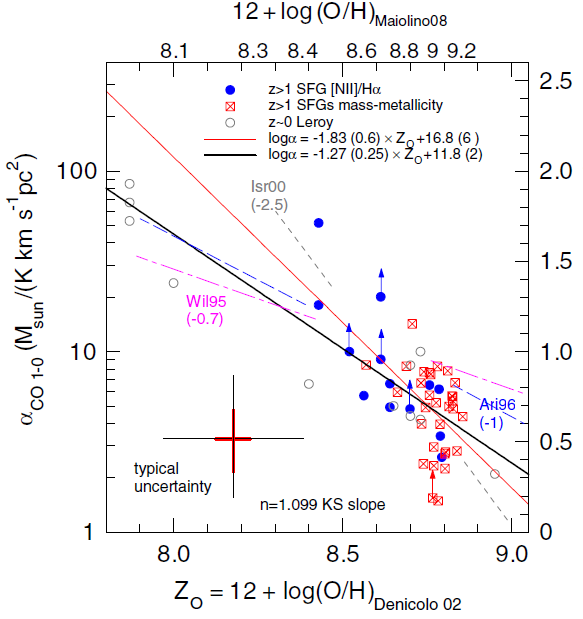}
\end{center}
\caption{The derived conversion factor between the CO(1-0) luminosity and 
the molecular gas mass, $\alpha_{CO}$ as a function of metallicity in the gas phase.
The molecular gas mass is computed independently of the CO luminosity from the 
SFR and the Kennicutt-Schmidt (KS) relation, with a slope n=1.1. 
The gray circles are for local galaxies \citep{Leroy2011}. 
The gas metallicity for high-z blue-symbol galaxies have been obtained from the
[NII]/H$\alpha$ flux ratio, converted to oxygen abundance scale
from \cite{Denicolo2002} and \cite{Maiolino2008}. The red symbols are 
galaxies where metallicity is derived from the total stellar mass.
Image reproduced with permission from \cite{Genzel2012}, copyright by AAS.}
\label{fig:alpha-CO}
\end{figure}

The derivation of the molecular content from the CO lines relies on
the CO-to-H$_2$ conversion factor, well calibrated at low redshift, and in 
particular in the Milky Way: clouds are then detected individually,
and their virial masses estimated. The conversion factor has a robust statistical value,
when averaged over a large cloud population, with a wide range of masses
and densities. The factor depends however strongly on gas metallicity
 \citep[e.g.][]{Bolatto2013}. It is possible to quantify the fraction of
   diffuse and clumpy components, with high density tracers such as HCO$^+$, HCN,
   allowing to refine the conversion factor.
   For example \cite{Oteo2017} have shown with ALMA that the excitation of HCO$^+$, HCN
   and HNC in two lensed dusty starbursts at z$\sim$ 2 is very similar to what is already known
   in local IR-bright galaxies.
   Due to the virial hypothesis, the conversion factor is thought
to vary as the square root of the H$_2$ volumic density, divided by the brightness
temperature of the clouds, $\propto$ n(H$_2$)$^{1/2}$/T$_B$. In local 
starbursts, both the cloud brightness and their density increases, which 
limits the variation of the ratio \citep[e.g.][]{Leroy2011}. The metallicity
is however a serious problem, since it is thought to decrease with redshift. 
 Using one of the first large surveys of star forming galaxies at z$>$1,
\cite{Genzel2012} have estimated the variation of the conversion factor 
with metallicity (see Figure \ref{fig:alpha-CO}).

\subsection{Absorption lines}
\label{sec:abs}

A complementary way to probe the interstellar medium of high redshift
galaxies is from absorption lines in front of a strong millimeter continuum source.
These molecular lines can provide information on the chemistry and its evolution
with z, and also the physical conditions of the gas (density, temperature). The lines
may be very narrow ($<$ 1km/s) and useful to constrain the variations
of fundamental constants. The absorbing systems before ALMA were only 4-5
\citep[e.g.][]{Combes2008}. With ALMA, it was possible to
carry on a molecular survey towards the system PKS1830-211, corresponding
to two gravitational images and an Einstein ring in the absorbing foreground lens
\citep{Muller2014}. In particular, isotopes of the chloronium were detected, and
OH$^+$, H$_2$O$^+$ have allowed to measure the molecular gas fraction and
the ionization rate of the gas, in several lines of sight \citep{Muller2016}.
With ALMA, new molecular lines are detected in z=0.5-1 absorbers \citep[e.g.][]{Wiklind2018},
but also a large number in more local radio sources \citep{David2014, Tremblay2016}, where
the absorptions are determinant to disentangle inflows from outflows.

\section{Dust emission as a molecular gas tracer}
\label{sec:dust}
Although the CO molecule is arguably the best tracer of the
H$_2$ content of a galaxy, it is paramount to gather results from
several tracers, to inter-compare them, and avoid some of the
main biases of any given diagnostic. At high redshift, this is even
more required, since the CO lines observed are from a high J-level,
and several CO lines are needed to derive the gas excitation, and
estimate the CO(1-0) intensity calibrated in terms of H$_2$ mass.
The main alternate tracer is the dust continuum emission in the Rayleigh Jeans
domain (i.e. close to the CO(4-3) to CO(7-6) if excited), where the dust emission
is linear in both temperature and column density. The assumed dust-to-gas ratio will
account for metallicity effect.
Other tracers at very high z are fine structure line emission
such as the [CII] at 158$\mu$m, or the [OIII] at 88$\mu$m, with some
specificity as gas tracers, as will be developed in Section \ref{sec:EoR}.

\cite{Casey2014} have written a detailed review of far-infrared and sub-millimeter
survey of high redshift galaxies with dust emission. Although the temperature of the 
dust heated by star formation in molecular clouds, is expected in average
of the order of 20-40K, in some
cases, nuclear starbursts or AGN, the dust can peak at 60K. This produces large 
uncertainties in the detection rates of continuum surveys, since their success 
rate depends in the SED distribution of the sources.

\cite{Scoville2014, Scoville2016} have proposed that the Rayleigh-Jeans tail of the dust
emission spectrum, peaking around 100$\mu$m, acts as a good tracer of the gas content
of galaxies: the Rayleigh-Jeans regime ( $\lambda >>$ 100 $\mu$m)
means that the dust temperature is involved
only linearly, and does not introduce too much uncertainty. Of course the dust abundance
is also proportional to metallicity, so the conversion factor between the dust emission and gas
mass, through the dust-to-gas ratio, is as incertain as the CO method. However, the detection
of the dust continuum might be easier than the line, and does not require specific tunings.
In compensation, there is more confusion and no redshift or kinematical information
on the detected objects. While it is relatively easy to detect actively star forming galaxies and
starbursts, main sequence objects at relatively low redshifts are more 
difficult to detect in dust continuum
than in the CO lines, given the non-linear L$_{FIR}$-L$_{CO}$ scaling relation. For instance,
in a sample of normal star-forming galaxies of the COSMOS field at z$\sim$ 3,
 about half of the galaxies are detected in continuum with ALMA, and
the rest of the undetected sources have to be stacked \citep{Schinnerer2016}.
Besides, the CO line observation provides more very useful information as the dynamical mass,
and gas excitation.

\begin{figure*}
\begin{center}
  \includegraphics[width=0.95\textwidth]{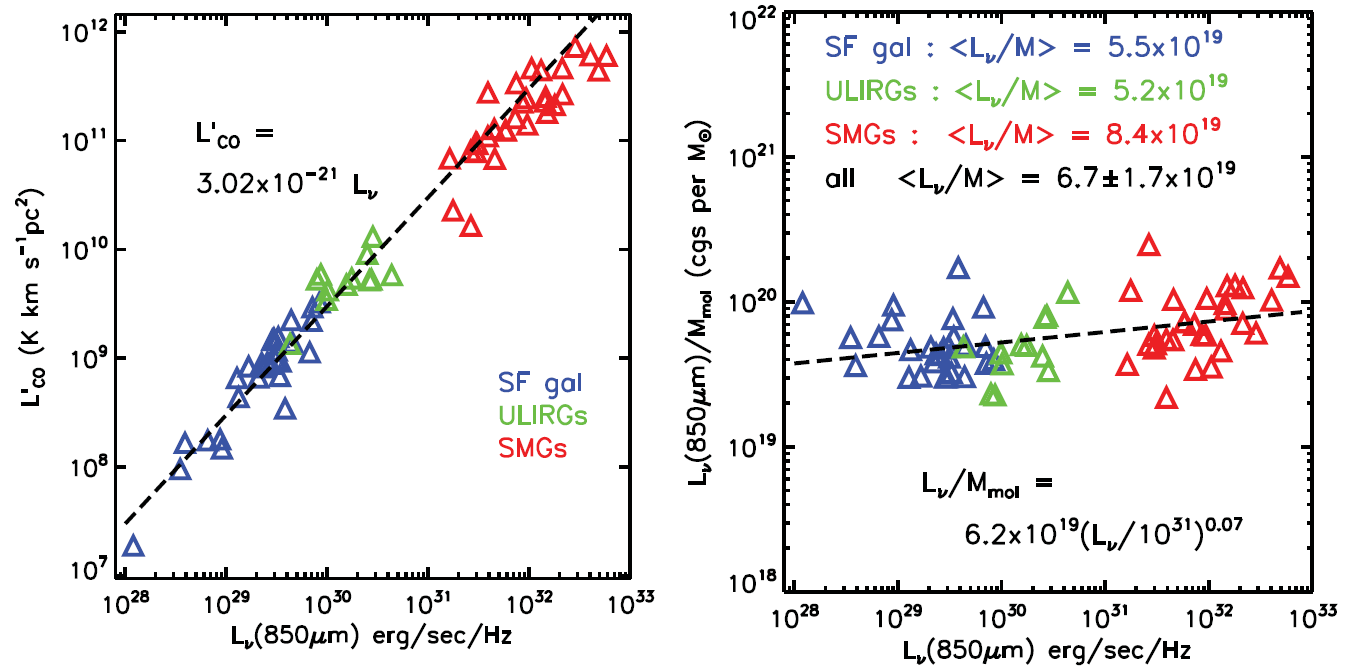}
\end{center}
  \caption{Comparison between the dust continuum and CO line tracers of the
    interstellar gas. The left panel compares the 850$\mu$m and CO luminosities for
    normal low-z star forming galaxies (SF gal), low-z ULIRGs, and z$\sim$ 2
    submillimeter galaxies (SMG, many of them are lensed).
    At right, the ratio between these two luminosities (L'$_{CO}$ being converted
    to M$_{mol}$) shows in more detail an almost constant proportionality factor.
    The conversion factor used is M$_{mol}$ = 6.5 L'$_{CO}$ [K km/s pc$^2$].
    Image reproduced with permission from \cite{Scoville2016}, copyright by AAS.}
\label{fig:Dust-CO} 
\end{figure*}

Another difference between dust and CO tracers, is that the former traces both
atomic and molecular gas. A calibration experiment has shown however 
that at low and high redshift the two main tracers of the molecular/interstellar gas
agree very well, see Figure \ref{fig:Dust-CO}.

\section{Galaxy surveys at high redshift}
\label{sec:stat}
Although ALMA is not a large survey facility, it is of critical importance
to gather the properties (gas content and excitation, star formation rate, etc.)
of a large number of objects, to gain a statistical significance, and to be able
to split the samples in several categories, to explore the influence of parameters.
Essential in the value of surveys are the selection criteria, for them to be representative
even if flux-limited.  Surveys have been done with carefully chosen
criteria from multi-wavelength studies: these are pointed surveys when
sources are already well-known (position, z, stellar mass, SFR).
Another type of surveys is the deep field method, unbiased, but sometimes less
successful, according to the choice of sensitivity required (shallow, deep), and the choice
of tuning (continuum, lines, etc.). Blind surveys have however the immense advantage
to be completely unbiased by other wavelengths, with the hope to detect brand new objects,
obscured in the optical and UV. Both are now described in turn.

\begin{figure*}
\begin{center}
  \includegraphics[width=0.95\textwidth]{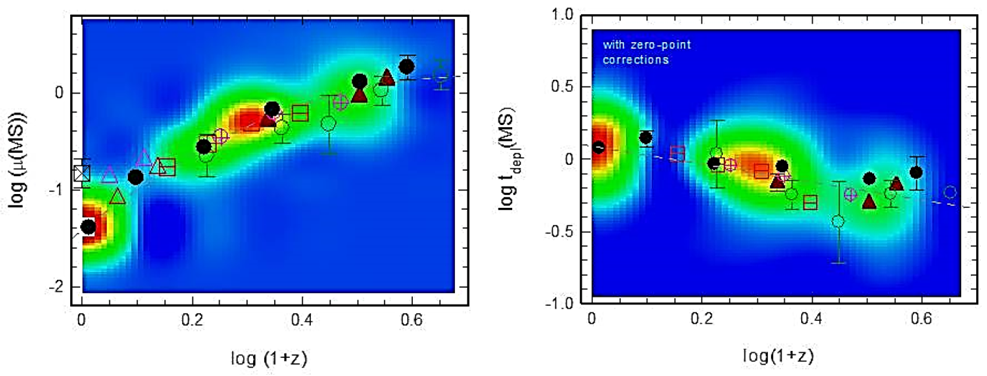}
\end{center}
\caption{Scaling relations of $\mu_{gas}$=M$_{molgas}$/M$_*$ with redshift 
(left), and depletion time t$_{dep}$ = M$_{molgas}$/SFR (right), 
for the binned data sets (large symbols) and the individual data points 
(colored distributions), from \cite{Tacconi2018}. All available data from
NOEMA and ALMA have been taken into account, with zero point corrections.
    Image reproduced with permission from \cite{Tacconi2018}, copyright by AAS.}
\label{fig:scaling-2018} 
\end{figure*}

\subsection{Pointed surveys}
\label{sec:pointed}

Optical/IR surveys have shown that the cosmic star formation density
had a peak at about z=2, and then has dropped by a factor $\sim$ 20
\citep[e.g.][]{Madau2014}. It was also discovered that, although
luminous and ultra-luminous infrared galaxies (ULIRGs) tend to dominate more and more the star formation
as z approaches 1 \citep{LeFloch2005}, the starburst mode is not the dominant star
forming mode at z$\sim$ 2, but contributes only by 10\%.
The confusion arose at the start because local ULIRGs are all starbursts, due to galaxy
interactions and mergers \citep[e.g.][]{Sanders1996}.
Starbursts can be defined as transient states, with an elevated  star formation rate (SFR)
which cannot be sustained during the "normal" time-scale to consume the gas content
of  a galaxy, which is 2 Gyr  \citep{Bigiel2008}. This time-scale is also called the depletion
time t$_{dep}$.
When the redshift increases, the average SFR increases, and LIRGs or ULIRGs do not
require anymore the starburst mode for their interpretation. It is then possible to define a
Main Sequence (MS) of ``normal'' star formation \citep{Whitaker2012, Whitaker2014, Speagle2014}.
The specific SFR, divided by the stellar mass, i.e. sSFR=SFR/M$_*$ increases strongly with
redshift, in (1+z)$^3$ up to z$\sim$ 2, and decreases only slightly with mass,
as M$_*^{-0.1,-0.4}$ \citep{Lilly2013}.  About 90\% of the star formation in the Universe occurs
on the main sequence, essentially in exponential galaxy disks  \citep[e.g.][]{Wuyts2011}.

The evolution of galaxies along the main sequence (MS), and the evolution of
the main sequence itself with redshift, have been the object of many models, implying
both gas accretion to re-fuel galaxies after the depletion times of the order of a few Gyr,
and the moderation of star formation due to the feedback. Some propose a quasi-stationary
state \citep{Bouche2010, Lilly2013}, and others a violent evolution across the MS, passing
through a starburst phase or a quiescent one, via violent instabilities and compaction
\citep{Dekel2009, Tacchella2016}.

It is crucial to investigate through the abundance of the molecular component
for galaxies on the MS, as a function of redshift, what are the main processes
regulating the star formation, and galaxy evolution. A large number of surveys have
been carried out, for instance the PHIBSS survey with NOEMA, or the
COSMOS with ALMA from dust emission, which demonstrate
a large increase of the molecular gas fraction with redshift \citep{Tacconi2010,Tacconi2013,
  Scoville2014, Scoville2016}, but also a slight decrease of the depletion time, i.e.
an increase of star formation efficiency.

In these surveys, two different tracers for the interstellar gas have been used: the most direct
one, the CO lines, depending on the CO-to-H$_2$ conversion factor, and the dust emission in the
Rayleigh-Jeans domain, which traces both atomic and molecular gas, but depends also
on metallicity, and on the assumed dust temperature, albeit in a linear way
\citep{Scoville2014, Schinnerer2016}.  The proportionality factor between dust emission and gas mass
is established from nearby galaxies and the Milky Way (through the Planck satellite data),
but several parameters may vary, as the slope of the dust opacity with frequency,
the metallicity and dust abundance, or the nature of dust.
In the literature, it was found that the gas masses derived at high redshift from the dust emission
are somewhat larger than that from the CO lines, at least by a factor 2.
Part of the explanation could be that dust emission traces both atomic
and molecular gas.
Although it is very difficult to have direct estimation of the HI gas at high z,
the best estimation comes from the Ly$\alpha$ absorbers along the line of sight
towards remote quasars  \citep[e.g.][]{Prochaska2005}, and it appears that the cosmic
density of HI gas is roughly constant. Given that the molecular gas strongly
increases with z, it is assumed that it will dominate as soon as z$>$ 0.5
\citep[e.g.][]{Lagos2012}.

From a compilation of all literature data on molecular gas at high z,
around the main sequence and slightly above, scaling relations have now been
derived for about 1400 objects \citep{Genzel2015,Tacconi2018}. The main
results are a quantification of the increase of gas fraction with z, and decrease
of depletion time, as shown in Fig. \ref{fig:scaling-2018}.
These scaling relations take into account all the various parameters
(distance from the MS, defined as $\delta$MS = sSFR/sSFR(MS,z,M$_*$),
redshift, stellar mass),
exploiting the fact that the dependence of gas fraction and depletion
time (or SFE, the Star Formation Efficiency= 1/t$_{dep}$) on 
these parameters is uncorrelated to some extent,
i.e. the variables are separable.

The variation of the depletion time with $\delta$MS is clear, at a given z and M$_*$,
t$_{dep}$ is decreasing for galaxies above the MS in the starburst phase, and
increasing below in the quenching phase. 
 The main results found by all surveys and analysis is that the gas content in galaxies
(i.e. the gas to stellar mass ratio)  increases significantly with z, as $\sim$ (1+z)$^2$, but still lower
than the SFR on the MS, which is increasing as $\sim$ (1+z)$^3$ up to z=3-4
\citep{Scoville2017,Tacconi2018}. In addition, the star formation efficiency (SFE)
is increasing with z, i.e. the depletion time varies as (1+z)$^{-0.6}$ to (1+z)$^{-1}$.
There is no dependency of SFE with stellar mass, on the main sequence.
The gas fraction, or the gas-to-stellar mass ratio decreases with stellar mass.
This explains also the variation of the slope of the MS with stellar mass:
if the SFR is almost linear with M$_*$ for small masses, it then saturates
and the slope is lower than 1. Since high mass galaxies have a more
massive bulge, and bulges do not participate to star formation, it is
tempting to subtract the bulge mass, to check the SFR variation of
disk only. \cite{Abramson2014} performed the bulge-disk decomposition
for large samples of low-z galaxies in the Sloan survey, and indeed, the MS
slope is almost vanishing (and vanishes completely in some samples). 
The sSFR of disks only is quasi independent of their stellar mass. The
remaining dependency could be related to the central bulge 
concentration \citep{Pan2016}.

\begin{figure*}
\begin{center}
  \includegraphics[width=0.45\textwidth]{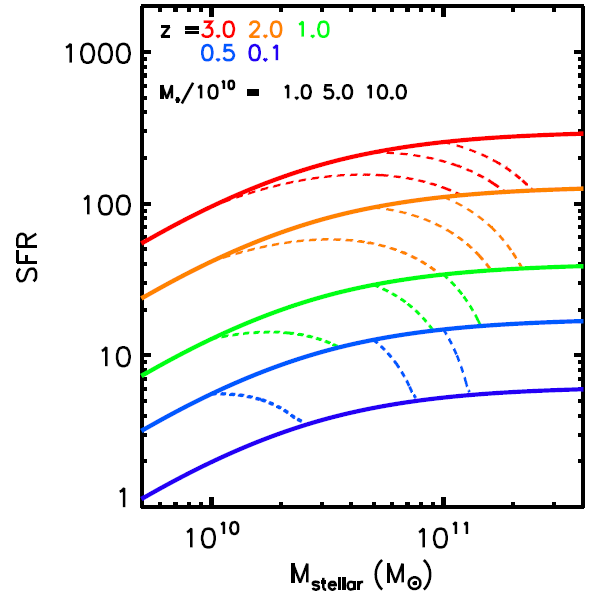}
  \includegraphics[width=0.45\textwidth]{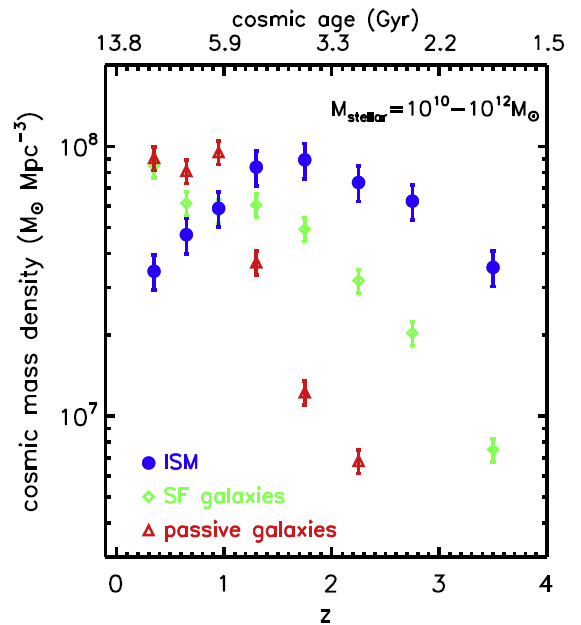}
\end{center}
\caption{{\it Left}: Evolution of galaxies on the main sequence (MS),
assuming they evolve continuously on the MS. The full 
lines show the MS relations at 5 different redshifts (colours) from the 
observed consensus compiled by \cite{Speagle2014}. The dash lines show
the time evolution of 1, 5, and 10$\times$10$^{10}$ M$_\odot$ galaxies in
between two redshifts. They evolve downward and rightward with time,
at a rate 0.7 SFR, taking into account 30\% of stellar mass loss.
{\it Right}: The gas (blue) and stellar mass (green and red) cosmic densities versus redshift
for galaxies in the range M$_*$ = 10$^{10}$ to 10$^{12}$ M$_\odot$.
These computations have used the observed scaling relations
between gas mass and the 3 parameters (z, $\delta$MS, M$_*$), and the
stellar mass functions from \cite{Ilbert2013}.
    Images reproduced with permission from \cite{Scoville2017}, copyright by AAS.}
\label{fig:Model-H2-z}
\end{figure*}

All these results are supporting models where galaxy evolution 
and star formation are mainly driven by external gas accretion.
Several interpretations have been elaborated \citep{Berta2013, Scoville2017}.
It is possible to trace the evolution of galaxies,
their star formation rate and gas content, assuming continuity,
 neglecting in a first step the contribution of starbursts (5-10\%), and the quenched galaxies, 
which must be of very high mass, and in dense environments \citep{Peng2010}.
 To maintain the evolution of the MS, it is necessary that galaxies are continuously
accreting gas, to refuel their SFR, given their low depletion time-scales,
lower than 1Gyr. This refueling can be mostly due to cold gas accretion \citep{Dekel2013},
but also may include minor mergers. The major mergers are considered to be
exceptional events, making galaxies to exit the MS from above for a transient
period. Using the empirically determined relations between gas content and SFE with
redshift, stellar mass and offset from MS, and assuming dM$_*$/dt =0.7 SFR 
(30\% of the gas is returned to the interstellar medium through stellar mass
loss), it is possible to trace the evolution of individual 
galaxies on the MS diagram (cf Figure \ref{fig:Model-H2-z}). 
The set of equations can be closed,
and the net accretion rates can be derived as a function of 
z and the main parameters. For the average considered mass, 
the accretion rate increases as $\sim$ (1+z)$^{3.5}$,
which may explain the high SFR in early galaxies, and is justified by the high 
gas density in the early universe.

Adopting these simplifying hypotheses, and summing over the 
stellar mass function \citep[e.g.][]{Ilbert2013}, the equations can yield the 
evolution of the cosmic density of the total gas in galaxies 
(H$_2$ +HI, traced by dust emission,
result given in Figure \ref{fig:Model-H2-z}).

Lensed galaxies allow to explore a lower mass regime (M$_*<$2.5 10$^{10}$ M$_\odot$), with
lower SFR ($<$ 40 M$_\odot$/yr) \citep{Dessauges2015}. It is now possible to see 
the star formation efficiency decrease with stellar mass. This low mass regime
reveals the same increase of gas fraction and SFE with redshift. With both CO lines and
dust continuum emission, it is possible to see large variations of the dust-to-gas 
ratio among the various types of star forming galaxies, even at a given metallicity. 

Observations of some gas-rich galaxies below the MS suggests that
quenching does not require the total removal or depletion of molecular gas, 
as many quenching models propose \citep{Suess2017}.
\cite{Spilker2018} detected with ALMA the CO line in 4 out of 8 
z$\sim$0.7 passive galaxies
3-10 times below the MS. Their gas fraction is below 10\%, small enough that
the depletion time is rather short. The gas rotation axis is aligned on 
the stellar one, implying no recent gas accretion. Even though the samples
are still not enough to draw firm conclusions, it appears that the quenching 
towards forming massive red and dead galaxies is rather slow, and due
to the cessation of gas accretion.

\begin{figure*}
\begin{center}
  \includegraphics[width=0.95\textwidth]{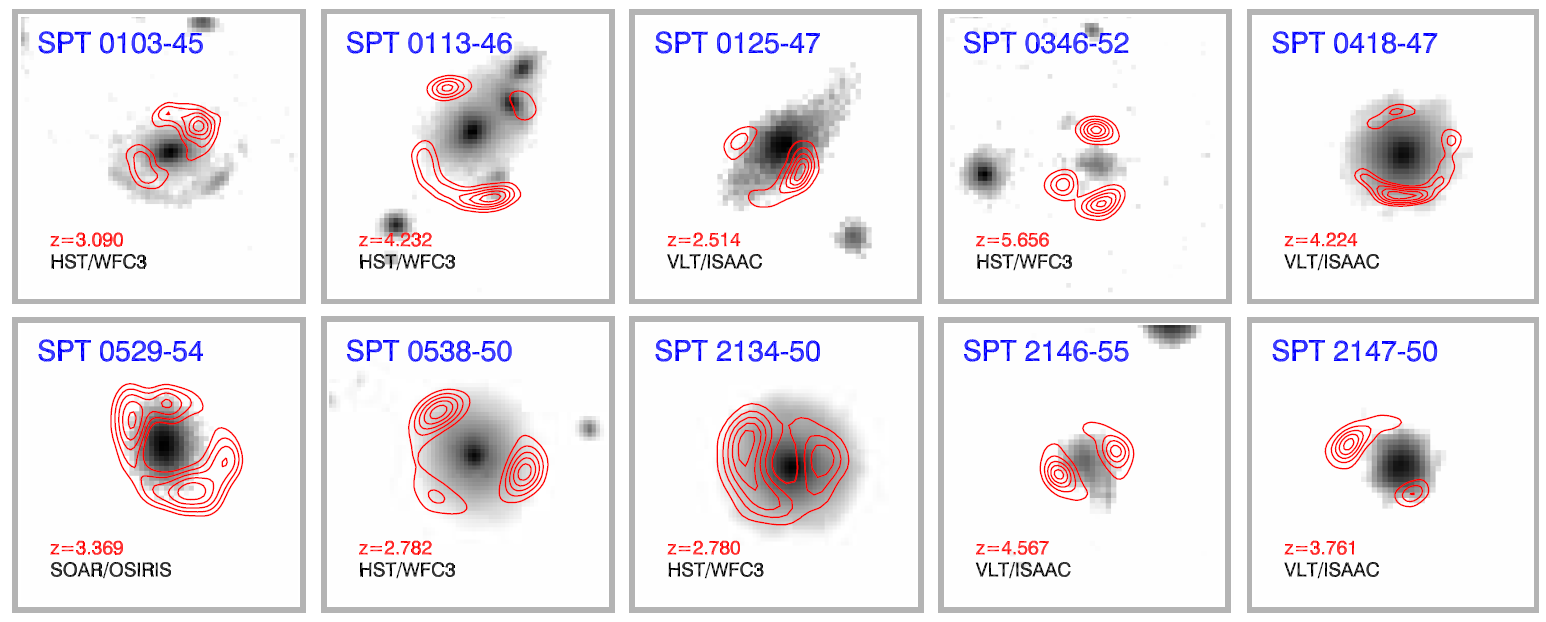}
\end{center}
\caption{ALMA 870$\mu$m images of 10 SPT (South Pole Telescope) sources
(red contours), superposed on near-infrared images (NIR, grey-scale) from HST,
VLT or SOAR telescopes. The NIR indicates the starlight from ths foreground lensing 
galaxies. The high-z galaxies are only seen by ALMA. Their spectroscopic redshifts 
were determined by ALMA CO line observations, and are
shown in red in each panel, of size 8"$\times$8". 
Image reproduced with permission from \cite{Vieira2013}, copyright by Springer.}
\label{fig:SPT2013} 
\end{figure*}

\begin{figure}
\begin{center}
  \includegraphics[width=0.5\textwidth]{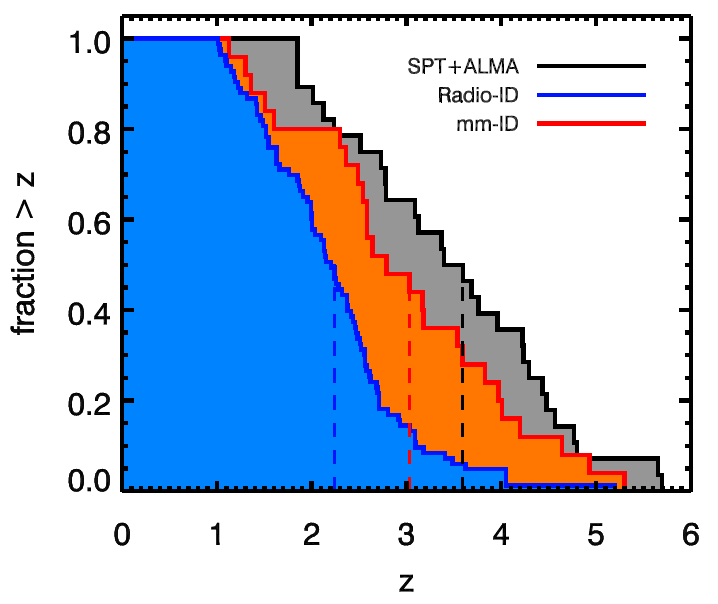}
\end{center}
\caption{The cumulative redshift distribution of luminous, dusty starburst 
galaxies: the SPT galaxies, with ALMA determined redshifts, are shown in black.
The blue sample objects have 
redshifts determined from rest-frame ultraviolet spectroscopy.
The orange sample galaxies in the COSMOS survey have only photometric
redshifts from optical/IR. ALMA detected a large fraction of high-z
dusty starburst galaxies, and previous surveys were  
biased to lower redshift than the underlying population.
    Image reproduced with permission from \cite{Vieira2013}, copyright by Springer.}
\label{fig:SPT-z} 
\end{figure}

One of the very successful surveys of ALMA in its first cycle (16 antenna) was to search for CO lines
in the high-z (z $>$1) sample of dusty continuum sources, assembled over 1300 square
degrees by the South Pole Telescope (SPT). The survey benefitted from the negative
K-correction, and therefore all redshifts were expected with minimum bias.
Most of the highest flux sources are lensed. Out of 26 sources, ALMA detected 23
in one CO line (among them 12 with multiple lines, so that the redshift is clearly determined).
In the continuum at 870$\mu$m, with spatial resolutions 0.5 - 1.5'', only one minute integration
per source was sufficient to reveal the arc and ring morphology of the lensed background
objects, as displayed in Figure \ref{fig:SPT2013}. Star formation rates larger than 500 M$_\odot$/yr
imply that the sources are ULIRGs \citep{Vieira2013}.
The spectroscopic survey in Band 3 more than doubled the known redshifts at this epoch,
and had a median redshift of z=3.5. The fraction of dusty and luminous starbursts at high z appears
higher than previously thought (see Figure  \ref{fig:SPT-z}).

\subsection{ALMA deep fields}
\label{sec:deep}

In the UV/optical/IR domains, considerable knowledge on galaxy evolution has
come from the study of blank fields, integrating deeply in selected regions of the sky
minimizing foregrounds, with the HST (HDF, UDF, XDF, \cite{Illingworth2013}),
and also with a multitude of instruments at all wavelengths, from the X-ray
(Chandra/XMM), to the far-infrared (Spitzer, Herschel) and radio (VLA). From 
the 11 HST filters, it has been possible to obtain nearly 10 000 photometric
redshifts \citep[e.g.][]{Rafelski2015}. Follow-up from the ground has obtained
also spectroscopic redshifts, namely with the VLT \citep{LeFevre2004, Bacon2017},
although the latter spectro-z still amount to less than 2\% of the total.
These surveys have allowed precious knowledge on galaxy properties
(sizes, stellar masses, star formation rates), and their evolution with 
redshift \citep[e.g.][]{Madau2014}.
However, to understand galaxy evolution, the fuel of star formation, the molecular
gas, has to be observed. Also optical surveys are biased against the
most obscured and dusty star forming galaxies, and sub-mm surveys are needed.
Already pointed observations have shown that indeed dusty starbursts exist 
up to z=6 \citep{Riechers2013}, and surveys with Herschel \citep{Elbaz2011}, or SCUBA-2 
 \citep{Coppin2015} have used priors to tackle blending, 
and stacking to explore just below the sensitivity limit of their instruments.

With ALMA gaining a large factor in sensitivity and spatial resolution,
deep surveys are now eagerly expected. The first survey of the Subaru-XMM
(SXDF) deep field (1.5 arcmin$^2$ with ALMA) reported by \cite{Tadaki2015}
has observed in 1.1mm continuum, with a sensitivity of $\sigma$ = 55$\mu$Jy/beam.
They targetted 12 H$\alpha$-selected star-forming galaxies (SFG) at z=2-2.5, but 
detected only 3 of them. The frequency corresponds to 300-400$\mu$m in the rest-frame,
so that dust emission should be easy to detect. It also corresponds to
100$\mu$m, the peak of dust emission, for z=10, and objects with the same
mass should be even more easy to detect with the K-correction, so that their
absence indicates a drop in the luminosity function with z. 
One of the objects detected is very
compact (R$_e$ = 0.7 kpc), with a high gas fraction of 44\%. In the same
ALMA field \cite{Hatsukade2016} conclude from the possibly detected
 23 sources above 0.2mJy that the source count is typical, and comparable
to all previous ALMA serendipitous detections.
 
\begin{figure*}
\begin{center}
  \includegraphics[width=0.95\textwidth]{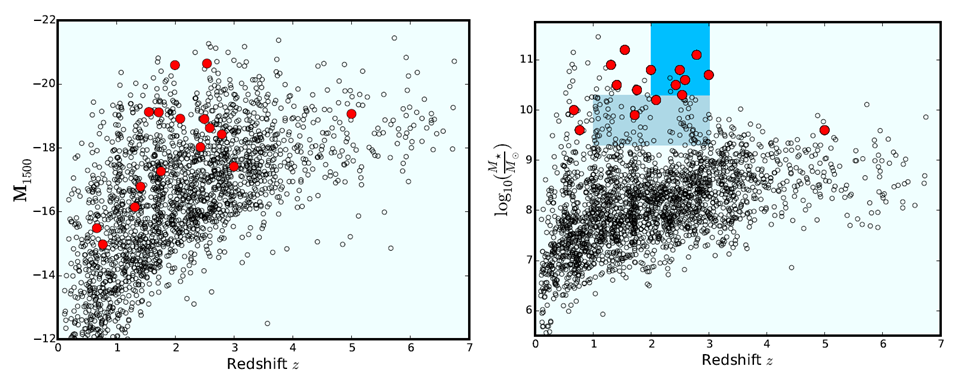}
\end{center}
\caption{{\it Left:} the UV absolute magnitude of all galaxies in the HUDF, as a 
function of redshift, with the ALMA 1.3mm detections in red. {\it Right:}
when stellar masses are considered, now the ALMA detections are at the top,
meaning that they are indeed the most obscured in UV. The bright blue box
emphasizes the ALMA detection of 80\% of the most massive galaxies 
(M$_* >$ 2 10$^{10}$ M$_\odot$) at z$>$2. Below z=1, the detection
rate of massive galaxies drops, which indicates possible quenching.
The grey-blue box gathers the galaxies which have been stacked and
lead to an ALMA global detection. The absence of very massive galaxies
above z=3 is clearly visible. 
    Image reproduced with permission from \cite{Dunlop2017}.}
\label{fig:HUDF-2017} 
\end{figure*}

\cite{Dunlop2017} reports about the 4.5 arcmin$^2$ ALMA survey of the HUDF
at 1.3mm at $\sigma$ = 30$\mu$Jy sensitivity. The extraction of reliable sources
in continuum is difficult. About 50 sources are first found above 3.5$\sigma$, but 
around 30 are also found in negative, i.e. below -3.5$\sigma$. Therefore most of the 50 sources 
must be spurious. Comparing with other data, 16 detections are then secured, 
through counterparts with HST, infrared and/or radio-cm, 13 of them 
having a spectroscopic redshift in the optical. The average redshift is z=2.15
and only one source has z $>$ 3. The lack of high-redshift detections confirms
 the rapid drop-off of high-mass galaxies in the field, above z=3.
Figure \ref{fig:HUDF-2017} shows clearly that the ALMA detections
are among the most UV-obscured objects in the HUDF.

\cite{Aravena2016} have carried out a deeper 1.2mm ALMA survey of the HUDF in a 
restricted region of 1 arcmin$^2$, with $\sigma$ = 13$\mu$Jy sensitivity.
They detect 9 sources at 3.5$\sigma$ with average z=1.6, and only one source above z=2,
which is significantly lower than the shallower survey of \cite{Dunlop2017}.
The detections correspond to 55\% of the extragalactic background light (EBL) 
at 1.2mm measured by the Planck satellite; when stacking all
the sources optically known in this region, it is possible to recover 80\% of
this EBL.

\begin{figure*}
\begin{center}
  \includegraphics[width=0.95\textwidth]{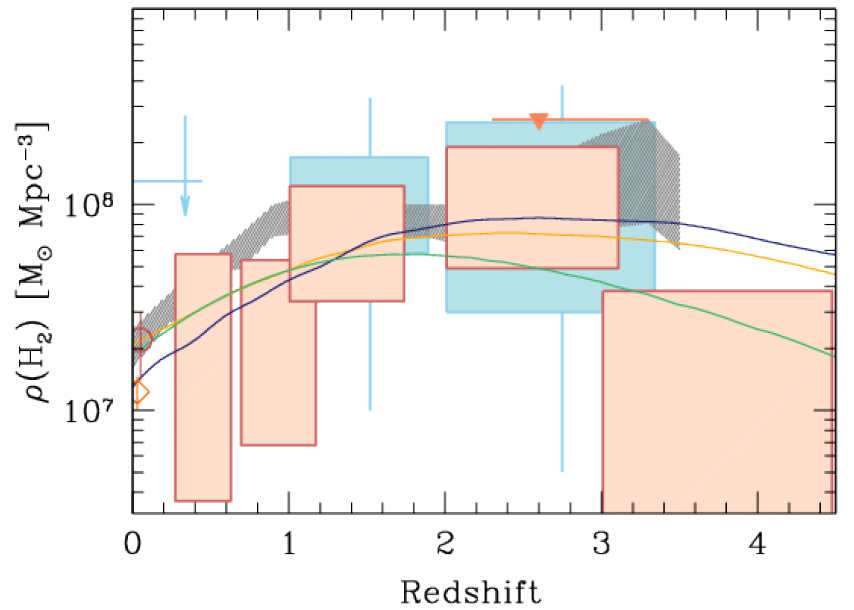}
\end{center}
  \caption{Comoving mass density of molecular gas in galaxies $\rho$(H$_2$)
    as a function of redshift. The ALMA spectroscopic survey (ASPECS) constraints
    are plotted in pink boxes, with vertical sizes corresponding to the uncertainties.
    The blue boxes represent the IRAM interferometer constraints  \citep{Walter2014}.
    The predictions of semi-analytical models are superposed as a yellow line
    \citep{Obreschkow2009}, a blue line  \citep{Lagos2012} and a green line
    \citep{Popping2014}. The compilation of literature data on MS galaxies
    is the grey area  \citep{Sargent2014}. The circle symbol at z=0 is from
    \cite{Keres2003}, and the losange from  \cite{Boselli2014}.
    \cite{Keating2016} have computed an upper limit (orange triangle)
    from CO intensity mapping at z$\sim$3. The global behaviour of the H$_2$
    cosmic density is very similar to the star formation density, with a peak
    around z=2. 
      Image reproduced with permission from \cite{Decarli2016a}, copyright by AAS.}
\label{fig:H2cosmic} 
\end{figure*}

In addition to this continuum survey of 1 arcmin$^2$ of the HUDF, the same team
carried out an ALMA spectroscopic survey (ASPECS, CO and [CII] lines) 
in two frequency bands at 3mm and 1mm,
covering the frequency ranges 84-115 GHz, and 212-272 GHz \citep{Walter2016}.
A blind search for lines have found 10 candidates in the 3mm band and 11 at 1mm.
The identification of the sources is then done searching for optical/NIR counterparts,
with a known redshift. This occurs in 9 out of the 21 candidates.
In one or two cases, other CO lines at higher J are also detected in the same survey, and
confirm the identification. Most of the times, the lack of other lines suggest that
the redshift of the object is large and/or the upper level J of the CO line is large.
In addition, stacking has been done for all sources with known redshifts for the
first 4 CO lines, but with no detection  \citep{Decarli2016a}.
Molecular masses were derived for each of the identified sources, and
found compatible with previous results for main sequence galaxies,
with a large scatter \citep{Decarli2016b}.
All results and constraints on the derived cosmic H$_2$ density are
gathered in Figure \ref{fig:H2cosmic}.

From the ASPECS survey, it is now possible to estimate the expected signal
from CO lines during an intensity mapping experiment. Based on individual
detections only, \cite{Carilli2016} estimate the mean surface brightness to
0.94 $\mu$K at 3mm and 0.55$\mu$K at 1.3mm, these values being lower
limits to take into account all the possible lines below detection.

\section{Individual galaxies at high redshift}
\label{sec:indiv}

Besides the large surveys with statistical value to explore galaxy evolution
with redshift, the discovery of special cases, pointed observations of high-z quasars,
and the study of over-densities, have provided a wealth of information.

 \subsection{Starburst and quasar associations}

 Objects already known as SMG with single dish continuum detectors
 were easily detected with ALMA, like this association of three LBG
 at z=5.3 studied by \cite{Riechers2014}. From the lines of [CII] and OH detected,
 an SFR surface density of 530 M$_\odot$/yr/kpc$^2$ was derived, implying
 a disk approaching the Eddington limit for radiation pressure on dust
 \citep{Miettinen2017, Crocker2018}. Since OH
 is slightly blue-shifted with respect to [CII], this might indicate a molecular
 outflow due to SN feedback.
 \cite{Swinbank2014} have made a survey with ALMA in the Extended Chandra Deep Field South (ECDFS)
 of 99 SMG, and found that they are all ULIRGs with 
SFR $\sim$ 300 M$_\odot$/yr and dust temperature of 32K.
 The contribution of these SMG to the cosmic star formation is about 20\% over z=1-4.

High redshift starbursts can be detected serendipitously, like the
bright z = 5.24 lensed submillimeter galaxy in the field of Abell 773 \citep{Combes2012},  
as part of the Herschel Lensing Survey (HLS, \cite{Egami2010}). This project surveyed
 a series of nearby galaxy clusters at z$\sim$0.1-0.5 playing the role of
 gravitational telescopes, amplifying background galaxies. These were selected by their
 very red SPIRE colours, implying a high redshift. Follow-up at millimeter wavelengths
 allows to discover the spectroscopic redshift, with the help of at least two detected lines.
 In this case several CO lines up to CO(7-6), CI, H$_2$O and the [NII]205$\mu$m lines were discovered, and
 allowed to constrain the variations of fundamental constants \citep{Levshakov2012}.
 With ALMA, spectroscopic redshifts are obtained routinely. 

An hyperluminous quasar at z=4.4 selected from WISE-SDSS was detected with ALMA by 
\cite{Bischetti2018} in dust continuum and [CII] line. It is at the center of a proto-cluster,
merging with two close companions. The quasar is actively forming stars with SFR $\sim$ 100 M$_\odot$/yr,
and the host galaxy will increase its stellar mass more rapidly than its black hole mass,
which is observed 2 orders of magnitude too massive, with respect to local relations.

 \cite{Venemans2017}  have detected several CO lines, CI and [CII] in z$\sim$ 7 quasars, and shown
 that these lines have excitation compatible with photodissociation regions, but not X-ray dominated regions.
 The properties of the molecular gas and dust in these quasars are dominated by an important star-formation activity,
 confirming that intense starbursts are co-existing with AGN activities.

\subsection{ Lensed high-z galaxies, high spatial resolution and GMC studies}
   
\begin{figure*}
\begin{center}
  \includegraphics[width=0.95\textwidth]{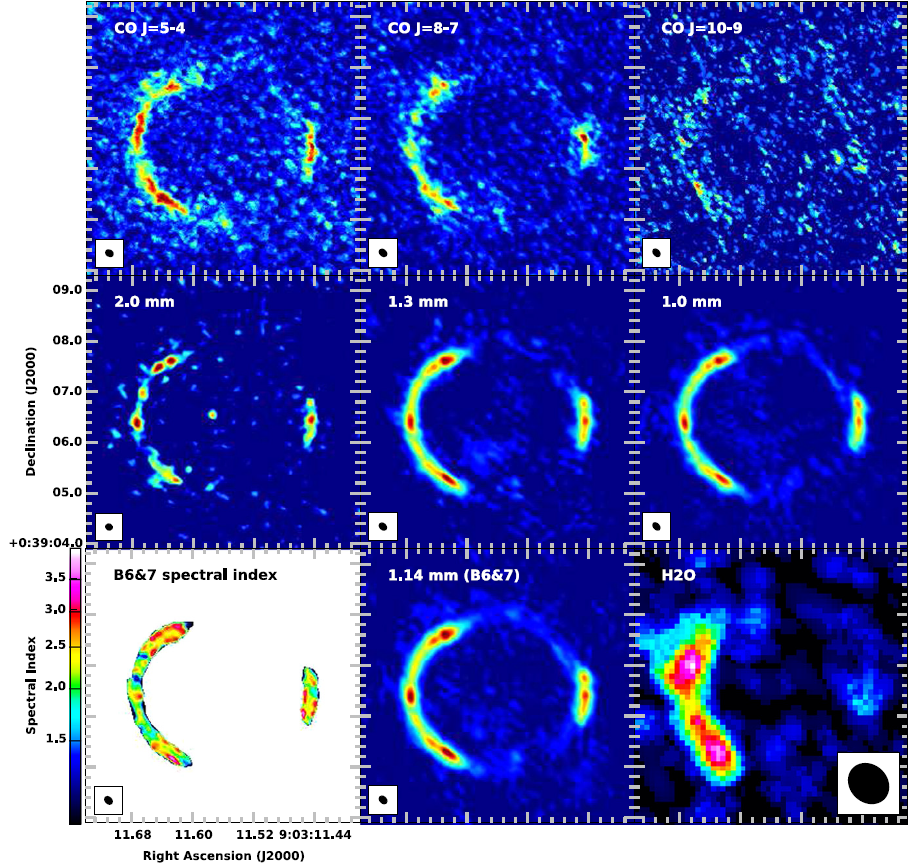}
\end{center}
\caption{ALMA images with high resolution (CO lines and continuum, 100-170mas) or 
lower resolution (H$_2$O line, 900mas). {\it Top:} CO J = 5 − 4, 8 − 7, and 10 − 9 integrated intensity.
{\it Middle:} 2.0, 1.3, and 1.0 mm continuum. {\it Bottom:} Band 6 and 7 spectral index, 
1.14 mm continuum (combined Band 6 and 7 data, and H$_2$O
integrated intensity. The beam sizes are indicated by the black ellipses at the
bottom of the panels.
  Image reproduced with permission from \cite{ALMA2015}, copyright by AAS.}
\label{fig:SDP81}      
\end{figure*}

ALMA can have very high spatial resolution in its extended configuration,
up to 15-20 milli-arcsec (mas).
A remarkable object was observed to demonstrate these capabilities,
with baselines up to 15km: SDP.81 \citep{ALMA2015}. This gravitationally
lensed galaxy at z=3.042 was discovered by the Herschel survey 
H-ATLAS \citep{Eales2010, Negrello2010}.
Its redshift was determined through CO lines; the lensing galaxy
is at z = 0.299, and the amplification factor is
$\mu$ = 11 \citep{Bussmann2013}.
With a beam of 25mas at 1mm (180pc at z=3.042), the ALMA continuum map reveals
the two gravitational arcs with unprecedented sharpness. 
Figure \ref{fig:SDP81} shows a tapered version of the maps in continuum, 
CO and H$_2$O lines; the images have been  tapered to lower resolution to gain
more signal to noise. The two arcs are part of an Einstein ring,
of radius 1.5". The foreground lensing galaxy is invisible
on these images, except for a weak continuum source
at the center of the ring, which has a spectral index consistent 
with synchrotron emission. The lensing galaxy is a massive elliptical
(3.6 10$^{11}$ M$_\odot$ inside the Einstein ring of 1.5"= 6.7 kpc,
at z=0.299, and no AGN is detected optically. But the 1.4 GHz flux
is compatible with the mm spectral index, and corresponds
to an AGN radio core.
The continuum from the arcs comes from dust emission in the 
high-z star forming galaxy, with an SFR = 527 M$_\odot$/yr.
The three CO lines detected (from J=5, 8 and 10) show regions
in the galaxy of different excitation, implying a complex structure.
The H$_2$O emission comes from a thermal line, which ratio with the
CO lines is rather weak, may be due to differential lensing.
The wealth of details acquired in $\sim$30h of telescope time 
in early science with only 22 to 36 antennae is quite impressive.

Given the enhanced spatial resolution due to lensing, it is possible to
explore the resolved Kennicutt-Schmidt relation (KS) in these high-z galaxies.
The surface densities of the molecular gas and star formation rate
have been compared in different regions of SDP.81 (z$\sim$3) \citep{Sharda2018}.
There is much more SFR than predicted from the linear KS relation,
and the authors propose another relation between gas and SFR, taking into
account the free-fall time of the clouds. Since the observed turbulence in the cloud
is much higher than for local galaxies, based on the observed high gas velocity dispersion,
a model of multifreefall based on turbulence \citep{Salim2015} is in better agreement with
observations.
Note that another attempt to derive the resolved KS relation for distant galaxies,
even without any lensing, has given results compatible with the linear KS relation
\citep{Freundlich2013}.

ALMA is now able to detect normal galaxies at z$\sim$ 7 \citep[e.g.][]{Maiolino2015}.
With the [CII] line and continuum dust emission, the detection of Lyman break galaxies
have been successful, implying SFR of 5-15 M$_\odot$/yr.  A spatial offset
of the order of 4 kpc has been observed between the [CII] emission and the
Ly$\alpha$ line and far UV, suggesting that stellar feedback rapidly
destroys/disperses the molecular clouds. \cite{Jones2017} have detected dust and [CII] emissions
in a Lyman break galaxy at z=6.1; they show that galaxies can form from the accretion of
small companions and gas both located in a filamentary structure.

\subsection{Black hole mass estimation at high-z}

A large number of quasars have been detected now at z$\sim$ 6 in molecules,
and their CO/[CII] kinematics can be used to derive the central dynamical mass.
From their broad lines detected in optical, and a widely known relation
between Broad Line Region (BLR) luminosity and radius calibrated from reverberation mapping
\citep[e.g.][]{Bentz2013}, it is possible to compare the black hole mass (M$_{BH}$), and
the host dynamical mass.
The high-z quasars appear all with a much higher black-hole mass 
than expected from their dynamical mass and the local M-$\sigma$ relation
\citep{Wang2013, Venemans2016}.
This surprising result could come from the large uncertainties of the mass
estimation. It is not possible to distinguish bulge and disk, so the M$_{BH}$ is compared
to the total host dynamical mass, but this is precisely conservative.  The inclination
of the rotating molecular disk is not well known, and the result is valid only statistically.
The M$_{BH}$ estimation has also a fudge factor for inclination. In most systems, it is
assumed that the [CII] or CO lines are centered on the systemic velocity, and can be reliably used
to derive the dynamical mass of the central stellar bulge. However, the
optical MgII broad emission lines are systematically blue-shifted. The average
blue-shift is of $\sim$ 500 km/s, but can be found up to 1700 km/s. This is
interpreted as an outflow due to the central AGN, given that the symmetrical
red-shifted region behind is too obscured to be seen.

\begin{figure}
\begin{center}
  \includegraphics[width=0.8\textwidth]{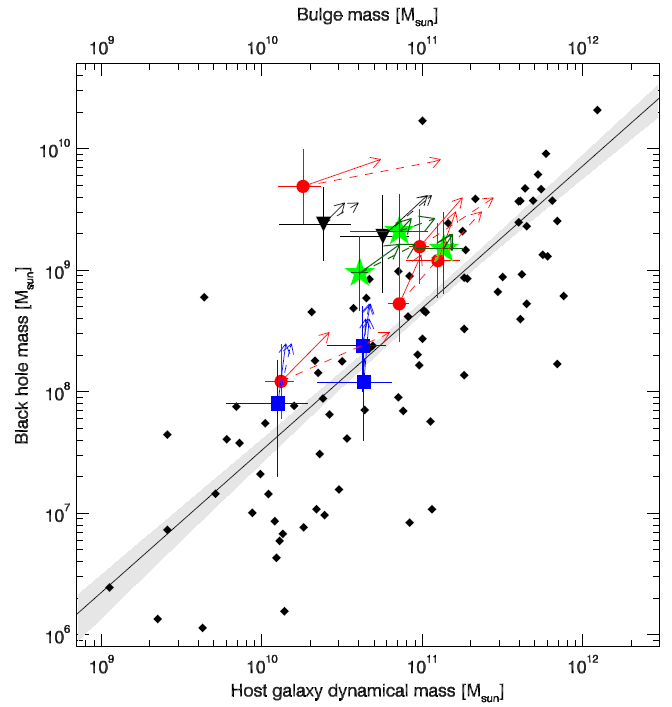}
\end{center}
\caption{Black hole mass versus the dynamical mass of z$\sim$6 quasar
host galaxies and the bulge mass of local galaxies. The black diamonds are
values obtained for local galaxies \citep{Kormendy2013}. Their 
M$_{BH}$ -- M$_{bulge}$ relation is represented by the solid line 
and the shaded area. 
The large and colored symbols are the high-z quasars.
The green stars are the z$>$6.5 quasars from \cite{Venemans2016}. For a given bulge
mass, the high-redshift quasars have a more massive black hole than local
galaxies. From the quasar luminosity (linked to its accretion rate), and from the
observed star formation rate, it is possible to extrapolate the trajectory of the 
points (arrows) during the next 50 Myr.
  Image reproduced with permission from \cite{Venemans2016}, copyright by AAS.}
\label{fig:MBH-16} 
\end{figure}

Even if the dynamical mass is not well estimated, it is possible to have a lower limit for it
with the mass of the gas, estimated from both the lines and the dust emission. In these high-z systems,
the gas fraction is often larger than 50\%, and the dynamical mass cannot be more
underestimated than by a factor 2. The derived black hole masses are then robustly 3-4 higher than expected
from the local relation (see Figure \ref{fig:MBH-16}).

\subsection{ Ly-alpha blobs and proto-clusters}
  
Protoclusters are overdensities in the early universe, where the growth of structures
and their accompanying black holes are accelerated. They are not yet virialised into
clusters, but are precious to understand why black holes might start growing
very quickly, and AGN feedback might shape the first galaxies. Narrowband
imaging at rest-frame Ly-$\alpha$ have revealed accumulation
of Ly-$\alpha$ emitters (LAE), but also extended 
($>$30 kpc) Ly-$\alpha$ emission (often termed Lyman-Alpha
Blobs, LAB)  \citep[e.g.][]{Steidel2000,  Matsuda2004}.
In these protoclusters, X-ray observations have revealed a significantly
higher (by a factor $\sim$ 5) fraction of AGN  \citep[e.g.][]{Lehmer2009},
suggesting a longer duty-cycle for black hole accretion in galaxies of rich environments.
\cite{Alexander2016} observed with ALMA such AGN in proto-clusters and
obtained a high detection rate, implying SFRs of 200-400 M$_\odot$/yr,
somewhat enhanced with respect to the field. This enhanced star formation
may explain the extended Ly-$\alpha$ emission of the LAB, given
a reasonable escape fraction for the continuum ionizing photons.

\begin{figure*}
   \begin{center}
     \includegraphics[width=0.45\textwidth]{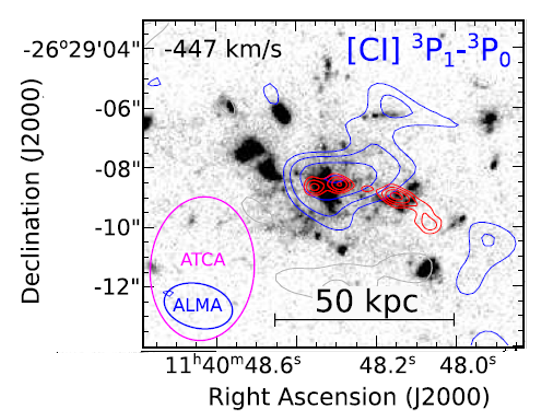}
       \includegraphics[width=0.49\textwidth]{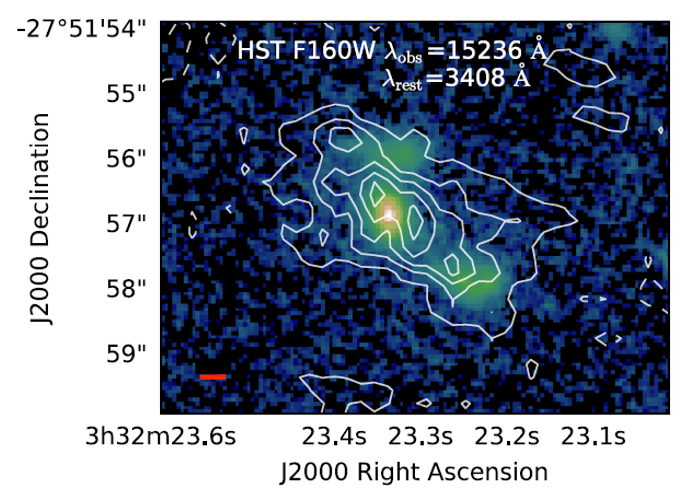}
\end{center}
  \caption{{\it Left:} One channel map (V=-447km/s, 90km/s wide) of the [CI] emission obtained with ALMA
    towards the Spiderweb proto-cluster of galaxies \citep{Emonts2018}. The blue contours are from
    [CI]$^3$P$_1$-$^3$P$_0$ emission, and the red contours from the 36 GHz radio continuum,
    both overlaid on the HST image.
    {\it Right:} Overlay of CO(4-3) contours on the HST F160W image of the Candels-5001 proto-cluster
    at z=3.47  \citep{Ginolfi2017}. The red bar is the HST PSF, at 0.34$\mu$m in the rest frame.
  Images reproduced with permission from \cite{Emonts2018} and \cite{Ginolfi2017}.}    
\label{fig:Spiderweb}      
\end{figure*}

Proto-clusters can also be the site of a colder gas phase,
which is extended as a circumgalactic medium (CGM) around the main galaxies.
 A striking example is the  Spiderweb, a conglomerate of merging galaxies at z=2.2
 \citep{Emonts2016}. Several CO lines and CI were observed with ALMA
 and ATCA, showing an extended network of clumps and filaments, with
 gas excitation similar to that of the Milky Way (see Figure \ref{fig:Spiderweb}).
The gas is metal enriched and dense, and
 most of it must have been recycled in galaxies coming from tidal and ram-pressure stripping,
and/or AGN and star formation feedback in the central region of the proto-cluster
 \citep{Emonts2018}.

 Molecular gas structures elongated on scales of $\sim$ 40 kpc are not rare in high-z proto-clusters, and
 a molecular mass of 2-6 10$^{11}$ M$_\odot$ has been detected in CO(4-3) and dust emission,
 with clumping and relatively high metallicity, at z=3.5 \citep{Ginolfi2017}. The extended structure is
 compatible with a tidal/ram pressure origin, but could also be fueled by some cold gas accretion
  (see Figure \ref{fig:Spiderweb}).

\section{Epoch of reionization}
\label{sec:EoR}

\begin{figure*}
  \includegraphics[width=0.95\textwidth]{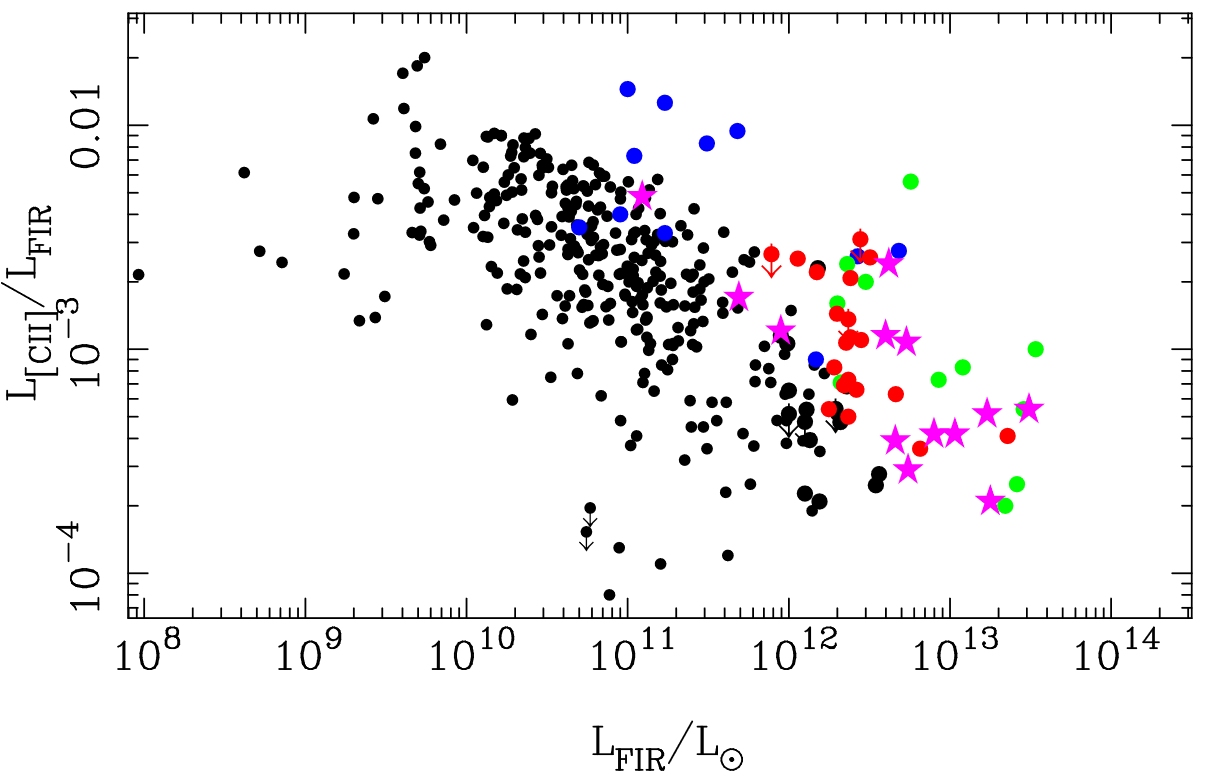}
\caption{The [CII] to FIR luminosity ratio versus the
FIR luminosity. The low-redshift galaxies are plotted as
black circles, from \cite{Malhotra2001}, \cite{Luhman2013}, and
\cite{Diaz2013}. Various ULIRGs at high redshift 
(z=1 to 6) detections from the literature
are in green circles, and the Hello sources (z=1-3) amplified
by lenses are in blue circles \citep{Malhotra2017}. The red circles
are the high-z SPT sources from \cite{Gullberg2015}, corrected for 
their amplification factor.
Quasars at z$>$4 are plotted as magenta stars \citep{Venemans2012, Venemans2016}.
At high z, the [C II]/FIR ratio still declines 
with FIR luminosity, but takes higher values than at z=0. 
}
\label{fig:plot-CII}      
\end{figure*}

\cite{Stark2016} has reviewed our knowledge of early galaxies,
in the first billion years after the Big-Bang. 
For z$>$6, ALMA yields a good opportunity to detect the dust emission,
provided that they are dusty enough. The peak of dust emission is indeed
shifted towards $\lambda >$ 0.7mm. For high-z objects, it
becomes difficult to obtain spectroscopic redshifts optically,
especially when obscured by dust. ALMA can then help to identify
the objects, thanks to the [CII] line at 158$\mu$m, redshifted to
 $\lambda >$ 1.1mm.
Models of the ISM had predicted that the main coolant would be through 
this [CII] line, however the observations reserved some surprises.
The photoelectric heating efficiency of the dust, measured by the 
ratio L$_{[CII]}$/L$_{FIR}$, varies by about 2 orders of magnitude,
and is decreasing at high L$_{FIR}$, for strong starbursts.
The main factor reducing this efficiency has been shown
to be the dust temperature, and the strong UV field \citep{Malhotra2017}:
indeed, the  L$_{[CII]}$/L$_{FIR}$ ratio is very well anti-correlated to the
dust temperature, whatever the redshift. Figure \ref{fig:plot-CII}
gathers a large fraction of the [CII] studies so far, and shows
that the [CII]/FIR ratio is higher at high redshift, although 
still declining with L$_{FIR}$. The high-z quasars detected reveal
a wide range of properties, sometimes behaving like starbursts,
while sometimes the quasar excitation may prevail \citep[e.g.][]{Venemans2016}.
 
 Many searches have been made with ALMA, with some surprising 
failures, indicating that galaxies are really "primordial",
with low metallicity (Z$<$0.1) and little dust. The typical
Ly$\alpha$ emitter Himiko was not detected in the continuum, nor in the [CII]
line \citep{Ouchi2013}. Several other upper limits
confirmed that most LBG between z=6 and 8 are
very difficult to detect, even with gravitational lensing
\citep[e.g.][]{Schaerer2015}. With more observations, Himiko is
now detected in the [CII] line, but not in dust emission
\citep{Carniani2018}, revealing a dust deficiency.

Evidence also exists of early galaxies, with their ISM extended
over kpc sizes, but weak or undetected in dust emission, while revealing 
strong [CII] line emission \citep{Capak2015}. These show some
similarity to low-metallicity dwarfs in the local Universe,
like the LMC, which have a very high L$_{[CII]}$/L$_{FIR}$ ratio.
However, these high-z objects are much more massive, which implies
on the contrary a rapid evolution in the ISM dust properties over
redshift, due to metal deficiency.
Selecting their objects with a relatively lower Ly$\alpha$
equivalent widths, indicating the presence of dust, at a given
UV luminosity, \cite{Willott2015} reports dust emission and
[CII] line detections at z$\sim$6. The velocity redshift of 
the Ly$\alpha$ with respect to the [CII] line is very prominent
at high-z, due to increased intergalactic gas (IGM) absorption
of the blue wing of Ly$\alpha$. The expected enhancement
of IGM absorption in the EoR, is not always
there \citep{Pentericci2016}, implying patchy reionization.

At high-z, galaxies are clumpy, and sometimes the [CII] line
and even the [OIII] line at 88$\mu$m are spatially offset from the
 Ly$\alpha$ or UV clumps \citep{Carniani2017, Matthee2017}. 
These offsets may be explained by obscuration, 
different excitation or metallicity of the different tracers.
Alternatively, strong feedback could have removed 
a large fraction of gas and dust, or several parts of the 
systems are interacting while assembling,
as suggested by theoretical models \citep{Katz2017}. 

Some of the highest redshifts found in the EoR with ALMA
are the z=8.38 gravitationally lensed galaxy selected from deep
HST imaging in the Frontier Field cluster Abell 2744
 \citep{Laporte2017}, or the Lyman Break galaxy at z=8.31 
behind the Frontier Field cluster MACS J0416.1-2403 \citep{Tamura2018}.
Dust emission and the [OIII] line have been
detected, raising the problem of forming such dust amounts
$\sim$ 600 Myr after the Big Bang. This would imply that each SN-II explosion
has been able to produce 0.5 M$_\odot$ of dust, during the SFR=15-20 M$_\odot$/yr
star forming phase, since z=10-12.

The highest redshift is MACS1149-JD1 at z=9.11, a lensed galaxy detected in the [OIII] line.
No redshift was known from the optical before, and the [OIII] line was used to measure the redshift.
The colors of its stellar population show that star formation began at z=15
in this galaxy \citep{Hashimoto2018}.

Contrary to many ALMA surveys, finding a drop in
their source number at high redshift,
\cite{Strandet2016} find a redshift distribution much more
weighted towards the high-z, because of a low-frequency selection.
Dusty sources were selected from the South Pole Telescope (SPT) survey,
from their 1.4mm continuum flux; eliminating the synchrotron sources,
by requiring 1.4mm flux being twice higher than the 2mm flux.

Although most of high-z star forming objects
selected optically have low dust content, 
exceptional objects exist, like HFLS3, at z=6.34,
with SFR=2900 M$_\odot$/yr, a gas mass of 10$^{11}$ M$_\odot$,
including 2 10$^{11}$ M$_\odot$ of atomic gas, and a depletion time
of 36 Myr \citep{Riechers2013}. These must be located in proto-clusters,
and are the progenitors of massive ellipticals in clusters today.

\section{Summary}
\label{sec:conclu}
ALMA has been working now for about 7 years since its
commissioning in 2011, and has accumulated a wealth of data, which have yet to
be digested and interpreted. In many domains, ALMA has provided impressive breakthroughs,
with unprecedented sensitivity and spatial resolution.

One of the main goals in galaxy evolution is to determine the molecular gas
properties of galaxies as a function of redshift, to better understand the cosmic
star formation history. This has been done through pointed observations,
with large sample of objects selected from their stellar mass and SFR, being on the main
sequence of star forming galaxies, where are born 90\% of the stars in the Universe.
Two main factors have been emphasized: the gas fraction increases steadily
on the main sequence, as (1+z)$^2$, and this is the main reason of the peak in the SFRD at z$\sim$ 2.
At a lower level, the star formation efficiency is also increasing with redshift,
as (1+z)$^\alpha$, with $\alpha$=0.6-1, (or the depletion time is decreasing,
as (1+z)$^{-\alpha}$), although the cause of this has not be
clearly identified: either due to smaller and denser galaxies, with a shorter dynamical
time, or due to a larger importance of galaxy interactions.
Although some galaxy molecular maps have been done, and resolved Kennicutt-Schmidt
relation explored, this is only the beginning, and the influence of morphology, kinematics and
dynamics of galaxies is not yet understood.

In parallel, deep blind surveys, completely unbiased by previous wavelengths, have been
carried out focussed either on the dust continuum, or the CO lines, with shallow or deeper
approaches, according to the surface covered. The hope was to detect dusty galaxies
at very high redshift, not suspected by other surveys. Eventually, dusty and massive galaxies
are not as frequent as previously hoped at z larger than 5, which confirms the high-z drop in  optical
surveys. ALMA is now opening clearly the windows of the epoch of re-ionization, and it is likely that
the main actors to reionize the Universe will turn out to be a large number of small and dwarf galaxies,
while the major starbursts and quasars have a minor influence.

ALMA surveys have begun to unveil the cosmic evolution of the H$_2$ content, but
this is only the beginning, with huge error-bars, not allowing to disentangle the various
theoretical models, as shown in Figure \ref{fig:H2cosmic}.
In the future, this cosmic evolution will be compared with the star formation
history, and also with the atomic gas content, to have a more precise budget of gas
and star formation at all epochs.

There remain a large number of unsolved issues, like the symbiotic evolution
of black hole and bulges in galaxies, which appear to be divergent at high redshift,
the importance of AGN feedback in the early universe, the influence of environment
in proto-clusters. The detection of important quantities of cold and dense gas in
the circumgalactic medium at early times might give some clues in the missing
baryon problem.

\begin{acknowledgements}
I thank Paul Ho for inviting me to write this review, and an anonymous referee
for constructive comments.
\end{acknowledgements}

\def\aj{AJ}%
          % Astronomical Journal
\def\actaa{Acta Astron.}%
          % Acta Astronomica
\def\araa{ARA\&A}%
          % Annual Review of Astron and Astrophys
\def\apj{ApJ}%
          % Astrophysical Journal
\def\apjl{ApJ}%
          % Astrophysical Journal, Letters
\def\apjs{ApJS}%
          % Astrophysical Journal, Supplement
\def\ao{Appl.~Opt.}%
          % Applied Optics
\def\apss{Ap\&SS}%
          % Astrophysics and Space Science
\def\aap{A\&A}%
          % Astronomy and Astrophysics
\def\aapr{A\&A~Rev.}%
          % Astronomy and Astrophysics Reviews
\def\aaps{A\&AS}%
          % Astronomy and Astrophysics, Supplement
\def\azh{AZh}%
          % Astronomicheskii Zhurnal
\def\baas{BAAS}%
          % Bulletin of the AAS
\def\bac{Bull. astr. Inst. Czechosl.}%
          % Bulletin of the Astronomical Institutes of Czechoslovakia
\def\caa{Chinese Astron. Astrophys.}%
          % Chinese Astronomy and Astrophysics
\def\cjaa{Chinese J. Astron. Astrophys.}%
          % Chinese Journal of Astronomy and Astrophysics
\def\icarus{Icarus}%
          % Icarus
\def\jcap{J. Cosmology Astropart. Phys.}%
          % Journal of Cosmology and Astroparticle Physics
\def\jrasc{JRASC}%
          % Journal of the RAS of Canada
\def\mnras{MNRAS}%
          % Monthly Notices of the RAS
\def\na{New A}%
          % New Astronomy
\def\nar{New A Rev.}%
          % New Astronomy Review
\def\physrep{Phys.~Rep.} % Physics Report
\def\pasa{PASA}%
          % Publications of the Astron. Soc. of Australia
\def\pra{Phys.~Rev.~A}%
          % Physical Review A: General Physics
\def\prb{Phys.~Rev.~B}%
          % Physical Review B: Solid State
\def\prc{Phys.~Rev.~C}%
          % Physical Review C
\def\prd{Phys.~Rev.~D}%
          % Physical Review D
\def\pre{Phys.~Rev.~E}%
          % Physical Review E
\def\prl{Phys.~Rev.~Lett.}%
          % Physical Review Letters
\def\pasp{PASP}%
          % Publications of the ASP
\def\pasj{PASJ}%
          % Publications of the ASJ
\def\qjras{QJRAS}%
          % Quarterly Journal of the RAS
\def\rmxaa{Rev. Mexicana Astron. Astrofis.}%
          % Revista Mexicana de Astronomia y Astrofisica
\def\skytel{S\&T}%
          % Sky and Telescope
\def\solphys{Sol.~Phys.}%
          % Solar Physics
\def\sovast{Soviet~Ast.}%
          % Soviet Astronomy
\def\ssr{Space~Sci.~Rev.}%
          % Space Science Reviews
\def\zap{ZAp}%
          % Zeitschrift fuer Astrophysik
\def\nat{Nature}%
          % Nature
\def\iaucirc{IAU~Circ.}%
          % IAU Circulars
\def\aplett{Astrophys.~Lett.}%
          % Astrophysics Letters
\def\fcp{Fund.~Cosmic~Phys.}%
          % Fundamental Cosmic Physics
\def\planss{Planet.~Space~Sci.}%
          % Planetary Space Science
\def\procspie{Proc.~SPIE}%
          % Proceedings of the SPIE

\bibliographystyle{aabasic}      % basic style, author-year citations
\bibliography{combes-alma-hiz.bib}   % name your BibTeX data base

\begin{thebibliography}{127}
\providecommand{\natexlab}[1]{#1}
\providecommand{\url}[1]{{#1}}
\providecommand{\urlprefix}{URL }
\expandafter\ifx\csname urlstyle\endcsname\relax
  \providecommand{\doi}[1]{DOI~\discretionary{}{}{}#1}\else
  \providecommand{\doi}{DOI~\discretionary{}{}{}\begingroup
  \urlstyle{rm}\Url}\fi
\providecommand{\eprint}[2][]{\url{#2}}

\bibitem[{{Abramson} et~al(2014){Abramson}, {Kelson}, {Dressler}, {Poggianti},
  {Gladders}, {Oemler}, and {Vulcani}}]{Abramson2014}
{Abramson} LE, {Kelson} DD, {Dressler} A, {Poggianti} B, {Gladders} MD,
  {Oemler} A Jr, {Vulcani} B (2014) {The Mass-independence of Specific Star
  Formation Rates in Galactic Disks}. \apjl 785:L36,
  \doi{10.1088/2041-8205/785/2/L36}, \eprint{1402.7076}

\bibitem[{{Alexander} et~al(2016){Alexander}, {Simpson}, {Harrison},
  {Mullaney}, {Smail}, {Geach}, {Hickox}, {Hine}, {Karim}, {Kubo}, {Lehmer},
  {Matsuda}, {Rosario}, {Stanley}, {Swinbank}, {Umehata}, and
  {Yamada}}]{Alexander2016}
{Alexander} DM, {Simpson} JM, {Harrison} CM, {Mullaney} JR, {Smail} I, {Geach}
  JE, {Hickox} RC, {Hine} NK, {Karim} A, {Kubo} M, {Lehmer} BD, {Matsuda} Y,
  {Rosario} DJ, {Stanley} F, {Swinbank} AM, {Umehata} H, {Yamada} T (2016)
  {ALMA observations of a z $\sim$ 3.1 protocluster: star formation from active
  galactic nuclei and Lyman-alpha blobs in an overdense environment}. \mnras
  461:2944--2952, \doi{10.1093/mnras/stw1509}, \eprint{1601.00682}

\bibitem[{{ALMA Partnership} et~al(2015){ALMA Partnership}, {Vlahakis},
  {Hunter}, {Hodge}, {P{\'e}rez}, {Andreani}, {Brogan}, {Cox}, {Martin},
  {Zwaan}, {Matsushita}, {Dent}, {Impellizzeri}, {Fomalont}, {Asaki},
  {Barkats}, {Hills}, {Hirota}, {Kneissl}, {Liuzzo}, {Lucas}, {Marcelino},
  {Nakanishi}, {Phillips}, {Richards}, {Toledo}, {Aladro}, {Broguiere},
  {Cortes}, {Cortes}, {Espada}, {Galarza}, {Garcia-Appadoo}, {Guzman-Ramirez},
  {Hales}, {Humphreys}, {Jung}, {Kameno}, {Laing}, {Leon}, {Marconi},
  {Mignano}, {Nikolic}, {Nyman}, {Radiszcz}, {Remijan}, {Rod{\'o}n}, {Sawada},
  {Takahashi}, {Tilanus}, {Vila Vilaro}, {Watson}, {Wiklind}, {Ao}, {Di
  Francesco}, {Hatsukade}, {Hatziminaoglou}, {Mangum}, {Matsuda}, {van Kampen},
  {Wootten}, {de Gregorio-Monsalvo}, {Dumas}, {Francke}, {Gallardo}, {Garcia},
  {Gonzalez}, {Hill}, {Iono}, {Kaminski}, {Karim}, {Krips}, {Kurono},
  {Lonsdale}, {Lopez}, {Morales}, {Plarre}, {Videla}, {Villard}, {Hibbard}, and
  {Tatematsu}}]{ALMA2015}
{ALMA Partnership}, {Vlahakis} C, {Hunter} TR, {Hodge} JA, {P{\'e}rez} LM,
  {Andreani} P, {Brogan} CL, {Cox} P, {Martin} S, {Zwaan} M, {Matsushita} S,
  {Dent} WRF, {Impellizzeri} CMV, {Fomalont} EB, {Asaki} Y, {Barkats} D,
  {Hills} RE, {Hirota} A, {Kneissl} R, {Liuzzo} E, {Lucas} R, {Marcelino} N,
  {Nakanishi} K, {Phillips} N, {Richards} AMS, {Toledo} I, {Aladro} R,
  {Broguiere} D, {Cortes} JR, {Cortes} PC, {Espada} D, {Galarza} F,
  {Garcia-Appadoo} D, {Guzman-Ramirez} L, {Hales} AS, {Humphreys} EM, {Jung} T,
  {Kameno} S, {Laing} RA, {Leon} S, {Marconi} G, {Mignano} A, {Nikolic} B,
  {Nyman} LA, {Radiszcz} M, {Remijan} A, {Rod{\'o}n} JA, {Sawada} T,
  {Takahashi} S, {Tilanus} RPJ, {Vila Vilaro} B, {Watson} LC, {Wiklind} T, {Ao}
  Y, {Di Francesco} J, {Hatsukade} B, {Hatziminaoglou} E, {Mangum} J, {Matsuda}
  Y, {van Kampen} E, {Wootten} A, {de Gregorio-Monsalvo} I, {Dumas} G,
  {Francke} H, {Gallardo} J, {Garcia} J, {Gonzalez} S, {Hill} T, {Iono} D,
  {Kaminski} T, {Karim} A, {Krips} M, {Kurono} Y, {Lonsdale} C, {Lopez} C,
  {Morales} F, {Plarre} K, {Videla} L, {Villard} E, {Hibbard} JE, {Tatematsu} K
  (2015) {The 2014 ALMA Long Baseline Campaign: Observations of the Strongly
  Lensed Submillimeter Galaxy HATLAS J090311.6+003906 at z = 3.042}. \apjl
  808:L4, \doi{10.1088/2041-8205/808/1/L4}, \eprint{1503.02652}

\bibitem[{{Aravena} et~al(2016){Aravena}, {Decarli}, {Walter}, {Da Cunha},
  {Bauer}, {Carilli}, {Daddi}, {Elbaz}, {Ivison}, {Riechers}, {Smail},
  {Swinbank}, {Weiss}, {Anguita}, {Assef}, {Bell}, {Bertoldi}, {Bacon},
  {Bouwens}, {Cortes}, {Cox}, {G{\'o}nzalez-L{\'o}pez}, {Hodge}, {Ibar},
  {Inami}, {Infante}, {Karim}, {Le Le F{\`e}vre}, {Magnelli}, {Ota}, {Popping},
  {Sheth}, {van der Werf}, and {Wagg}}]{Aravena2016}
{Aravena} M, {Decarli} R, {Walter} F, {Da Cunha} E, {Bauer} FE, {Carilli} CL,
  {Daddi} E, {Elbaz} D, {Ivison} RJ, {Riechers} DA, {Smail} I, {Swinbank} AM,
  {Weiss} A, {Anguita} T, {Assef} RJ, {Bell} E, {Bertoldi} F, {Bacon} R,
  {Bouwens} R, {Cortes} P, {Cox} P, {G{\'o}nzalez-L{\'o}pez} J, {Hodge} J,
  {Ibar} E, {Inami} H, {Infante} L, {Karim} A, {Le Le F{\`e}vre} O, {Magnelli}
  B, {Ota} K, {Popping} G, {Sheth} K, {van der Werf} P, {Wagg} J (2016) {The
  ALMA Spectroscopic Survey in the Hubble Ultra Deep Field: Continuum Number
  Counts, Resolved 1.2 mm Extragalactic Background, and Properties of the
  Faintest Dusty Star-forming Galaxies}. \apj 833:68,
  \doi{10.3847/1538-4357/833/1/68}, \eprint{1607.06769}

\bibitem[{{Bacon} et~al(2017){Bacon}, {Conseil}, {Mary}, {Brinchmann},
  {Shepherd}, {Akhlaghi}, {Weilbacher}, {Piqueras}, {Wisotzki}, {Lagattuta},
  {Epinat}, {Guerou}, {Inami}, {Cantalupo}, {Courbot}, {Contini}, {Richard},
  {Maseda}, {Bouwens}, {Bouch{\'e}}, {Kollatschny}, {Schaye}, {Marino},
  {Pello}, {Herenz}, {Guiderdoni}, and {Carollo}}]{Bacon2017}
{Bacon} R, {Conseil} S, {Mary} D, {Brinchmann} J, {Shepherd} M, {Akhlaghi} M,
  {Weilbacher} PM, {Piqueras} L, {Wisotzki} L, {Lagattuta} D, {Epinat} B,
  {Guerou} A, {Inami} H, {Cantalupo} S, {Courbot} JB, {Contini} T, {Richard} J,
  {Maseda} M, {Bouwens} R, {Bouch{\'e}} N, {Kollatschny} W, {Schaye} J,
  {Marino} RA, {Pello} R, {Herenz} C, {Guiderdoni} B, {Carollo} M (2017) {The
  MUSE Hubble Ultra Deep Field Survey. I. Survey description, data reduction,
  and source detection}. \aap 608:A1, \doi{10.1051/0004-6361/201730833},
  \eprint{1710.03002}

\bibitem[{{Bentz} et~al(2013){Bentz}, {Denney}, {Grier}, {Barth}, {Peterson},
  {Vestergaard}, {Bennert}, {Canalizo}, {De Rosa}, {Filippenko}, {Gates},
  {Greene}, {Li}, {Malkan}, {Pogge}, {Stern}, {Treu}, and {Woo}}]{Bentz2013}
{Bentz} MC, {Denney} KD, {Grier} CJ, {Barth} AJ, {Peterson} BM, {Vestergaard}
  M, {Bennert} VN, {Canalizo} G, {De Rosa} G, {Filippenko} AV, {Gates} EL,
  {Greene} JE, {Li} W, {Malkan} MA, {Pogge} RW, {Stern} D, {Treu} T, {Woo} JH
  (2013) {The Low-luminosity End of the Radius-Luminosity Relationship for
  Active Galactic Nuclei}. \apj 767:149, \doi{10.1088/0004-637X/767/2/149},
  \eprint{1303.1742}

\bibitem[{{Berta} et~al(2013){Berta}, {Lutz}, {Nordon}, {Genzel}, {Magnelli},
  {Popesso}, {Rosario}, {Saintonge}, {Wuyts}, and {Tacconi}}]{Berta2013}
{Berta} S, {Lutz} D, {Nordon} R, {Genzel} R, {Magnelli} B, {Popesso} P,
  {Rosario} D, {Saintonge} A, {Wuyts} S, {Tacconi} LJ (2013) {Molecular gas
  mass functions of normal star-forming galaxies since z \~{} 3}. \aap 555:L8,
  \doi{10.1051/0004-6361/201321776}, \eprint{1304.7771}

\bibitem[{{Bigiel} et~al(2008){Bigiel}, {Leroy}, {Walter}, {Brinks}, {de Blok},
  {Madore}, and {Thornley}}]{Bigiel2008}
{Bigiel} F, {Leroy} A, {Walter} F, {Brinks} E, {de Blok} WJG, {Madore} B,
  {Thornley} MD (2008) {The Star Formation Law in Nearby Galaxies on Sub-Kpc
  Scales}. \aj 136:2846--2871, \doi{10.1088/0004-6256/136/6/2846},
  \eprint{0810.2541}

\bibitem[{{Bischetti} et~al(2018){Bischetti}, {Piconcelli}, {Feruglio},
  {Duras}, {Bongiorno}, {Carniani}, {Marconi}, {Pappalardo}, {Schneider},
  {Travascio}, {Valiante}, {Vietri}, {Zappacosta}, and {Fiore}}]{Bischetti2018}
{Bischetti} M, {Piconcelli} E, {Feruglio} C, {Duras} F, {Bongiorno} A,
  {Carniani} S, {Marconi} A, {Pappalardo} C, {Schneider} R, {Travascio} A,
  {Valiante} R, {Vietri} G, {Zappacosta} L, {Fiore} F (2018) {The WISSH quasars
  project V. ALMA reveals the assembly of a giant galaxy around a z=4.4
  hyper-luminous QSO}. ArXiv e-prints \eprint{1804.06399}

\bibitem[{{Blain} et~al(2002){Blain}, {Smail}, {Ivison}, {Kneib}, and
  {Frayer}}]{Blain2002}
{Blain} AW, {Smail} I, {Ivison} RJ, {Kneib} JP, {Frayer} DT (2002)
  {Submillimeter galaxies}. \physrep 369:111--176,
  \doi{10.1016/S0370-1573(02)00134-5}, \eprint{astro-ph/0202228}

\bibitem[{{Bolatto} et~al(2013){Bolatto}, {Wolfire}, and {Leroy}}]{Bolatto2013}
{Bolatto} AD, {Wolfire} M, {Leroy} AK (2013) {The CO-to-H$_{2}$ Conversion
  Factor}. \araa 51:207--268, \doi{10.1146/annurev-astro-082812-140944},
  \eprint{1301.3498}

\bibitem[{{Boselli} et~al(2014){Boselli}, {Cortese}, {Boquien}, {Boissier},
  {Catinella}, {Lagos}, and {Saintonge}}]{Boselli2014}
{Boselli} A, {Cortese} L, {Boquien} M, {Boissier} S, {Catinella} B, {Lagos} C,
  {Saintonge} A (2014) {Cold gas properties of the Herschel Reference Survey.
  II. Molecular and total gas scaling relations}. \aap 564:A66,
  \doi{10.1051/0004-6361/201322312}, \eprint{1401.8101}

\bibitem[{{Bouch{\'e}} et~al(2010){Bouch{\'e}}, {Dekel}, {Genzel}, {Genel},
  {Cresci}, {F{\"o}rster Schreiber}, {Shapiro}, {Davies}, and
  {Tacconi}}]{Bouche2010}
{Bouch{\'e}} N, {Dekel} A, {Genzel} R, {Genel} S, {Cresci} G, {F{\"o}rster
  Schreiber} NM, {Shapiro} KL, {Davies} RI, {Tacconi} L (2010) {The Impact of
  Cold Gas Accretion Above a Mass Floor on Galaxy Scaling Relations}. \apj
  718:1001--1018, \doi{10.1088/0004-637X/718/2/1001}, \eprint{0912.1858}

\bibitem[{{Brown} and {Vanden Bout}(1992)}]{Brown1992}
{Brown} RL, {Vanden Bout} PA (1992) {IRAS F10214 + 4724 - an extended CO
  emission source at Z = 2.2867}. \apjl 397:L19--L22, \doi{10.1086/186534}

\bibitem[{{Bussmann} et~al(2013){Bussmann}, {P{\'e}rez-Fournon}, {Amber},
  {Calanog}, {Gurwell}, {Dannerbauer}, {De Bernardis}, {Fu}, {Harris}, {Krips},
  {Lapi}, {Maiolino}, {Omont}, {Riechers}, {Wardlow}, {Baker}, {Birkinshaw},
  {Bock}, {Bourne}, {Clements}, {Cooray}, {De Zotti}, {Dunne}, {Dye}, {Eales},
  {Farrah}, {Gavazzi}, {Gonz{\'a}lez Nuevo}, {Hopwood}, {Ibar}, {Ivison},
  {Laporte}, {Maddox}, {Mart{\'{\i}}nez-Navajas}, {Michalowski}, {Negrello},
  {Oliver}, {Roseboom}, {Scott}, {Serjeant}, {Smith}, {Smith}, {Streblyanska},
  {Valiante}, {van der Werf}, {Verma}, {Vieira}, {Wang}, and
  {Wilner}}]{Bussmann2013}
{Bussmann} RS, {P{\'e}rez-Fournon} I, {Amber} S, {Calanog} J, {Gurwell} MA,
  {Dannerbauer} H, {De Bernardis} F, {Fu} H, {Harris} AI, {Krips} M, {Lapi} A,
  {Maiolino} R, {Omont} A, {Riechers} D, {Wardlow} J, {Baker} AJ, {Birkinshaw}
  M, {Bock} J, {Bourne} N, {Clements} DL, {Cooray} A, {De Zotti} G, {Dunne} L,
  {Dye} S, {Eales} S, {Farrah} D, {Gavazzi} R, {Gonz{\'a}lez Nuevo} J,
  {Hopwood} R, {Ibar} E, {Ivison} RJ, {Laporte} N, {Maddox} S,
  {Mart{\'{\i}}nez-Navajas} P, {Michalowski} M, {Negrello} M, {Oliver} SJ,
  {Roseboom} IG, {Scott} D, {Serjeant} S, {Smith} AJ, {Smith} M, {Streblyanska}
  A, {Valiante} E, {van der Werf} P, {Verma} A, {Vieira} JD, {Wang} L, {Wilner}
  D (2013) {Gravitational Lens Models Based on Submillimeter Array Imaging of
  Herschel-selected Strongly Lensed Sub-millimeter Galaxies at z $>$ 1.5}. \apj
  779:25, \doi{10.1088/0004-637X/779/1/25}, \eprint{1309.0836}

\bibitem[{{Capak} et~al(2015){Capak}, {Carilli}, {Jones}, {Casey}, {Riechers},
  {Sheth}, {Carollo}, {Ilbert}, {Karim}, {Lefevre}, {Lilly}, {Scoville},
  {Smolcic}, and {Yan}}]{Capak2015}
{Capak} PL, {Carilli} C, {Jones} G, {Casey} CM, {Riechers} D, {Sheth} K,
  {Carollo} CM, {Ilbert} O, {Karim} A, {Lefevre} O, {Lilly} S, {Scoville} N,
  {Smolcic} V, {Yan} L (2015) {Galaxies at redshifts 5 to 6 with systematically
  low dust content and high [C II] emission}. \nat 522:455--458,
  \doi{10.1038/nature14500}, \eprint{1503.07596}

\bibitem[{{Carilli} and {Walter}(2013)}]{Carilli2013}
{Carilli} CL, {Walter} F (2013) {Cool Gas in High-Redshift Galaxies}. \araa
  51:105--161, \doi{10.1146/annurev-astro-082812-140953}, \eprint{1301.0371}

\bibitem[{{Carilli} et~al(2016){Carilli}, {Chluba}, {Decarli}, {Walter},
  {Aravena}, {Wagg}, {Popping}, {Cortes}, {Hodge}, {Weiss}, {Bertoldi}, and
  {Riechers}}]{Carilli2016}
{Carilli} CL, {Chluba} J, {Decarli} R, {Walter} F, {Aravena} M, {Wagg} J,
  {Popping} G, {Cortes} P, {Hodge} J, {Weiss} A, {Bertoldi} F, {Riechers} D
  (2016) {The ALMA Spectroscopic Survey in the Hubble Ultra Deep Field:
  Implications for Spectral Line Intensity Mapping at Millimeter Wavelengths
  and CMB Spectral Distortions}. \apj 833:73, \doi{10.3847/1538-4357/833/1/73},
  \eprint{1607.06773}

\bibitem[{{Carniani} et~al(2017){Carniani}, {Maiolino}, {Pallottini},
  {Vallini}, {Pentericci}, {Ferrara}, {Castellano}, {Vanzella}, {Grazian},
  {Gallerani}, {Santini}, {Wagg}, and {Fontana}}]{Carniani2017}
{Carniani} S, {Maiolino} R, {Pallottini} A, {Vallini} L, {Pentericci} L,
  {Ferrara} A, {Castellano} M, {Vanzella} E, {Grazian} A, {Gallerani} S,
  {Santini} P, {Wagg} J, {Fontana} A (2017) {Extended ionised and clumpy gas in
  a normal galaxy at z = 7.1 revealed by ALMA}. \aap 605:A42,
  \doi{10.1051/0004-6361/201630366}, \eprint{1701.03468}

\bibitem[{{Carniani} et~al(2018){Carniani}, {Maiolino}, {Smit}, and
  {Amor{\'{\i}}n}}]{Carniani2018}
{Carniani} S, {Maiolino} R, {Smit} R, {Amor{\'{\i}}n} R (2018) {ALMA Detection
  of Extended [C II] Emission in Himiko at z=6.6}. \apjl 854:L7,
  \doi{10.3847/2041-8213/aaab45}, \eprint{1712.01890}

\bibitem[{{Casey} et~al(2014){Casey}, {Narayanan}, and {Cooray}}]{Casey2014}
{Casey} CM, {Narayanan} D, {Cooray} A (2014) {Dusty star-forming galaxies at
  high redshift}. \physrep 541:45--161, \doi{10.1016/j.physrep.2014.02.009},
  \eprint{1402.1456}

\bibitem[{{Chapman} et~al(2005){Chapman}, {Blain}, {Smail}, and
  {Ivison}}]{Chapman2005}
{Chapman} SC, {Blain} AW, {Smail} I, {Ivison} RJ (2005) {A Redshift Survey of
  the Submillimeter Galaxy Population}. \apj 622:772--796,
  \doi{10.1086/428082}, \eprint{astro-ph/0412573}

\bibitem[{{Combes}(2008)}]{Combes2008}
{Combes} F (2008) {Molecular absorptions in high-z objects}. \apss
  313:321--326, \doi{10.1007/s10509-007-9632-3}, \eprint{astro-ph/0701894}

\bibitem[{{Combes} et~al(1999){Combes}, {Maoli}, and {Omont}}]{Combes1999}
{Combes} F, {Maoli} R, {Omont} A (1999) {CO lines in high redshift galaxies:
  perspective for future MM instruments}. \aap 345:369--379,
  \eprint{astro-ph/9902286}

\bibitem[{{Combes} et~al(2012){Combes}, {Rex}, {Rawle}, {Egami}, {Boone},
  {Smail}, {Richard}, {Ivison}, {Gurwell}, {Casey}, {Omont}, {Berciano Alba},
  {Dessauges-Zavadsky}, {Edge}, {Fazio}, {Kneib}, {Okabe}, {Pell{\'o}},
  {P{\'e}rez-Gonz{\'a}lez}, {Schaerer}, {Smith}, {Swinbank}, and {van der
  Werf}}]{Combes2012}
{Combes} F, {Rex} M, {Rawle} TD, {Egami} E, {Boone} F, {Smail} I, {Richard} J,
  {Ivison} RJ, {Gurwell} M, {Casey} CM, {Omont} A, {Berciano Alba} A,
  {Dessauges-Zavadsky} M, {Edge} AC, {Fazio} GG, {Kneib} JP, {Okabe} N,
  {Pell{\'o}} R, {P{\'e}rez-Gonz{\'a}lez} PG, {Schaerer} D, {Smith} GP,
  {Swinbank} AM, {van der Werf} P (2012) {A bright z = 5.2 lensed submillimeter
  galaxy in the field of Abell 773. HLSJ091828.6+514223}. \aap 538:L4,
  \doi{10.1051/0004-6361/201118750}, \eprint{1201.2908}

\bibitem[{{Coppin} et~al(2015){Coppin}, {Geach}, {Almaini}, {Arumugam},
  {Dunlop}, {Hartley}, {Ivison}, {Simpson}, {Smith}, {Swinbank}, {Blain},
  {Bourne}, {Bremer}, {Conselice}, {Harrison}, {Mortlock}, {Chapman}, {Davies},
  {Farrah}, {Gibb}, {Jenness}, {Karim}, {Knudsen}, {Ibar}, {Micha{\l}owski},
  {Peacock}, {Rigopoulou}, {Robson}, {Scott}, {Stevens}, and {van der
  Werf}}]{Coppin2015}
{Coppin} KEK, {Geach} JE, {Almaini} O, {Arumugam} V, {Dunlop} JS, {Hartley} WG,
  {Ivison} RJ, {Simpson} CJ, {Smith} DJB, {Swinbank} AM, {Blain} AW, {Bourne}
  N, {Bremer} M, {Conselice} C, {Harrison} CM, {Mortlock} A, {Chapman} SC,
  {Davies} LJM, {Farrah} D, {Gibb} A, {Jenness} T, {Karim} A, {Knudsen} KK,
  {Ibar} E, {Micha{\l}owski} MJ, {Peacock} JA, {Rigopoulou} D, {Robson} EI,
  {Scott} D, {Stevens} J, {van der Werf} PP (2015) {The SCUBA-2 Cosmology
  Legacy Survey: the submillimetre properties of Lyman-break galaxies at z =
  3-5}. \mnras 446:1293--1304, \doi{10.1093/mnras/stu2185}, \eprint{1407.6712}

\bibitem[{{Cox}(2005)}]{Cox2005}
{Cox} P (2005) {Molecular Gas at High Redshift}. In: {Lis} DC, {Blake} GA,
  {Herbst} E (eds) Astrochemistry: Recent Successes and Current Challenges, IAU
  Symposium, vol 231, pp 291--300, \doi{10.1017/S1743921306007289}

\bibitem[{{da Cunha} et~al(2013){da Cunha}, {Groves}, {Walter}, {Decarli},
  {Weiss}, {Bertoldi}, {Carilli}, {Daddi}, {Elbaz}, {Ivison}, {Maiolino},
  {Riechers}, {Rix}, {Sargent}, and {Smail}}]{daCunha2013}
{da Cunha} E, {Groves} B, {Walter} F, {Decarli} R, {Weiss} A, {Bertoldi} F,
  {Carilli} C, {Daddi} E, {Elbaz} D, {Ivison} R, {Maiolino} R, {Riechers} D,
  {Rix} HW, {Sargent} M, {Smail} I (2013) {On the Effect of the Cosmic
  Microwave Background in High-redshift (Sub-)millimeter Observations}. \apj
  766:13, \doi{10.1088/0004-637X/766/1/13}, \eprint{1302.0844}

\bibitem[{{David} et~al(2014){David}, {Lim}, {Forman}, {Vrtilek}, {Combes},
  {Salome}, {Edge}, {Hamer}, {Jones}, {Sun}, {O'Sullivan}, {Gastaldello},
  {Bardelli}, {Temi}, {Schmitt}, {Ohyama}, {Mathews}, {Brighenti},
  {Giacintucci}, and {Trung}}]{David2014}
{David} LP, {Lim} J, {Forman} W, {Vrtilek} J, {Combes} F, {Salome} P, {Edge} A,
  {Hamer} S, {Jones} C, {Sun} M, {O'Sullivan} E, {Gastaldello} F, {Bardelli} S,
  {Temi} P, {Schmitt} H, {Ohyama} Y, {Mathews} W, {Brighenti} F, {Giacintucci}
  S, {Trung} DV (2014) {Molecular Gas in the X-Ray Bright Group NGC 5044 as
  Revealed by ALMA}. \apj 792:94, \doi{10.1088/0004-637X/792/2/94},
  \eprint{1407.3235}

\bibitem[{{Decarli} et~al(2016{\natexlab{a}}){Decarli}, {Walter}, {Aravena},
  {Carilli}, {Bouwens}, {da Cunha}, {Daddi}, {Elbaz}, {Riechers}, {Smail},
  {Swinbank}, {Weiss}, {Bacon}, {Bauer}, {Bell}, {Bertoldi}, {Chapman},
  {Colina}, {Cortes}, {Cox}, {G{\'o}nzalez-L{\'o}pez}, {Inami}, {Ivison},
  {Hodge}, {Karim}, {Magnelli}, {Ota}, {Popping}, {Rix}, {Sargent}, {van der
  Wel}, and {van der Werf}}]{Decarli2016b}
{Decarli} R, {Walter} F, {Aravena} M, {Carilli} C, {Bouwens} R, {da Cunha} E,
  {Daddi} E, {Elbaz} D, {Riechers} D, {Smail} I, {Swinbank} M, {Weiss} A,
  {Bacon} R, {Bauer} F, {Bell} EF, {Bertoldi} F, {Chapman} S, {Colina} L,
  {Cortes} PC, {Cox} P, {G{\'o}nzalez-L{\'o}pez} J, {Inami} H, {Ivison} R,
  {Hodge} J, {Karim} A, {Magnelli} B, {Ota} K, {Popping} G, {Rix} HW, {Sargent}
  M, {van der Wel} A, {van der Werf} P (2016{\natexlab{a}}) {The ALMA
  Spectroscopic Survey in the Hubble Ultra Deep Field: Molecular Gas Reservoirs
  in High-redshift Galaxies}. \apj 833:70, \doi{10.3847/1538-4357/833/1/70},
  \eprint{1607.06771}

\bibitem[{{Decarli} et~al(2016{\natexlab{b}}){Decarli}, {Walter}, {Aravena},
  {Carilli}, {Bouwens}, {da Cunha}, {Daddi}, {Ivison}, {Popping}, {Riechers},
  {Smail}, {Swinbank}, {Weiss}, {Anguita}, {Assef}, {Bauer}, {Bell},
  {Bertoldi}, {Chapman}, {Colina}, {Cortes}, {Cox}, {Dickinson}, {Elbaz},
  {G{\'o}nzalez-L{\'o}pez}, {Ibar}, {Infante}, {Hodge}, {Karim}, {Le Fevre},
  {Magnelli}, {Neri}, {Oesch}, {Ota}, {Rix}, {Sargent}, {Sheth}, {van der Wel},
  {van der Werf}, and {Wagg}}]{Decarli2016a}
{Decarli} R, {Walter} F, {Aravena} M, {Carilli} C, {Bouwens} R, {da Cunha} E,
  {Daddi} E, {Ivison} RJ, {Popping} G, {Riechers} D, {Smail} IR, {Swinbank} M,
  {Weiss} A, {Anguita} T, {Assef} RJ, {Bauer} FE, {Bell} EF, {Bertoldi} F,
  {Chapman} S, {Colina} L, {Cortes} PC, {Cox} P, {Dickinson} M, {Elbaz} D,
  {G{\'o}nzalez-L{\'o}pez} J, {Ibar} E, {Infante} L, {Hodge} J, {Karim} A, {Le
  Fevre} O, {Magnelli} B, {Neri} R, {Oesch} P, {Ota} K, {Rix} HW, {Sargent} M,
  {Sheth} K, {van der Wel} A, {van der Werf} P, {Wagg} J (2016{\natexlab{b}})
  {ALMA Spectroscopic Survey in the Hubble Ultra Deep Field: CO Luminosity
  Functions and the Evolution of the Cosmic Density of Molecular Gas}. \apj
  833:69, \doi{10.3847/1538-4357/833/1/69}, \eprint{1607.06770}

\bibitem[{{Dekel} et~al(2009){Dekel}, {Birnboim}, {Engel}, {Freundlich},
  {Goerdt}, {Mumcuoglu}, {Neistein}, {Pichon}, {Teyssier}, and
  {Zinger}}]{Dekel2009}
{Dekel} A, {Birnboim} Y, {Engel} G, {Freundlich} J, {Goerdt} T, {Mumcuoglu} M,
  {Neistein} E, {Pichon} C, {Teyssier} R, {Zinger} E (2009) {Cold streams in
  early massive hot haloes as the main mode of galaxy formation}. \nat
  457:451--454, \doi{10.1038/nature07648}, \eprint{0808.0553}

\bibitem[{{Dekel} et~al(2013){Dekel}, {Zolotov}, {Tweed}, {Cacciato},
  {Ceverino}, and {Primack}}]{Dekel2013}
{Dekel} A, {Zolotov} A, {Tweed} D, {Cacciato} M, {Ceverino} D, {Primack} JR
  (2013) {Toy models for galaxy formation versus simulations}. \mnras
  435:999--1019, \doi{10.1093/mnras/stt1338}, \eprint{1303.3009}

\bibitem[{{Denicol{\'o}} et~al(2002){Denicol{\'o}}, {Terlevich}, and
  {Terlevich}}]{Denicolo2002}
{Denicol{\'o}} G, {Terlevich} R, {Terlevich} E (2002) {New light on the search
  for low-metallicity galaxies - I. The N2 calibrator}. \mnras 330:69--74,
  \doi{10.1046/j.1365-8711.2002.05041.x}, \eprint{astro-ph/0110356}

\bibitem[{{Dessauges-Zavadsky} et~al(2015){Dessauges-Zavadsky}, {Zamojski},
  {Schaerer}, {Combes}, {Egami}, {Swinbank}, {Richard}, {Sklias}, {Rawle},
  {Rex}, {Kneib}, {Boone}, and {Blain}}]{Dessauges2015}
{Dessauges-Zavadsky} M, {Zamojski} M, {Schaerer} D, {Combes} F, {Egami} E,
  {Swinbank} AM, {Richard} J, {Sklias} P, {Rawle} TD, {Rex} M, {Kneib} JP,
  {Boone} F, {Blain} A (2015) {Molecular gas content in strongly lensed z \~{}
  1.5-3 star-forming galaxies with low infrared luminosities}. \aap 577:A50,
  \doi{10.1051/0004-6361/201424661}, \eprint{1408.0816}

\bibitem[{{D{\'{\i}}az-Santos} et~al(2013){D{\'{\i}}az-Santos}, {Armus},
  {Charmandaris}, {Stierwalt}, {Murphy}, {Haan}, {Inami}, {Malhotra},
  {Meijerink}, {Stacey}, {Petric}, {Evans}, {Veilleux}, {van der Werf}, {Lord},
  {Lu}, {Howell}, {Appleton}, {Mazzarella}, {Surace}, {Xu}, {Schulz},
  {Sanders}, {Bridge}, {Chan}, {Frayer}, {Iwasawa}, {Melbourne}, and
  {Sturm}}]{Diaz2013}
{D{\'{\i}}az-Santos} T, {Armus} L, {Charmandaris} V, {Stierwalt} S, {Murphy}
  EJ, {Haan} S, {Inami} H, {Malhotra} S, {Meijerink} R, {Stacey} G, {Petric}
  AO, {Evans} AS, {Veilleux} S, {van der Werf} PP, {Lord} S, {Lu} N, {Howell}
  JH, {Appleton} P, {Mazzarella} JM, {Surace} JA, {Xu} CK, {Schulz} B,
  {Sanders} DB, {Bridge} C, {Chan} BHP, {Frayer} DT, {Iwasawa} K, {Melbourne}
  J, {Sturm} E (2013) {Explaining the [C II]157.7 {$\mu$}m Deficit in Luminous
  Infrared Galaxies -- First Results from a Herschel/PACS Study of the GOALS
  Sample}. \apj 774:68, \doi{10.1088/0004-637X/774/1/68}, \eprint{1307.2635}

\bibitem[{{Downes} et~al(1995){Downes}, {Solomon}, and {Radford}}]{Downes1995}
{Downes} D, {Solomon} PM, {Radford} SJE (1995) {New Observations and a New
  Interpretation of CO(3--2) in IRAS F10214+4724}. \apjl 453:L65,
  \doi{10.1086/309754}, \eprint{astro-ph/9508130}

\bibitem[{{Dunlop} et~al(2017){Dunlop}, {McLure}, {Biggs}, {Geach},
  {Micha{\l}owski}, {Ivison}, {Rujopakarn}, {van Kampen}, {Kirkpatrick},
  {Pope}, {Scott}, {Swinbank}, {Targett}, {Aretxaga}, {Austermann}, {Best},
  {Bruce}, {Chapin}, {Charlot}, {Cirasuolo}, {Coppin}, {Ellis}, {Finkelstein},
  {Hayward}, {Hughes}, {Ibar}, {Jagannathan}, {Khochfar}, {Koprowski},
  {Narayanan}, {Nyland}, {Papovich}, {Peacock}, {Rieke}, {Robertson},
  {Vernstrom}, {Werf}, {Wilson}, and {Yun}}]{Dunlop2017}
{Dunlop} JS, {McLure} RJ, {Biggs} AD, {Geach} JE, {Micha{\l}owski} MJ, {Ivison}
  RJ, {Rujopakarn} W, {van Kampen} E, {Kirkpatrick} A, {Pope} A, {Scott} D,
  {Swinbank} AM, {Targett} TA, {Aretxaga} I, {Austermann} JE, {Best} PN,
  {Bruce} VA, {Chapin} EL, {Charlot} S, {Cirasuolo} M, {Coppin} K, {Ellis} RS,
  {Finkelstein} SL, {Hayward} CC, {Hughes} DH, {Ibar} E, {Jagannathan} P,
  {Khochfar} S, {Koprowski} MP, {Narayanan} D, {Nyland} K, {Papovich} C,
  {Peacock} JA, {Rieke} GH, {Robertson} B, {Vernstrom} T, {Werf} PPvd, {Wilson}
  GW, {Yun} M (2017) {A deep ALMA image of the Hubble Ultra Deep Field}. \mnras
  466:861--883, \doi{10.1093/mnras/stw3088}, \eprint{1606.00227}

\bibitem[{{Eales} et~al(2010){Eales}, {Dunne}, {Clements}, {Cooray}, {De
  Zotti}, {Dye}, {Ivison}, {Jarvis}, {Lagache}, {Maddox}, {Negrello},
  {Serjeant}, {Thompson}, {Van Kampen}, {Amblard}, {Andreani}, {Baes},
  {Beelen}, {Bendo}, {Benford}, {Bertoldi}, {Bock}, {Bonfield}, {Boselli},
  {Bridge}, {Buat}, {Burgarella}, {Carlberg}, {Cava}, {Chanial}, {Charlot},
  {Christopher}, {Coles}, {Cortese}, {Dariush}, {da Cunha}, {Dalton}, {Danese},
  {Dannerbauer}, {Driver}, {Dunlop}, {Fan}, {Farrah}, {Frayer}, {Frenk},
  {Geach}, {Gardner}, {Gomez}, {Gonz{\'a}lez-Nuevo}, {Gonz{\'a}lez-Solares},
  {Griffin}, {Hardcastle}, {Hatziminaoglou}, {Herranz}, {Hughes}, {Ibar},
  {Jeong}, {Lacey}, {Lapi}, {Lawrence}, {Lee}, {Leeuw}, {Liske},
  {L{\'o}pez-Caniego}, {M{\"u}ller}, {Nandra}, {Panuzzo}, {Papageorgiou},
  {Patanchon}, {Peacock}, {Pearson}, {Phillipps}, {Pohlen}, {Popescu},
  {Rawlings}, {Rigby}, {Rigopoulou}, {Robotham}, {Rodighiero}, {Sansom},
  {Schulz}, {Scott}, {Smith}, {Sibthorpe}, {Smail}, {Stevens}, {Sutherland},
  {Takeuchi}, {Tedds}, {Temi}, {Tuffs}, {Trichas}, {Vaccari}, {Valtchanov},
  {van der Werf}, {Verma}, {Vieria}, {Vlahakis}, and {White}}]{Eales2010}
{Eales} S, {Dunne} L, {Clements} D, {Cooray} A, {De Zotti} G, {Dye} S, {Ivison}
  R, {Jarvis} M, {Lagache} G, {Maddox} S, {Negrello} M, {Serjeant} S,
  {Thompson} MA, {Van Kampen} E, {Amblard} A, {Andreani} P, {Baes} M, {Beelen}
  A, {Bendo} GJ, {Benford} D, {Bertoldi} F, {Bock} J, {Bonfield} D, {Boselli}
  A, {Bridge} C, {Buat} V, {Burgarella} D, {Carlberg} R, {Cava} A, {Chanial} P,
  {Charlot} S, {Christopher} N, {Coles} P, {Cortese} L, {Dariush} A, {da Cunha}
  E, {Dalton} G, {Danese} L, {Dannerbauer} H, {Driver} S, {Dunlop} J, {Fan} L,
  {Farrah} D, {Frayer} D, {Frenk} C, {Geach} J, {Gardner} J, {Gomez} H,
  {Gonz{\'a}lez-Nuevo} J, {Gonz{\'a}lez-Solares} E, {Griffin} M, {Hardcastle}
  M, {Hatziminaoglou} E, {Herranz} D, {Hughes} D, {Ibar} E, {Jeong} WS, {Lacey}
  C, {Lapi} A, {Lawrence} A, {Lee} M, {Leeuw} L, {Liske} J, {L{\'o}pez-Caniego}
  M, {M{\"u}ller} T, {Nandra} K, {Panuzzo} P, {Papageorgiou} A, {Patanchon} G,
  {Peacock} J, {Pearson} C, {Phillipps} S, {Pohlen} M, {Popescu} C, {Rawlings}
  S, {Rigby} E, {Rigopoulou} M, {Robotham} A, {Rodighiero} G, {Sansom} A,
  {Schulz} B, {Scott} D, {Smith} DJB, {Sibthorpe} B, {Smail} I, {Stevens} J,
  {Sutherland} W, {Takeuchi} T, {Tedds} J, {Temi} P, {Tuffs} R, {Trichas} M,
  {Vaccari} M, {Valtchanov} I, {van der Werf} P, {Verma} A, {Vieria} J,
  {Vlahakis} C, {White} GJ (2010) {The Herschel ATLAS}. \pasp 122:499,
  \doi{10.1086/653086}, \eprint{0910.4279}

\bibitem[{{Egami} et~al(2010){Egami}, {Rex}, {Rawle}, {P{\'e}rez-Gonz{\'a}lez},
  {Richard}, {Kneib}, {Schaerer}, {Altieri}, {Valtchanov}, {Blain}, {Fadda},
  {Zemcov}, {Bock}, {Boone}, {Bridge}, {Clement}, {Combes},
  {Dessauges-Zavadsky}, {Dowell}, {Ilbert}, {Ivison}, {Jauzac}, {Lutz},
  {Metcalfe}, {Omont}, {Pell{\'o}}, {Pereira}, {Rieke}, {Rodighiero}, {Smail},
  {Smith}, {Tramoy}, {Walth}, {van der Werf}, and {Werner}}]{Egami2010}
{Egami} E, {Rex} M, {Rawle} TD, {P{\'e}rez-Gonz{\'a}lez} PG, {Richard} J,
  {Kneib} JP, {Schaerer} D, {Altieri} B, {Valtchanov} I, {Blain} AW, {Fadda} D,
  {Zemcov} M, {Bock} JJ, {Boone} F, {Bridge} CR, {Clement} B, {Combes} F,
  {Dessauges-Zavadsky} M, {Dowell} CD, {Ilbert} O, {Ivison} RJ, {Jauzac} M,
  {Lutz} D, {Metcalfe} L, {Omont} A, {Pell{\'o}} R, {Pereira} MJ, {Rieke} GH,
  {Rodighiero} G, {Smail} I, {Smith} GP, {Tramoy} G, {Walth} GL, {van der Werf}
  P, {Werner} MW (2010) {The Herschel Lensing Survey (HLS): Overview}. \aap
  518:L12, \doi{10.1051/0004-6361/201014696}, \eprint{1005.3820}

\bibitem[{{Elbaz} et~al(2011){Elbaz}, {Dickinson}, {Hwang},
  {D{\'{\i}}az-Santos}, {Magdis}, {Magnelli}, {Le Borgne}, {Galliano},
  {Pannella}, {Chanial}, {Armus}, {Charmandaris}, {Daddi}, {Aussel}, {Popesso},
  {Kartaltepe}, {Altieri}, {Valtchanov}, {Coia}, {Dannerbauer}, {Dasyra},
  {Leiton}, {Mazzarella}, {Alexander}, {Buat}, {Burgarella}, {Chary}, {Gilli},
  {Ivison}, {Juneau}, {Le Floc'h}, {Lutz}, {Morrison}, {Mullaney}, {Murphy},
  {Pope}, {Scott}, {Brodwin}, {Calzetti}, {Cesarsky}, {Charlot}, {Dole},
  {Eisenhardt}, {Ferguson}, {F{\"o}rster Schreiber}, {Frayer}, {Giavalisco},
  {Huynh}, {Koekemoer}, {Papovich}, {Reddy}, {Surace}, {Teplitz}, {Yun}, and
  {Wilson}}]{Elbaz2011}
{Elbaz} D, {Dickinson} M, {Hwang} HS, {D{\'{\i}}az-Santos} T, {Magdis} G,
  {Magnelli} B, {Le Borgne} D, {Galliano} F, {Pannella} M, {Chanial} P, {Armus}
  L, {Charmandaris} V, {Daddi} E, {Aussel} H, {Popesso} P, {Kartaltepe} J,
  {Altieri} B, {Valtchanov} I, {Coia} D, {Dannerbauer} H, {Dasyra} K, {Leiton}
  R, {Mazzarella} J, {Alexander} DM, {Buat} V, {Burgarella} D, {Chary} RR,
  {Gilli} R, {Ivison} RJ, {Juneau} S, {Le Floc'h} E, {Lutz} D, {Morrison} GE,
  {Mullaney} JR, {Murphy} E, {Pope} A, {Scott} D, {Brodwin} M, {Calzetti} D,
  {Cesarsky} C, {Charlot} S, {Dole} H, {Eisenhardt} P, {Ferguson} HC,
  {F{\"o}rster Schreiber} N, {Frayer} D, {Giavalisco} M, {Huynh} M, {Koekemoer}
  AM, {Papovich} C, {Reddy} N, {Surace} C, {Teplitz} H, {Yun} MS, {Wilson} G
  (2011) {GOODS-Herschel: an infrared main sequence for star-forming galaxies}.
  \aap 533:A119, \doi{10.1051/0004-6361/201117239}, \eprint{1105.2537}

\bibitem[{{Emonts} et~al(2016){Emonts}, {Lehnert}, {Villar-Mart{\'{\i}}n},
  {Norris}, {Ekers}, {van Moorsel}, {Dannerbauer}, {Pentericci}, {Miley},
  {Allison}, {Sadler}, {Guillard}, {Carilli}, {Mao}, {R{\"o}ttgering}, {De
  Breuck}, {Seymour}, {Gullberg}, {Ceverino}, {Jagannathan}, {Vernet}, and
  {Indermuehle}}]{Emonts2016}
{Emonts} BHC, {Lehnert} MD, {Villar-Mart{\'{\i}}n} M, {Norris} RP, {Ekers} RD,
  {van Moorsel} GA, {Dannerbauer} H, {Pentericci} L, {Miley} GK, {Allison} JR,
  {Sadler} EM, {Guillard} P, {Carilli} CL, {Mao} MY, {R{\"o}ttgering} HJA, {De
  Breuck} C, {Seymour} N, {Gullberg} B, {Ceverino} D, {Jagannathan} P, {Vernet}
  J, {Indermuehle} BT (2016) {Molecular gas in the halo fuels the growth of a
  massive cluster galaxy at high redshift}. Science 354:1128--1130,
  \doi{10.1126/science.aag0512}, \eprint{1612.00387}

\bibitem[{{Emonts} et~al(2018){Emonts}, {Lehnert}, {Dannerbauer}, {De Breuck},
  {Villar-Mart{\'{\i}}n}, {Miley}, {Allison}, {Gullberg}, {Hatch}, {Guillard},
  {Mao}, and {Norris}}]{Emonts2018}
{Emonts} BHC, {Lehnert} MD, {Dannerbauer} H, {De Breuck} C,
  {Villar-Mart{\'{\i}}n} M, {Miley} GK, {Allison} JR, {Gullberg} B, {Hatch} NA,
  {Guillard} P, {Mao} MY, {Norris} RP (2018) {Giant galaxy growing from
  recycled gas: ALMA maps the circumgalactic molecular medium of the Spiderweb
  in [C I]}. \mnras 477:L60--L65, \doi{10.1093/mnrasl/sly034},
  \eprint{1802.08742}

\bibitem[{{Freundlich} et~al(2013){Freundlich}, {Combes}, {Tacconi}, {Cooper},
  {Genzel}, {Neri}, {Bolatto}, {Bournaud}, {Burkert}, {Cox}, {Davis},
  {F{\"o}rster Schreiber}, {Garcia-Burillo}, {Gracia-Carpio}, {Lutz}, {Naab},
  {Newman}, {Sternberg}, and {Weiner}}]{Freundlich2013}
{Freundlich} J, {Combes} F, {Tacconi} LJ, {Cooper} MC, {Genzel} R, {Neri} R,
  {Bolatto} A, {Bournaud} F, {Burkert} A, {Cox} P, {Davis} M, {F{\"o}rster
  Schreiber} NM, {Garcia-Burillo} S, {Gracia-Carpio} J, {Lutz} D, {Naab} T,
  {Newman} S, {Sternberg} A, {Weiner} B (2013) {Towards a resolved
  Kennicutt-Schmidt law at high redshift}. \aap 553:A130,
  \doi{10.1051/0004-6361/201220981}, \eprint{1301.0628}

\bibitem[{{Genzel} et~al(2012){Genzel}, {Tacconi}, {Combes}, {Bolatto}, {Neri},
  {Sternberg}, {Cooper}, {Bouch{\'e}}, {Bournaud}, {Burkert}, {Comerford},
  {Cox}, {Davis}, {F{\"o}rster Schreiber}, {Garcia-Burillo}, {Gracia-Carpio},
  {Lutz}, {Naab}, {Newman}, {Saintonge}, {Shapiro}, {Shapley}, and
  {Weiner}}]{Genzel2012}
{Genzel} R, {Tacconi} LJ, {Combes} F, {Bolatto} A, {Neri} R, {Sternberg} A,
  {Cooper} MC, {Bouch{\'e}} N, {Bournaud} F, {Burkert} A, {Comerford} J, {Cox}
  P, {Davis} M, {F{\"o}rster Schreiber} NM, {Garcia-Burillo} S, {Gracia-Carpio}
  J, {Lutz} D, {Naab} T, {Newman} S, {Saintonge} A, {Shapiro} K, {Shapley} A,
  {Weiner} B (2012) {The Metallicity Dependence of the CO to H$_{2}$ Conversion
  Factor in z $>$= 1 Star-forming Galaxies}. \apj 746:69,
  \doi{10.1088/0004-637X/746/1/69}, \eprint{1106.2098}

\bibitem[{{Genzel} et~al(2015){Genzel}, {Tacconi}, {Lutz}, {Saintonge},
  {Berta}, {Magnelli}, {Combes}, {Garc{\'{\i}}a-Burillo}, {Neri}, {Bolatto},
  {Contini}, {Lilly}, {Boissier}, {Boone}, {Bouch{\'e}}, {Bournaud}, {Burkert},
  {Carollo}, {Colina}, {Cooper}, {Cox}, {Feruglio}, {F{\"o}rster Schreiber},
  {Freundlich}, {Gracia-Carpio}, {Juneau}, {Kovac}, {Lippa}, {Naab}, {Salome},
  {Renzini}, {Sternberg}, {Walter}, {Weiner}, {Weiss}, and
  {Wuyts}}]{Genzel2015}
{Genzel} R, {Tacconi} LJ, {Lutz} D, {Saintonge} A, {Berta} S, {Magnelli} B,
  {Combes} F, {Garc{\'{\i}}a-Burillo} S, {Neri} R, {Bolatto} A, {Contini} T,
  {Lilly} S, {Boissier} J, {Boone} F, {Bouch{\'e}} N, {Bournaud} F, {Burkert}
  A, {Carollo} M, {Colina} L, {Cooper} MC, {Cox} P, {Feruglio} C, {F{\"o}rster
  Schreiber} NM, {Freundlich} J, {Gracia-Carpio} J, {Juneau} S, {Kovac} K,
  {Lippa} M, {Naab} T, {Salome} P, {Renzini} A, {Sternberg} A, {Walter} F,
  {Weiner} B, {Weiss} A, {Wuyts} S (2015) {Combined CO and Dust Scaling
  Relations of Depletion Time and Molecular Gas Fractions with Cosmic Time,
  Specific Star-formation Rate, and Stellar Mass}. \apj 800:20,
  \doi{10.1088/0004-637X/800/1/20}, \eprint{1409.1171}

\bibitem[{{Ginolfi} et~al(2017){Ginolfi}, {Maiolino}, {Nagao}, {Carniani},
  {Belfiore}, {Cresci}, {Hatsukade}, {Mannucci}, {Marconi}, {Pallottini},
  {Schneider}, and {Santini}}]{Ginolfi2017}
{Ginolfi} M, {Maiolino} R, {Nagao} T, {Carniani} S, {Belfiore} F, {Cresci} G,
  {Hatsukade} B, {Mannucci} F, {Marconi} A, {Pallottini} A, {Schneider} R,
  {Santini} P (2017) {Molecular gas on large circumgalactic scales at z =
  3.47}. \mnras 468:3468--3483, \doi{10.1093/mnras/stx712}, \eprint{1611.07026}

\bibitem[{{Gullberg} et~al(2015){Gullberg}, {De Breuck}, {Vieira}, {Wei{\ss}},
  {Aguirre}, {Aravena}, {B{\'e}thermin}, {Bradford}, {Bothwell}, {Carlstrom},
  {Chapman}, {Fassnacht}, {Gonzalez}, {Greve}, {Hezaveh}, {Holzapfel},
  {Husband}, {Ma}, {Malkan}, {Marrone}, {Menten}, {Murphy}, {Reichardt},
  {Spilker}, {Stark}, {Strandet}, and {Welikala}}]{Gullberg2015}
{Gullberg} B, {De Breuck} C, {Vieira} JD, {Wei{\ss}} A, {Aguirre} JE, {Aravena}
  M, {B{\'e}thermin} M, {Bradford} CM, {Bothwell} MS, {Carlstrom} JE, {Chapman}
  SC, {Fassnacht} CD, {Gonzalez} AH, {Greve} TR, {Hezaveh} Y, {Holzapfel} WL,
  {Husband} K, {Ma} J, {Malkan} M, {Marrone} DP, {Menten} K, {Murphy} EJ,
  {Reichardt} CL, {Spilker} JS, {Stark} AA, {Strandet} M, {Welikala} N (2015)
  {The nature of the [C II] emission in dusty star-forming galaxies from the
  SPT survey}. \mnras 449:2883--2900, \doi{10.1093/mnras/stv372},
  \eprint{1501.06909}

\bibitem[{{Hashimoto} et~al(2018){Hashimoto}, {Laporte}, {Mawatari}, {Ellis},
  {Inoue}, {Zackrisson}, {Roberts-Borsani}, {Zheng}, {Tamura}, {Bauer},
  {Fletcher}, {Harikane}, {Hatsukade}, {Hayatsu}, {Matsuda}, {Matsuo},
  {Okamoto}, {Ouchi}, {Pello}, {Rydberg}, {Shimizu}, {Taniguchi}, {Umehata},
  and {Yoshida}}]{Hashimoto2018}
{Hashimoto} T, {Laporte} N, {Mawatari} K, {Ellis} RS, {Inoue} AK, {Zackrisson}
  E, {Roberts-Borsani} G, {Zheng} W, {Tamura} Y, {Bauer} FE, {Fletcher} T,
  {Harikane} Y, {Hatsukade} B, {Hayatsu} NH, {Matsuda} Y, {Matsuo} H, {Okamoto}
  T, {Ouchi} M, {Pello} R, {Rydberg} CE, {Shimizu} I, {Taniguchi} Y, {Umehata}
  H, {Yoshida} N (2018) {The onset of star formation 250 million years after
  the Big Bang}. ArXiv e-prints \eprint{1805.05966}

\bibitem[{{Hatsukade} et~al(2016){Hatsukade}, {Kohno}, {Umehata}, {Aretxaga},
  {Caputi}, {Dunlop}, {Ikarashi}, {Iono}, {Ivison}, {Lee}, {Makiya}, {Matsuda},
  {Motohara}, {Nakanishi}, {Ohta}, {Tadaki}, {Tamura}, {Wang}, {Wilson},
  {Yamaguchi}, and {Yun}}]{Hatsukade2016}
{Hatsukade} B, {Kohno} K, {Umehata} H, {Aretxaga} I, {Caputi} KI, {Dunlop} JS,
  {Ikarashi} S, {Iono} D, {Ivison} RJ, {Lee} M, {Makiya} R, {Matsuda} Y,
  {Motohara} K, {Nakanishi} K, {Ohta} K, {Tadaki} Ki, {Tamura} Y, {Wang} WH,
  {Wilson} GW, {Yamaguchi} Y, {Yun} MS (2016) {SXDF-ALMA 2-arcmin$^{2}$ deep
  survey: 1.1-mm number counts}. \pasj 68:36, \doi{10.1093/pasj/psw026},
  \eprint{1602.08167}

\bibitem[{{Ilbert} et~al(2013){Ilbert}, {McCracken}, {Le F{\`e}vre}, {Capak},
  {Dunlop}, {Karim}, {Renzini}, {Caputi}, {Boissier}, {Arnouts}, {Aussel},
  {Comparat}, {Guo}, {Hudelot}, {Kartaltepe}, {Kneib}, {Krogager}, {Le Floc'h},
  {Lilly}, {Mellier}, {Milvang-Jensen}, {Moutard}, {Onodera}, {Richard},
  {Salvato}, {Sanders}, {Scoville}, {Silverman}, {Taniguchi}, {Tasca},
  {Thomas}, {Toft}, {Tresse}, {Vergani}, {Wolk}, and {Zirm}}]{Ilbert2013}
{Ilbert} O, {McCracken} HJ, {Le F{\`e}vre} O, {Capak} P, {Dunlop} J, {Karim} A,
  {Renzini} MA, {Caputi} K, {Boissier} S, {Arnouts} S, {Aussel} H, {Comparat}
  J, {Guo} Q, {Hudelot} P, {Kartaltepe} J, {Kneib} JP, {Krogager} JK, {Le
  Floc'h} E, {Lilly} S, {Mellier} Y, {Milvang-Jensen} B, {Moutard} T, {Onodera}
  M, {Richard} J, {Salvato} M, {Sanders} DB, {Scoville} N, {Silverman} JD,
  {Taniguchi} Y, {Tasca} L, {Thomas} R, {Toft} S, {Tresse} L, {Vergani} D,
  {Wolk} M, {Zirm} A (2013) {Mass assembly in quiescent and star-forming
  galaxies since z {$\sim$} 4 from UltraVISTA}. \aap 556:A55,
  \doi{10.1051/0004-6361/201321100}, \eprint{1301.3157}

\bibitem[{{Illingworth} et~al(2013){Illingworth}, {Magee}, {Oesch}, {Bouwens},
  {Labb{\'e}}, {Stiavelli}, {van Dokkum}, {Franx}, {Trenti}, {Carollo}, and
  {Gonzalez}}]{Illingworth2013}
{Illingworth} GD, {Magee} D, {Oesch} PA, {Bouwens} RJ, {Labb{\'e}} I,
  {Stiavelli} M, {van Dokkum} PG, {Franx} M, {Trenti} M, {Carollo} CM,
  {Gonzalez} V (2013) {The HST eXtreme Deep Field (XDF): Combining All ACS and
  WFC3/IR Data on the HUDF Region into the Deepest Field Ever}. \apjs 209:6,
  \doi{10.1088/0067-0049/209/1/6}, \eprint{1305.1931}

\bibitem[{{Katz} et~al(2017){Katz}, {Kimm}, {Sijacki}, and
  {Haehnelt}}]{Katz2017}
{Katz} H, {Kimm} T, {Sijacki} D, {Haehnelt} MG (2017) {Interpreting ALMA
  observations of the ISM during the epoch of reionization}. \mnras
  468:4831--4861, \doi{10.1093/mnras/stx608}, \eprint{1612.01786}

\bibitem[{{Keating} et~al(2016){Keating}, {Marrone}, {Bower}, {Leitch},
  {Carlstrom}, and {DeBoer}}]{Keating2016}
{Keating} GK, {Marrone} DP, {Bower} GC, {Leitch} E, {Carlstrom} JE, {DeBoer} DR
  (2016) {COPSS II: The Molecular Gas Content of Ten Million Cubic Megaparsecs
  at Redshift z $\sim$ 3}. \apj 830:34, \doi{10.3847/0004-637X/830/1/34},
  \eprint{1605.03971}

\bibitem[{{Keres} et~al(2003){Keres}, {Yun}, and {Young}}]{Keres2003}
{Keres} D, {Yun} MS, {Young} JS (2003) {CO Luminosity Functions for
  Far-Infrared- and B-Band-selected Galaxies and the First Estimate for
  {$\Omega$}$_{HI}$+H$_{2}$}. \apj 582:659--667, \doi{10.1086/344820},
  \eprint{astro-ph/0209413}

\bibitem[{{Kormendy} and {Ho}(2013)}]{Kormendy2013}
{Kormendy} J, {Ho} LC (2013) {Coevolution (Or Not) of Supermassive Black Holes
  and Host Galaxies}. \araa 51:511--653,
  \doi{10.1146/annurev-astro-082708-101811}, \eprint{1304.7762}

\bibitem[{{Lagos} et~al(2012){Lagos}, {Bayet}, {Baugh}, {Lacey}, {Bell},
  {Fanidakis}, and {Geach}}]{Lagos2012}
{Lagos} CdP, {Bayet} E, {Baugh} CM, {Lacey} CG, {Bell} TA, {Fanidakis} N,
  {Geach} JE (2012) {Predictions for the CO emission of galaxies from a coupled
  simulation of galaxy formation and photon-dominated regions}. \mnras
  426:2142--2165, \doi{10.1111/j.1365-2966.2012.21905.x}, \eprint{1204.0795}

\bibitem[{{Laporte} et~al(2017){Laporte}, {Ellis}, {Boone}, {Bauer},
  {Qu{\'e}nard}, {Roberts-Borsani}, {Pell{\'o}}, {P{\'e}rez-Fournon}, and
  {Streblyanska}}]{Laporte2017}
{Laporte} N, {Ellis} RS, {Boone} F, {Bauer} FE, {Qu{\'e}nard} D,
  {Roberts-Borsani} GW, {Pell{\'o}} R, {P{\'e}rez-Fournon} I, {Streblyanska} A
  (2017) {Dust in the Reionization Era: ALMA Observations of a z = 8.38
  Gravitationally Lensed Galaxy}. \apjl 837:L21,
  \doi{10.3847/2041-8213/aa62aa}, \eprint{1703.02039}

\bibitem[{{Le F{\`e}vre} et~al(2004){Le F{\`e}vre}, {Vettolani}, {Paltani},
  {Tresse}, {Zamorani}, {Le Brun}, {Moreau}, {Bottini}, {Maccagni}, {Picat},
  {Scaramella}, {Scodeggio}, {Zanichelli}, {Adami}, {Arnouts}, {Bardelli},
  {Bolzonella}, {Cappi}, {Charlot}, {Contini}, {Foucaud}, {Franzetti},
  {Garilli}, {Gavignaud}, {Guzzo}, {Ilbert}, {Iovino}, {McCracken}, {Mancini},
  {Marano}, {Marinoni}, {Mathez}, {Mazure}, {Meneux}, {Merighi}, {Pell{\`o}},
  {Pollo}, {Pozzetti}, {Radovich}, {Zucca}, {Arnaboldi}, {Bondi}, {Bongiorno},
  {Busarello}, {Ciliegi}, {Gregorini}, {Mellier}, {Merluzzi}, {Ripepi}, and
  {Rizzo}}]{LeFevre2004}
{Le F{\`e}vre} O, {Vettolani} G, {Paltani} S, {Tresse} L, {Zamorani} G, {Le
  Brun} V, {Moreau} C, {Bottini} D, {Maccagni} D, {Picat} JP, {Scaramella} R,
  {Scodeggio} M, {Zanichelli} A, {Adami} C, {Arnouts} S, {Bardelli} S,
  {Bolzonella} M, {Cappi} A, {Charlot} S, {Contini} T, {Foucaud} S, {Franzetti}
  P, {Garilli} B, {Gavignaud} I, {Guzzo} L, {Ilbert} O, {Iovino} A, {McCracken}
  HJ, {Mancini} D, {Marano} B, {Marinoni} C, {Mathez} G, {Mazure} A, {Meneux}
  B, {Merighi} R, {Pell{\`o}} R, {Pollo} A, {Pozzetti} L, {Radovich} M, {Zucca}
  E, {Arnaboldi} M, {Bondi} M, {Bongiorno} A, {Busarello} G, {Ciliegi} P,
  {Gregorini} L, {Mellier} Y, {Merluzzi} P, {Ripepi} V, {Rizzo} D (2004) {The
  VIMOS VLT Deep Survey. Public release of 1599 redshifts to I$_{AB}{<}$24
  across the Chandra Deep Field South}. \aap 428:1043--1049,
  \doi{10.1051/0004-6361:20048072}, \eprint{astro-ph/0403628}

\bibitem[{{Le Floc'h} et~al(2005){Le Floc'h}, {Papovich}, {Dole}, {Bell},
  {Lagache}, {Rieke}, {Egami}, {P{\'e}rez-Gonz{\'a}lez}, {Alonso-Herrero},
  {Rieke}, {Blaylock}, {Engelbracht}, {Gordon}, {Hines}, {Misselt}, {Morrison},
  and {Mould}}]{LeFloch2005}
{Le Floc'h} E, {Papovich} C, {Dole} H, {Bell} EF, {Lagache} G, {Rieke} GH,
  {Egami} E, {P{\'e}rez-Gonz{\'a}lez} PG, {Alonso-Herrero} A, {Rieke} MJ,
  {Blaylock} M, {Engelbracht} CW, {Gordon} KD, {Hines} DC, {Misselt} KA,
  {Morrison} JE, {Mould} J (2005) {Infrared Luminosity Functions from the
  Chandra Deep Field-South: The Spitzer View on the History of Dusty Star
  Formation at 0 $<$ z $<$ 1}. \apj 632:169--190, \doi{10.1086/432789},
  \eprint{astro-ph/0506462}

\bibitem[{{Lehmer} et~al(2009){Lehmer}, {Alexander}, {Geach}, {Smail},
  {Basu-Zych}, {Bauer}, {Chapman}, {Matsuda}, {Scharf}, {Volonteri}, and
  {Yamada}}]{Lehmer2009}
{Lehmer} BD, {Alexander} DM, {Geach} JE, {Smail} I, {Basu-Zych} A, {Bauer} FE,
  {Chapman} SC, {Matsuda} Y, {Scharf} CA, {Volonteri} M, {Yamada} T (2009) {The
  Chandra Deep Protocluster Survey: Evidence for an Enhancement of AGN Activity
  in the SSA22 Protocluster at z = 3.09}. \apj 691:687--695,
  \doi{10.1088/0004-637X/691/1/687}, \eprint{0809.5058}

\bibitem[{{Leroy} et~al(2011){Leroy}, {Bolatto}, {Gordon}, {Sandstrom},
  {Gratier}, {Rosolowsky}, {Engelbracht}, {Mizuno}, {Corbelli}, {Fukui}, and
  {Kawamura}}]{Leroy2011}
{Leroy} AK, {Bolatto} A, {Gordon} K, {Sandstrom} K, {Gratier} P, {Rosolowsky}
  E, {Engelbracht} CW, {Mizuno} N, {Corbelli} E, {Fukui} Y, {Kawamura} A (2011)
  {The CO-to-H$_{2}$ Conversion Factor from Infrared Dust Emission across the
  Local Group}. \apj 737:12, \doi{10.1088/0004-637X/737/1/12},
  \eprint{1102.4618}

\bibitem[{{Levshakov} et~al(2012){Levshakov}, {Combes}, {Boone}, {Agafonova},
  {Reimers}, and {Kozlov}}]{Levshakov2012}
{Levshakov} SA, {Combes} F, {Boone} F, {Agafonova} II, {Reimers} D, {Kozlov} MG
  (2012) {An upper limit to the variation in the fundamental constants at
  redshift z = 5.2}. \aap 540:L9, \doi{10.1051/0004-6361/201219042},
  \eprint{1203.3649}

\bibitem[{{Lilly} et~al(2013){Lilly}, {Carollo}, {Pipino}, {Renzini}, and
  {Peng}}]{Lilly2013}
{Lilly} SJ, {Carollo} CM, {Pipino} A, {Renzini} A, {Peng} Y (2013) {Gas
  Regulation of Galaxies: The Evolution of the Cosmic Specific Star Formation
  Rate, the Metallicity-Mass-Star-formation Rate Relation, and the Stellar
  Content of Halos}. \apj 772:119, \doi{10.1088/0004-637X/772/2/119},
  \eprint{1303.5059}

\bibitem[{{Luhman}(2013)}]{Luhman2013}
{Luhman} KL (2013) {Discovery of a Binary Brown Dwarf at 2 pc from the Sun}.
  \apjl 767:L1, \doi{10.1088/2041-8205/767/1/L1}, \eprint{1303.2401}

\bibitem[{{Madau} and {Dickinson}(2014)}]{Madau2014}
{Madau} P, {Dickinson} M (2014) {Cosmic Star-Formation History}. \araa
  52:415--486, \doi{10.1146/annurev-astro-081811-125615}, \eprint{1403.0007}

\bibitem[{{Maiolino} et~al(2005){Maiolino}, {Cox}, {Caselli}, {Beelen},
  {Bertoldi}, {Carilli}, {Kaufman}, {Menten}, {Nagao}, {Omont}, {Wei{\ss}},
  {Walmsley}, and {Walter}}]{Maiolino2005}
{Maiolino} R, {Cox} P, {Caselli} P, {Beelen} A, {Bertoldi} F, {Carilli} CL,
  {Kaufman} MJ, {Menten} KM, {Nagao} T, {Omont} A, {Wei{\ss}} A, {Walmsley} CM,
  {Walter} F (2005) {First detection of [CII]158 {$\mu$}m at high redshift:
  vigorous star formation in the early universe}. \aap 440:L51--L54,
  \doi{10.1051/0004-6361:200500165}, \eprint{astro-ph/0508064}

\bibitem[{{Maiolino} et~al(2008){Maiolino}, {Nagao}, {Grazian}, {Cocchia},
  {Marconi}, {Mannucci}, {Cimatti}, {Pipino}, {Ballero}, {Calura}, {Chiappini},
  {Fontana}, {Granato}, {Matteucci}, {Pastorini}, {Pentericci}, {Risaliti},
  {Salvati}, and {Silva}}]{Maiolino2008}
{Maiolino} R, {Nagao} T, {Grazian} A, {Cocchia} F, {Marconi} A, {Mannucci} F,
  {Cimatti} A, {Pipino} A, {Ballero} S, {Calura} F, {Chiappini} C, {Fontana} A,
  {Granato} GL, {Matteucci} F, {Pastorini} G, {Pentericci} L, {Risaliti} G,
  {Salvati} M, {Silva} L (2008) {AMAZE. I. The evolution of the
  mass-metallicity relation at z $>$ 3}. \aap 488:463--479,
  \doi{10.1051/0004-6361:200809678}, \eprint{0806.2410}

\bibitem[{{Maiolino} et~al(2015){Maiolino}, {Carniani}, {Fontana}, {Vallini},
  {Pentericci}, {Ferrara}, {Vanzella}, {Grazian}, {Gallerani}, {Castellano},
  {Cristiani}, {Brammer}, {Santini}, {Wagg}, and {Williams}}]{Maiolino2015}
{Maiolino} R, {Carniani} S, {Fontana} A, {Vallini} L, {Pentericci} L, {Ferrara}
  A, {Vanzella} E, {Grazian} A, {Gallerani} S, {Castellano} M, {Cristiani} S,
  {Brammer} G, {Santini} P, {Wagg} J, {Williams} R (2015) {The assembly of
  `normal' galaxies at z {$\sim$} 7 probed by ALMA}. \mnras 452:54--68,
  \doi{10.1093/mnras/stv1194}, \eprint{1502.06634}

\bibitem[{{Malhotra} et~al(2001){Malhotra}, {Kaufman}, {Hollenbach}, {Helou},
  {Rubin}, {Brauher}, {Dale}, {Lu}, {Lord}, {Stacey}, {Contursi}, {Hunter}, and
  {Dinerstein}}]{Malhotra2001}
{Malhotra} S, {Kaufman} MJ, {Hollenbach} D, {Helou} G, {Rubin} RH, {Brauher} J,
  {Dale} D, {Lu} NY, {Lord} S, {Stacey} G, {Contursi} A, {Hunter} DA,
  {Dinerstein} H (2001) {Far-Infrared Spectroscopy of Normal Galaxies: Physical
  Conditions in the Interstellar Medium}. \apj 561:766--786,
  \doi{10.1086/323046}, \eprint{astro-ph/0106485}

\bibitem[{{Malhotra} et~al(2017){Malhotra}, {Rhoads}, {Finkelstein}, {Yang},
  {Carilli}, {Combes}, {Dassas}, {Finkelstein}, {Frye}, {Gerin}, {Guillard},
  {Nesvadba}, {Rigby}, {Shin}, {Spaans}, {Strauss}, and
  {Papovich}}]{Malhotra2017}
{Malhotra} S, {Rhoads} JE, {Finkelstein} K, {Yang} H, {Carilli} C, {Combes} F,
  {Dassas} K, {Finkelstein} S, {Frye} B, {Gerin} M, {Guillard} P, {Nesvadba} N,
  {Rigby} J, {Shin} MS, {Spaans} M, {Strauss} MA, {Papovich} C (2017) {Herschel
  Extreme Lensing Line Observations: [CII] Variations in Galaxies at Redshifts
  z=1-3}. \apj 835:110, \doi{10.3847/1538-4357/835/1/110}

\bibitem[{{Matsuda} et~al(2004){Matsuda}, {Yamada}, {Hayashino}, {Tamura},
  {Yamauchi}, {Ajiki}, {Fujita}, {Murayama}, {Nagao}, {Ohta}, {Okamura},
  {Ouchi}, {Shimasaku}, {Shioya}, and {Taniguchi}}]{Matsuda2004}
{Matsuda} Y, {Yamada} T, {Hayashino} T, {Tamura} H, {Yamauchi} R, {Ajiki} M,
  {Fujita} SS, {Murayama} T, {Nagao} T, {Ohta} K, {Okamura} S, {Ouchi} M,
  {Shimasaku} K, {Shioya} Y, {Taniguchi} Y (2004) {A Subaru Search for
  Ly{$\alpha$} Blobs in and around the Protocluster Region At Redshift z =
  3.1}. \aj 128:569--584, \doi{10.1086/422020}, \eprint{astro-ph/0405221}

\bibitem[{{Matthee} et~al(2017){Matthee}, {Sobral}, {Boone}, {R{\"o}ttgering},
  {Schaerer}, {Girard}, {Pallottini}, {Vallini}, {Ferrara}, {Darvish}, and
  {Mobasher}}]{Matthee2017}
{Matthee} J, {Sobral} D, {Boone} F, {R{\"o}ttgering} H, {Schaerer} D, {Girard}
  M, {Pallottini} A, {Vallini} L, {Ferrara} A, {Darvish} B, {Mobasher} B (2017)
  {ALMA Reveals Metals yet No Dust within Multiple Components in CR7}. \apj
  851:145, \doi{10.3847/1538-4357/aa9931}, \eprint{1709.06569}

\bibitem[{{Muller} et~al(2014){Muller}, {Combes}, {Gu{\'e}lin}, {G{\'e}rin},
  {Aalto}, {Beelen}, {Black}, {Curran}, {Darling}, {V-Trung},
  {Garc{\'{\i}}a-Burillo}, {Henkel}, {Horellou}, {Mart{\'{\i}}n},
  {Mart{\'{\i}}-Vidal}, {Menten}, {Murphy}, {Ott}, {Wiklind}, and
  {Zwaan}}]{Muller2014}
{Muller} S, {Combes} F, {Gu{\'e}lin} M, {G{\'e}rin} M, {Aalto} S, {Beelen} A,
  {Black} JH, {Curran} SJ, {Darling} J, {V-Trung} D, {Garc{\'{\i}}a-Burillo} S,
  {Henkel} C, {Horellou} C, {Mart{\'{\i}}n} S, {Mart{\'{\i}}-Vidal} I, {Menten}
  KM, {Murphy} MT, {Ott} J, {Wiklind} T, {Zwaan} MA (2014) {An ALMA Early
  Science survey of molecular absorption lines toward PKS 1830-211. Analysis of
  the absorption profiles}. \aap 566:A112, \doi{10.1051/0004-6361/201423646},
  \eprint{1404.7667}

\bibitem[{{Muller} et~al(2016){Muller}, {M{\"u}ller}, {Black}, {Beelen},
  {Combes}, {Curran}, {G{\'e}rin}, {Gu{\'e}lin}, {Henkel}, {Mart{\'{\i}}n},
  {Aalto}, {Falgarone}, {Menten}, {Schilke}, {Wiklind}, and
  {Zwaan}}]{Muller2016}
{Muller} S, {M{\"u}ller} HSP, {Black} JH, {Beelen} A, {Combes} F, {Curran} S,
  {G{\'e}rin} M, {Gu{\'e}lin} M, {Henkel} C, {Mart{\'{\i}}n} S, {Aalto} S,
  {Falgarone} E, {Menten} KM, {Schilke} P, {Wiklind} T, {Zwaan} MA (2016)
  {OH$^{+}$ and H$_{2}$O$^{+}$ absorption toward PKS 1830-211}. \aap 595:A128,
  \doi{10.1051/0004-6361/201629073}, \eprint{1609.01060}

\bibitem[{{Negrello} et~al(2010){Negrello}, {Hopwood}, {De Zotti}, {Cooray},
  {Verma}, {Bock}, {Frayer}, {Gurwell}, {Omont}, {Neri}, {Dannerbauer},
  {Leeuw}, {Barton}, {Cooke}, {Kim}, {da Cunha}, {Rodighiero}, {Cox},
  {Bonfield}, {Jarvis}, {Serjeant}, {Ivison}, {Dye}, {Aretxaga}, {Hughes},
  {Ibar}, {Bertoldi}, {Valtchanov}, {Eales}, {Dunne}, {Driver}, {Auld},
  {Buttiglione}, {Cava}, {Grady}, {Clements}, {Dariush}, {Fritz}, {Hill},
  {Hornbeck}, {Kelvin}, {Lagache}, {Lopez-Caniego}, {Gonzalez-Nuevo}, {Maddox},
  {Pascale}, {Pohlen}, {Rigby}, {Robotham}, {Simpson}, {Smith}, {Temi},
  {Thompson}, {Woodgate}, {York}, {Aguirre}, {Beelen}, {Blain}, {Baker},
  {Birkinshaw}, {Blundell}, {Bradford}, {Burgarella}, {Danese}, {Dunlop},
  {Fleuren}, {Glenn}, {Harris}, {Kamenetzky}, {Lupu}, {Maddalena}, {Madore},
  {Maloney}, {Matsuhara}, {Micha{\l}owski}, {Murphy}, {Naylor}, {Nguyen},
  {Popescu}, {Rawlings}, {Rigopoulou}, {Scott}, {Scott}, {Seibert}, {Smail},
  {Tuffs}, {Vieira}, {van der Werf}, and {Zmuidzinas}}]{Negrello2010}
{Negrello} M, {Hopwood} R, {De Zotti} G, {Cooray} A, {Verma} A, {Bock} J,
  {Frayer} DT, {Gurwell} MA, {Omont} A, {Neri} R, {Dannerbauer} H, {Leeuw} LL,
  {Barton} E, {Cooke} J, {Kim} S, {da Cunha} E, {Rodighiero} G, {Cox} P,
  {Bonfield} DG, {Jarvis} MJ, {Serjeant} S, {Ivison} RJ, {Dye} S, {Aretxaga} I,
  {Hughes} DH, {Ibar} E, {Bertoldi} F, {Valtchanov} I, {Eales} S, {Dunne} L,
  {Driver} SP, {Auld} R, {Buttiglione} S, {Cava} A, {Grady} CA, {Clements} DL,
  {Dariush} A, {Fritz} J, {Hill} D, {Hornbeck} JB, {Kelvin} L, {Lagache} G,
  {Lopez-Caniego} M, {Gonzalez-Nuevo} J, {Maddox} S, {Pascale} E, {Pohlen} M,
  {Rigby} EE, {Robotham} A, {Simpson} C, {Smith} DJB, {Temi} P, {Thompson} MA,
  {Woodgate} BE, {York} DG, {Aguirre} JE, {Beelen} A, {Blain} A, {Baker} AJ,
  {Birkinshaw} M, {Blundell} R, {Bradford} CM, {Burgarella} D, {Danese} L,
  {Dunlop} JS, {Fleuren} S, {Glenn} J, {Harris} AI, {Kamenetzky} J, {Lupu} RE,
  {Maddalena} RJ, {Madore} BF, {Maloney} PR, {Matsuhara} H, {Micha{\l}owski}
  MJ, {Murphy} EJ, {Naylor} BJ, {Nguyen} H, {Popescu} C, {Rawlings} S,
  {Rigopoulou} D, {Scott} D, {Scott} KS, {Seibert} M, {Smail} I, {Tuffs} RJ,
  {Vieira} JD, {van der Werf} PP, {Zmuidzinas} J (2010) {The Detection of a
  Population of Submillimeter-Bright, Strongly Lensed Galaxies}. Science
  330:800, \doi{10.1126/science.1193420}, \eprint{1011.1255}

\bibitem[{{Obreschkow} et~al(2009){Obreschkow}, {Croton}, {De Lucia},
  {Khochfar}, and {Rawlings}}]{Obreschkow2009}
{Obreschkow} D, {Croton} D, {De Lucia} G, {Khochfar} S, {Rawlings} S (2009)
  {Simulation of the Cosmic Evolution of Atomic and Molecular Hydrogen in
  Galaxies}. \apj 698:1467--1484, \doi{10.1088/0004-637X/698/2/1467},
  \eprint{0904.2221}

\bibitem[{{Omont}(2007)}]{Omont2007}
{Omont} A (2007) {Molecules in galaxies}. Reports on Progress in Physics
  70:1099--1176, \doi{10.1088/0034-4885/70/7/R03}, \eprint{0709.3814}

\bibitem[{{Oteo} et~al(2017){Oteo}, {Zhang}, {Yang}, {Ivison}, {Omont},
  {Bremer}, {Bussmann}, {Cooray}, {Cox}, {Dannerbauer}, {Dunne}, {Eales},
  {Furlanetto}, {Gavazzi}, {Gao}, {Greve}, {Nayyeri}, {Negrello}, {Neri},
  {Riechers}, {Tunnard}, {Wagg}, and {Van der Werf}}]{Oteo2017}
{Oteo} I, {Zhang} ZY, {Yang} C, {Ivison} RJ, {Omont} A, {Bremer} M, {Bussmann}
  S, {Cooray} A, {Cox} P, {Dannerbauer} H, {Dunne} L, {Eales} S, {Furlanetto}
  C, {Gavazzi} R, {Gao} Y, {Greve} TR, {Nayyeri} H, {Negrello} M, {Neri} R,
  {Riechers} D, {Tunnard} R, {Wagg} J, {Van der Werf} P (2017) {High Dense Gas
  Fraction in Intensely Star-forming Dusty Galaxies}. \apj 850:170,
  \doi{10.3847/1538-4357/aa8ee3}, \eprint{1701.05901}

\bibitem[{{Ouchi} et~al(2013){Ouchi}, {Ellis}, {Ono}, {Nakanishi}, {Kohno},
  {Momose}, {Kurono}, {Ashby}, {Shimasaku}, {Willner}, {Fazio}, {Tamura}, and
  {Iono}}]{Ouchi2013}
{Ouchi} M, {Ellis} R, {Ono} Y, {Nakanishi} K, {Kohno} K, {Momose} R, {Kurono}
  Y, {Ashby} MLN, {Shimasaku} K, {Willner} SP, {Fazio} GG, {Tamura} Y, {Iono} D
  (2013) {An Intensely Star-forming Galaxy at z \~{} 7 with Low Dust and Metal
  Content Revealed by Deep ALMA and HST Observations}. \apj 778:102,
  \doi{10.1088/0004-637X/778/2/102}, \eprint{1306.3572}

\bibitem[{{Pan} et~al(2016){Pan}, {Zheng}, {Lin}, {Li}, {Wang}, {Fan}, and
  {Kong}}]{Pan2016}
{Pan} Z, {Zheng} X, {Lin} W, {Li} J, {Wang} J, {Fan} L, {Kong} X (2016) {The
  Spatially Resolved NUV-r Color of Local Star-forming Galaxies and Clues for
  Quenching}. \apj 819:91, \doi{10.3847/0004-637X/819/2/91},
  \eprint{1601.05503}

\bibitem[{{Papadopoulos} et~al(2010){Papadopoulos}, {van der Werf}, {Isaak},
  and {Xilouris}}]{Papadopoulos2010}
{Papadopoulos} PP, {van der Werf} P, {Isaak} K, {Xilouris} EM (2010) {CO
  Spectral Line Energy Distributions of Infrared-Luminous Galaxies and Active
  Galactic Nuclei}. \apj 715:775--792, \doi{10.1088/0004-637X/715/2/775},
  \eprint{1003.5889}

\bibitem[{{Peng} et~al(2010){Peng}, {Lilly}, {Kova{\v c}}, {Bolzonella},
  {Pozzetti}, {Renzini}, {Zamorani}, {Ilbert}, {Knobel}, {Iovino}, {Maier},
  {Cucciati}, {Tasca}, {Carollo}, {Silverman}, {Kampczyk}, {de Ravel},
  {Sanders}, {Scoville}, {Contini}, {Mainieri}, {Scodeggio}, {Kneib}, {Le
  F{\`e}vre}, {Bardelli}, {Bongiorno}, {Caputi}, {Coppa}, {de la Torre},
  {Franzetti}, {Garilli}, {Lamareille}, {Le Borgne}, {Le Brun}, {Mignoli},
  {Perez Montero}, {Pello}, {Ricciardelli}, {Tanaka}, {Tresse}, {Vergani},
  {Welikala}, {Zucca}, {Oesch}, {Abbas}, {Barnes}, {Bordoloi}, {Bottini},
  {Cappi}, {Cassata}, {Cimatti}, {Fumana}, {Hasinger}, {Koekemoer},
  {Leauthaud}, {Maccagni}, {Marinoni}, {McCracken}, {Memeo}, {Meneux}, {Nair},
  {Porciani}, {Presotto}, and {Scaramella}}]{Peng2010}
{Peng} Yj, {Lilly} SJ, {Kova{\v c}} K, {Bolzonella} M, {Pozzetti} L, {Renzini}
  A, {Zamorani} G, {Ilbert} O, {Knobel} C, {Iovino} A, {Maier} C, {Cucciati} O,
  {Tasca} L, {Carollo} CM, {Silverman} J, {Kampczyk} P, {de Ravel} L, {Sanders}
  D, {Scoville} N, {Contini} T, {Mainieri} V, {Scodeggio} M, {Kneib} JP, {Le
  F{\`e}vre} O, {Bardelli} S, {Bongiorno} A, {Caputi} K, {Coppa} G, {de la
  Torre} S, {Franzetti} P, {Garilli} B, {Lamareille} F, {Le Borgne} JF, {Le
  Brun} V, {Mignoli} M, {Perez Montero} E, {Pello} R, {Ricciardelli} E,
  {Tanaka} M, {Tresse} L, {Vergani} D, {Welikala} N, {Zucca} E, {Oesch} P,
  {Abbas} U, {Barnes} L, {Bordoloi} R, {Bottini} D, {Cappi} A, {Cassata} P,
  {Cimatti} A, {Fumana} M, {Hasinger} G, {Koekemoer} A, {Leauthaud} A,
  {Maccagni} D, {Marinoni} C, {McCracken} H, {Memeo} P, {Meneux} B, {Nair} P,
  {Porciani} C, {Presotto} V, {Scaramella} R (2010) {Mass and Environment as
  Drivers of Galaxy Evolution in SDSS and zCOSMOS and the Origin of the
  Schechter Function}. \apj 721:193--221, \doi{10.1088/0004-637X/721/1/193},
  \eprint{1003.4747}

\bibitem[{{Pentericci} et~al(2016){Pentericci}, {Carniani}, {Castellano},
  {Fontana}, {Maiolino}, {Guaita}, {Vanzella}, {Grazian}, {Santini}, {Yan},
  {Cristiani}, {Conselice}, {Giavalisco}, {Hathi}, and
  {Koekemoer}}]{Pentericci2016}
{Pentericci} L, {Carniani} S, {Castellano} M, {Fontana} A, {Maiolino} R,
  {Guaita} L, {Vanzella} E, {Grazian} A, {Santini} P, {Yan} H, {Cristiani} S,
  {Conselice} C, {Giavalisco} M, {Hathi} N, {Koekemoer} A (2016) {Tracing the
  Reionization Epoch with ALMA: [C II] Emission in z {$\sim$} 7 Galaxies}.
  \apjl 829:L11, \doi{10.3847/2041-8205/829/1/L11}, \eprint{1608.08837}

\bibitem[{{Popping} et~al(2014){Popping}, {Somerville}, and
  {Trager}}]{Popping2014}
{Popping} G, {Somerville} RS, {Trager} SC (2014) {Evolution of the atomic and
  molecular gas content of galaxies}. \mnras 442:2398--2418,
  \doi{10.1093/mnras/stu991}, \eprint{1308.6764}

\bibitem[{{Prochaska} et~al(2005){Prochaska}, {Herbert-Fort}, and
  {Wolfe}}]{Prochaska2005}
{Prochaska} JX, {Herbert-Fort} S, {Wolfe} AM (2005) {The SDSS Damped
  Ly{$\alpha$} Survey: Data Release 3}. \apj 635:123--142,
  \doi{10.1086/497287}, \eprint{astro-ph/0508361}

\bibitem[{{Rafelski} et~al(2015){Rafelski}, {Teplitz}, {Gardner}, {Coe},
  {Bond}, {Koekemoer}, {Grogin}, {Kurczynski}, {McGrath}, {Bourque}, {Atek},
  {Brown}, {Colbert}, {Codoreanu}, {Ferguson}, {Finkelstein}, {Gawiser},
  {Giavalisco}, {Gronwall}, {Hanish}, {Lee}, {Mehta}, {de Mello},
  {Ravindranath}, {Ryan}, {Scarlata}, {Siana}, {Soto}, and
  {Voyer}}]{Rafelski2015}
{Rafelski} M, {Teplitz} HI, {Gardner} JP, {Coe} D, {Bond} NA, {Koekemoer} AM,
  {Grogin} N, {Kurczynski} P, {McGrath} EJ, {Bourque} M, {Atek} H, {Brown} TM,
  {Colbert} JW, {Codoreanu} A, {Ferguson} HC, {Finkelstein} SL, {Gawiser} E,
  {Giavalisco} M, {Gronwall} C, {Hanish} DJ, {Lee} KS, {Mehta} V, {de Mello}
  DF, {Ravindranath} S, {Ryan} RE, {Scarlata} C, {Siana} B, {Soto} E, {Voyer}
  EN (2015) {UVUDF: Ultraviolet Through Near-infrared Catalog and Photometric
  Redshifts of Galaxies in the Hubble Ultra Deep Field}. \aj 150:31,
  \doi{10.1088/0004-6256/150/1/31}, \eprint{1505.01160}

\bibitem[{{Riechers} et~al(2013){Riechers}, {Bradford}, {Clements}, {Dowell},
  {P{\'e}rez-Fournon}, {Ivison}, {Bridge}, {Conley}, {Fu}, {Vieira}, {Wardlow},
  {Calanog}, {Cooray}, {Hurley}, {Neri}, {Kamenetzky}, {Aguirre}, {Altieri},
  {Arumugam}, {Benford}, {B{\'e}thermin}, {Bock}, {Burgarella},
  {Cabrera-Lavers}, {Chapman}, {Cox}, {Dunlop}, {Earle}, {Farrah}, {Ferrero},
  {Franceschini}, {Gavazzi}, {Glenn}, {Solares}, {Gurwell}, {Halpern},
  {Hatziminaoglou}, {Hyde}, {Ibar}, {Kov{\'a}cs}, {Krips}, {Lupu}, {Maloney},
  {Martinez-Navajas}, {Matsuhara}, {Murphy}, {Naylor}, {Nguyen}, {Oliver},
  {Omont}, {Page}, {Petitpas}, {Rangwala}, {Roseboom}, {Scott}, {Smith},
  {Staguhn}, {Streblyanska}, {Thomson}, {Valtchanov}, {Viero}, {Wang},
  {Zemcov}, and {Zmuidzinas}}]{Riechers2013}
{Riechers} DA, {Bradford} CM, {Clements} DL, {Dowell} CD, {P{\'e}rez-Fournon}
  I, {Ivison} RJ, {Bridge} C, {Conley} A, {Fu} H, {Vieira} JD, {Wardlow} J,
  {Calanog} J, {Cooray} A, {Hurley} P, {Neri} R, {Kamenetzky} J, {Aguirre} JE,
  {Altieri} B, {Arumugam} V, {Benford} DJ, {B{\'e}thermin} M, {Bock} J,
  {Burgarella} D, {Cabrera-Lavers} A, {Chapman} SC, {Cox} P, {Dunlop} JS,
  {Earle} L, {Farrah} D, {Ferrero} P, {Franceschini} A, {Gavazzi} R, {Glenn} J,
  {Solares} EAG, {Gurwell} MA, {Halpern} M, {Hatziminaoglou} E, {Hyde} A,
  {Ibar} E, {Kov{\'a}cs} A, {Krips} M, {Lupu} RE, {Maloney} PR,
  {Martinez-Navajas} P, {Matsuhara} H, {Murphy} EJ, {Naylor} BJ, {Nguyen} HT,
  {Oliver} SJ, {Omont} A, {Page} MJ, {Petitpas} G, {Rangwala} N, {Roseboom} IG,
  {Scott} D, {Smith} AJ, {Staguhn} JG, {Streblyanska} A, {Thomson} AP,
  {Valtchanov} I, {Viero} M, {Wang} L, {Zemcov} M, {Zmuidzinas} J (2013) {A
  dust-obscured massive maximum-starburst galaxy at a redshift of 6.34}. \nat
  496:329--333, \doi{10.1038/nature12050}, \eprint{1304.4256}

\bibitem[{{Riechers} et~al(2014){Riechers}, {Carilli}, {Capak}, {Scoville},
  {Smol{\v c}i{\'c}}, {Schinnerer}, {Yun}, {Cox}, {Bertoldi}, {Karim}, and
  {Yan}}]{Riechers2014}
{Riechers} DA, {Carilli} CL, {Capak} PL, {Scoville} NZ, {Smol{\v c}i{\'c}} V,
  {Schinnerer} E, {Yun} M, {Cox} P, {Bertoldi} F, {Karim} A, {Yan} L (2014)
  {ALMA Imaging of Gas and Dust in a Galaxy Protocluster at Redshift 5.3: [C
  II] Emission in ``Typical'' Galaxies and Dusty Starbursts {$\sim$}1 Billion
  Years after the Big Bang}. \apj 796:84, \doi{10.1088/0004-637X/796/2/84},
  \eprint{1404.7159}

\bibitem[{{Salim} et~al(2015){Salim}, {Federrath}, and {Kewley}}]{Salim2015}
{Salim} DM, {Federrath} C, {Kewley} LJ (2015) {A Universal,
  Turbulence-regulated Star Formation Law: From Milky Way Clouds to
  High-redshift Disk and Starburst Galaxies}. \apjl 806:L36,
  \doi{10.1088/2041-8205/806/2/L36}, \eprint{1505.03144}

\bibitem[{{Sanders} and {Mirabel}(1996)}]{Sanders1996}
{Sanders} DB, {Mirabel} IF (1996) {Luminous Infrared Galaxies}. \araa 34:749,
  \doi{10.1146/annurev.astro.34.1.749}

\bibitem[{{Sargent} et~al(2014){Sargent}, {Daddi}, {B{\'e}thermin}, {Aussel},
  {Magdis}, {Hwang}, {Juneau}, {Elbaz}, and {da Cunha}}]{Sargent2014}
{Sargent} MT, {Daddi} E, {B{\'e}thermin} M, {Aussel} H, {Magdis} G, {Hwang} HS,
  {Juneau} S, {Elbaz} D, {da Cunha} E (2014) {Regularity Underlying Complexity:
  A Redshift-independent Description of the Continuous Variation of
  Galaxy-scale Molecular Gas Properties in the Mass-star Formation Rate Plane}.
  \apj 793:19, \doi{10.1088/0004-637X/793/1/19}, \eprint{1303.4392}

\bibitem[{{Schaerer} et~al(2015){Schaerer}, {Boone}, {Zamojski}, {Staguhn},
  {Dessauges-Zavadsky}, {Finkelstein}, and {Combes}}]{Schaerer2015}
{Schaerer} D, {Boone} F, {Zamojski} M, {Staguhn} J, {Dessauges-Zavadsky} M,
  {Finkelstein} S, {Combes} F (2015) {New constraints on dust emission and UV
  attenuation of z = 6.5-7.5 galaxies from millimeter observations}. \aap
  574:A19, \doi{10.1051/0004-6361/201424649}, \eprint{1407.5793}

\bibitem[{{Schinnerer} et~al(2016){Schinnerer}, {Groves}, {Sargent}, {Karim},
  {Oesch}, {Magnelli}, {LeFevre}, {Tasca}, {Civano}, {Cassata}, and {Smol{\v
  c}i{\'c}}}]{Schinnerer2016}
{Schinnerer} E, {Groves} B, {Sargent} MT, {Karim} A, {Oesch} PA, {Magnelli} B,
  {LeFevre} O, {Tasca} L, {Civano} F, {Cassata} P, {Smol{\v c}i{\'c}} V (2016)
  {Gas Fraction and Depletion Time of Massive Star-forming Galaxies at z \~{}
  3.2: No Change in Global Star Formation Process out to z $>$ 3}. \apj
  833:112, \doi{10.3847/1538-4357/833/1/112}, \eprint{1610.03656}

\bibitem[{{Scoville} et~al(2014){Scoville}, {Aussel}, {Sheth}, {Scott},
  {Sanders}, {Ivison}, {Pope}, {Capak}, {Vanden Bout}, {Manohar}, {Kartaltepe},
  {Robertson}, and {Lilly}}]{Scoville2014}
{Scoville} N, {Aussel} H, {Sheth} K, {Scott} KS, {Sanders} D, {Ivison} R,
  {Pope} A, {Capak} P, {Vanden Bout} P, {Manohar} S, {Kartaltepe} J,
  {Robertson} B, {Lilly} S (2014) {The Evolution of Interstellar Medium Mass
  Probed by Dust Emission: ALMA Observations at z = 0.3-2}. \apj 783:84,
  \doi{10.1088/0004-637X/783/2/84}, \eprint{1401.2987}

\bibitem[{{Scoville} et~al(2016){Scoville}, {Sheth}, {Aussel}, {Vanden Bout},
  {Capak}, {Bongiorno}, {Casey}, {Murchikova}, {Koda},
  {{\'A}lvarez-M{\'a}rquez}, {Lee}, {Laigle}, {McCracken}, {Ilbert}, {Pope},
  {Sanders}, {Chu}, {Toft}, {Ivison}, and {Manohar}}]{Scoville2016}
{Scoville} N, {Sheth} K, {Aussel} H, {Vanden Bout} P, {Capak} P, {Bongiorno} A,
  {Casey} CM, {Murchikova} L, {Koda} J, {{\'A}lvarez-M{\'a}rquez} J, {Lee} N,
  {Laigle} C, {McCracken} HJ, {Ilbert} O, {Pope} A, {Sanders} D, {Chu} J,
  {Toft} S, {Ivison} RJ, {Manohar} S (2016) {ISM Masses and the Star formation
  Law at Z = 1 to 6: ALMA Observations of Dust Continuum in 145 Galaxies in the
  COSMOS Survey Field}. \apj 820:83, \doi{10.3847/0004-637X/820/2/83},
  \eprint{1511.05149}

\bibitem[{{Scoville} et~al(2017){Scoville}, {Lee}, {Vanden Bout},
  {Diaz-Santos}, {Sanders}, {Darvish}, {Bongiorno}, {Casey}, {Murchikova},
  {Koda}, {Capak}, {Vlahakis}, {Ilbert}, {Sheth}, {Morokuma-Matsui}, {Ivison},
  {Aussel}, {Laigle}, {McCracken}, {Armus}, {Pope}, {Toft}, and
  {Masters}}]{Scoville2017}
{Scoville} N, {Lee} N, {Vanden Bout} P, {Diaz-Santos} T, {Sanders} D, {Darvish}
  B, {Bongiorno} A, {Casey} CM, {Murchikova} L, {Koda} J, {Capak} P, {Vlahakis}
  C, {Ilbert} O, {Sheth} K, {Morokuma-Matsui} K, {Ivison} RJ, {Aussel} H,
  {Laigle} C, {McCracken} HJ, {Armus} L, {Pope} A, {Toft} S, {Masters} D (2017)
  {Evolution of Interstellar Medium, Star Formation, and Accretion at High
  Redshift}. \apj 837:150, \doi{10.3847/1538-4357/aa61a0}, \eprint{1702.04729}

\bibitem[{{Sharda} et~al(2018){Sharda}, {Federrath}, {da{\^A} Cunha},
  {Swinbank}, and {Dye}}]{Sharda2018}
{Sharda} P, {Federrath} C, {da{\^A} Cunha} E, {Swinbank} AM, {Dye} S (2018)
  {Testing star formation laws in a starburst galaxy at redshift 3 resolved
  with ALMA}. \mnras 477:4380--4390, \doi{10.1093/mnras/sty886},
  \eprint{1712.03661}

\bibitem[{{Solomon} and {Vanden Bout}(2005)}]{Solomon2005}
{Solomon} PM, {Vanden Bout} PA (2005) {Molecular Gas at High Redshift}. \araa
  43:677--725, \doi{10.1146/annurev.astro.43.051804.102221},
  \eprint{astro-ph/0508481}

\bibitem[{{Speagle} et~al(2014){Speagle}, {Steinhardt}, {Capak}, and
  {Silverman}}]{Speagle2014}
{Speagle} JS, {Steinhardt} CL, {Capak} PL, {Silverman} JD (2014) {A Highly
  Consistent Framework for the Evolution of the Star-Forming ``Main Sequence''
  from z \~{} 0-6}. \apjs 214:15, \doi{10.1088/0067-0049/214/2/15},
  \eprint{1405.2041}

\bibitem[{{Spilker} et~al(2018){Spilker}, {Bezanson}, {Barisic}, {Bell},
  {Lagos}, {Maseda}, {Muzzin}, {Pacifici}, {Sobral}, {Straatman}, {van der
  Wel}, {van Dokkum}, {Weiner}, {Whitaker}, {Williams}, and {Wu}}]{Spilker2018}
{Spilker} J, {Bezanson} R, {Barisic} I, {Bell} E, {Lagos} CdP, {Maseda} M,
  {Muzzin} A, {Pacifici} C, {Sobral} D, {Straatman} C, {van der Wel} A, {van
  Dokkum} P, {Weiner} B, {Whitaker} K, {Williams} CC, {Wu} PF (2018) {Molecular
  Gas Contents and Scaling Relations for Massive Passive Galaxies at
  Intermediate Redshifts from the LEGA-C Survey}. ArXiv e-prints
  \eprint{1805.02667}

\bibitem[{{Stark}(2016)}]{Stark2016}
{Stark} DP (2016) {Galaxies in the First Billion Years After the Big Bang}.
  \araa 54:761--803, \doi{10.1146/annurev-astro-081915-023417}

\bibitem[{{Steidel} et~al(2000){Steidel}, {Adelberger}, {Shapley}, {Pettini},
  {Dickinson}, and {Giavalisco}}]{Steidel2000}
{Steidel} CC, {Adelberger} KL, {Shapley} AE, {Pettini} M, {Dickinson} M,
  {Giavalisco} M (2000) {Ly{$\alpha$} Imaging of a Proto-Cluster Region at
  z=3.09}. \apj 532:170--182, \doi{10.1086/308568}, \eprint{astro-ph/9910144}

\bibitem[{{Strandet} et~al(2016){Strandet}, {Weiss}, {Vieira}, {de Breuck},
  {Aguirre}, {Aravena}, {Ashby}, {B{\'e}thermin}, {Bradford}, {Carlstrom},
  {Chapman}, {Crawford}, {Everett}, {Fassnacht}, {Furstenau}, {Gonzalez},
  {Greve}, {Gullberg}, {Hezaveh}, {Kamenetzky}, {Litke}, {Ma}, {Malkan},
  {Marrone}, {Menten}, {Murphy}, {Nadolski}, {Rotermund}, {Spilker}, {Stark},
  and {Welikala}}]{Strandet2016}
{Strandet} ML, {Weiss} A, {Vieira} JD, {de Breuck} C, {Aguirre} JE, {Aravena}
  M, {Ashby} MLN, {B{\'e}thermin} M, {Bradford} CM, {Carlstrom} JE, {Chapman}
  SC, {Crawford} TM, {Everett} W, {Fassnacht} CD, {Furstenau} RM, {Gonzalez}
  AH, {Greve} TR, {Gullberg} B, {Hezaveh} Y, {Kamenetzky} JR, {Litke} K, {Ma}
  J, {Malkan} M, {Marrone} DP, {Menten} KM, {Murphy} EJ, {Nadolski} A,
  {Rotermund} KM, {Spilker} JS, {Stark} AA, {Welikala} N (2016) {The Redshift
  Distribution of Dusty Star-forming Galaxies from the SPT Survey}. \apj
  822:80, \doi{10.3847/0004-637X/822/2/80}, \eprint{1603.05094}

\bibitem[{{Suess} et~al(2017){Suess}, {Bezanson}, {Spilker}, {Kriek}, {Greene},
  {Feldmann}, {Hunt}, and {Narayanan}}]{Suess2017}
{Suess} KA, {Bezanson} R, {Spilker} JS, {Kriek} M, {Greene} JE, {Feldmann} R,
  {Hunt} Q, {Narayanan} D (2017) {Massive Quenched Galaxies at z {$\sim$} 0.7
  Retain Large Molecular Gas Reservoirs}. \apjl 846:L14,
  \doi{10.3847/2041-8213/aa85dc}, \eprint{1708.03337}

\bibitem[{{Swinbank} et~al(2014){Swinbank}, {Simpson}, {Smail}, {Harrison},
  {Hodge}, {Karim}, {Walter}, {Alexander}, {Brandt}, {de Breuck}, {da Cunha},
  {Chapman}, {Coppin}, {Danielson}, {Dannerbauer}, {Decarli}, {Greve},
  {Ivison}, {Knudsen}, {Lagos}, {Schinnerer}, {Thomson}, {Wardlow}, {Wei{\ss}},
  and {van der Werf}}]{Swinbank2014}
{Swinbank} AM, {Simpson} JM, {Smail} I, {Harrison} CM, {Hodge} JA, {Karim} A,
  {Walter} F, {Alexander} DM, {Brandt} WN, {de Breuck} C, {da Cunha} E,
  {Chapman} SC, {Coppin} KEK, {Danielson} ALR, {Dannerbauer} H, {Decarli} R,
  {Greve} TR, {Ivison} RJ, {Knudsen} KK, {Lagos} CDP, {Schinnerer} E, {Thomson}
  AP, {Wardlow} JL, {Wei{\ss}} A, {van der Werf} P (2014) {An ALMA survey of
  sub-millimetre Galaxies in the Extended Chandra Deep Field South: the
  far-infrared properties of SMGs}. \mnras 438:1267--1287,
  \doi{10.1093/mnras/stt2273}, \eprint{1310.6362}

\bibitem[{{Tacchella} et~al(2016){Tacchella}, {Dekel}, {Carollo}, {Ceverino},
  {DeGraf}, {Lapiner}, {Mandelker}, and {Primack Joel}}]{Tacchella2016}
{Tacchella} S, {Dekel} A, {Carollo} CM, {Ceverino} D, {DeGraf} C, {Lapiner} S,
  {Mandelker} N, {Primack Joel} R (2016) {The confinement of star-forming
  galaxies into a main sequence through episodes of gas compaction, depletion
  and replenishment}. \mnras 457:2790--2813, \doi{10.1093/mnras/stw131},
  \eprint{1509.02529}

\bibitem[{{Tacconi} et~al(2010){Tacconi}, {Genzel}, {Neri}, {Cox}, {Cooper},
  {Shapiro}, {Bolatto}, {Bouch{\'e}}, {Bournaud}, {Burkert}, {Combes},
  {Comerford}, {Davis}, {Schreiber}, {Garcia-Burillo}, {Gracia-Carpio}, {Lutz},
  {Naab}, {Omont}, {Shapley}, {Sternberg}, and {Weiner}}]{Tacconi2010}
{Tacconi} LJ, {Genzel} R, {Neri} R, {Cox} P, {Cooper} MC, {Shapiro} K,
  {Bolatto} A, {Bouch{\'e}} N, {Bournaud} F, {Burkert} A, {Combes} F,
  {Comerford} J, {Davis} M, {Schreiber} NMF, {Garcia-Burillo} S,
  {Gracia-Carpio} J, {Lutz} D, {Naab} T, {Omont} A, {Shapley} A, {Sternberg} A,
  {Weiner} B (2010) {High molecular gas fractions in normal massive
  star-forming galaxies in the young Universe}. \nat 463:781--784,
  \doi{10.1038/nature08773}, \eprint{1002.2149}

\bibitem[{{Tacconi} et~al(2013){Tacconi}, {Neri}, {Genzel}, {Combes},
  {Bolatto}, {Cooper}, {Wuyts}, {Bournaud}, {Burkert}, {Comerford}, {Cox},
  {Davis}, {F{\"o}rster Schreiber}, {Garc{\'{\i}}a-Burillo}, {Gracia-Carpio},
  {Lutz}, {Naab}, {Newman}, {Omont}, {Saintonge}, {Shapiro Griffin}, {Shapley},
  {Sternberg}, and {Weiner}}]{Tacconi2013}
{Tacconi} LJ, {Neri} R, {Genzel} R, {Combes} F, {Bolatto} A, {Cooper} MC,
  {Wuyts} S, {Bournaud} F, {Burkert} A, {Comerford} J, {Cox} P, {Davis} M,
  {F{\"o}rster Schreiber} NM, {Garc{\'{\i}}a-Burillo} S, {Gracia-Carpio} J,
  {Lutz} D, {Naab} T, {Newman} S, {Omont} A, {Saintonge} A, {Shapiro Griffin}
  K, {Shapley} A, {Sternberg} A, {Weiner} B (2013) {Phibss: Molecular Gas
  Content and Scaling Relations in z \~{} 1-3 Massive, Main-sequence
  Star-forming Galaxies}. \apj 768:74, \doi{10.1088/0004-637X/768/1/74},
  \eprint{1211.5743}

\bibitem[{{Tacconi} et~al(2018){Tacconi}, {Genzel}, {Saintonge}, {Combes},
  {Garc{\'{\i}}a-Burillo}, {Neri}, {Bolatto}, {Contini}, {F{\"o}rster
  Schreiber}, {Lilly}, {Lutz}, {Wuyts}, {Accurso}, {Boissier}, {Boone},
  {Bouch{\'e}}, {Bournaud}, {Burkert}, {Carollo}, {Cooper}, {Cox}, {Feruglio},
  {Freundlich}, {Herrera-Camus}, {Juneau}, {Lippa}, {Naab}, {Renzini},
  {Salome}, {Sternberg}, {Tadaki}, {{\"U}bler}, {Walter}, {Weiner}, and
  {Weiss}}]{Tacconi2018}
{Tacconi} LJ, {Genzel} R, {Saintonge} A, {Combes} F, {Garc{\'{\i}}a-Burillo} S,
  {Neri} R, {Bolatto} A, {Contini} T, {F{\"o}rster Schreiber} NM, {Lilly} S,
  {Lutz} D, {Wuyts} S, {Accurso} G, {Boissier} J, {Boone} F, {Bouch{\'e}} N,
  {Bournaud} F, {Burkert} A, {Carollo} M, {Cooper} M, {Cox} P, {Feruglio} C,
  {Freundlich} J, {Herrera-Camus} R, {Juneau} S, {Lippa} M, {Naab} T, {Renzini}
  A, {Salome} P, {Sternberg} A, {Tadaki} K, {{\"U}bler} H, {Walter} F, {Weiner}
  B, {Weiss} A (2018) {PHIBSS: Unified Scaling Relations of Gas Depletion Time
  and Molecular Gas Fractions}. \apj 853:179, \doi{10.3847/1538-4357/aaa4b4},
  \eprint{1702.01140}

\bibitem[{{Tadaki} et~al(2015){Tadaki}, {Kohno}, {Kodama}, {Ikarashi},
  {Aretxaga}, {Berta}, {Caputi}, {Dunlop}, {Hatsukade}, {Hayashi}, {Hughes},
  {Ivison}, {Izumi}, {Koyama}, {Lutz}, {Makiya}, {Matsuda}, {Nakanishi},
  {Rujopakarn}, {Tamura}, {Umehata}, {Wang}, {Wilson}, {Wuyts}, {Yamaguchi},
  and {Yun}}]{Tadaki2015}
{Tadaki} Ki, {Kohno} K, {Kodama} T, {Ikarashi} S, {Aretxaga} I, {Berta} S,
  {Caputi} KI, {Dunlop} JS, {Hatsukade} B, {Hayashi} M, {Hughes} DH, {Ivison}
  R, {Izumi} T, {Koyama} Y, {Lutz} D, {Makiya} R, {Matsuda} Y, {Nakanishi} K,
  {Rujopakarn} W, {Tamura} Y, {Umehata} H, {Wang} WH, {Wilson} GW, {Wuyts} S,
  {Yamaguchi} Y, {Yun} MS (2015) {SXDF-ALMA 1.5 arcmin$^{2}$ Deep Survey: A
  Compact Dusty Star-forming Galaxy at z = 2.5}. \apjl 811:L3,
  \doi{10.1088/2041-8205/811/1/L3}, \eprint{1508.05950}

\bibitem[{{Tamura} et~al(2018){Tamura}, {Mawatari}, {Hashimoto}, {Inoue},
  {Zackrisson}, {Christensen}, {Binggeli}, {Matsuda}, {Matsuo}, {Takeuchi},
  {Asano}, {Shimizu}, {Okamoto}, {Yoshida}, {Lee}, {Shibuya}, {Taniguchi},
  {Umehata}, {Hatsukade}, {Kohno}, and {Ota}}]{Tamura2018}
{Tamura} Y, {Mawatari} K, {Hashimoto} T, {Inoue} AK, {Zackrisson} E,
  {Christensen} L, {Binggeli} C, {Matsuda} Y, {Matsuo} H, {Takeuchi} TT,
  {Asano} RS, {Shimizu} I, {Okamoto} T, {Yoshida} N, {Lee} M, {Shibuya} T,
  {Taniguchi} Y, {Umehata} H, {Hatsukade} B, {Kohno} K, {Ota} K (2018)
  {Detection of the Far-infrared [O III] and Dust Emission in a Galaxy at
  Redshift 8.312: Early Metal Enrichment in the Heart of the Reionization Era}.
  ArXiv e-prints \eprint{1806.04132}

\bibitem[{{Tremblay} et~al(2016){Tremblay}, {Oonk}, {Combes}, {Salom{\'e}},
  {O'Dea}, {Baum}, {Voit}, {Donahue}, {McNamara}, {Davis}, {McDonald}, {Edge},
  {Clarke}, {Galv{\'a}n-Madrid}, {Bremer}, {Edwards}, {Fabian}, {Hamer}, {Li},
  {Maury}, {Russell}, {Quillen}, {Urry}, {Sanders}, and {Wise}}]{Tremblay2016}
{Tremblay} GR, {Oonk} JBR, {Combes} F, {Salom{\'e}} P, {O'Dea} CP, {Baum} SA,
  {Voit} GM, {Donahue} M, {McNamara} BR, {Davis} TA, {McDonald} MA, {Edge} AC,
  {Clarke} TE, {Galv{\'a}n-Madrid} R, {Bremer} MN, {Edwards} LOV, {Fabian} AC,
  {Hamer} S, {Li} Y, {Maury} A, {Russell} HR, {Quillen} AC, {Urry} CM,
  {Sanders} JS, {Wise} MW (2016) {Cold, clumpy accretion onto an active
  supermassive black hole}. \nat 534:218--221, \doi{10.1038/nature17969},
  \eprint{1606.02304}

\bibitem[{{van der Werf} et~al(2010){van der Werf}, {Isaak}, {Meijerink},
  {Spaans}, {Rykala}, {Fulton}, {Loenen}, {Walter}, {Wei{\ss}}, {Armus},
  {Fischer}, {Israel}, {Harris}, {Veilleux}, {Henkel}, {Savini}, {Lord},
  {Smith}, {Gonz{\'a}lez-Alfonso}, {Naylor}, {Aalto}, {Charmandaris}, {Dasyra},
  {Evans}, {Gao}, {Greve}, {G{\"u}sten}, {Kramer}, {Mart{\'{\i}}n-Pintado},
  {Mazzarella}, {Papadopoulos}, {Sanders}, {Spinoglio}, {Stacey}, {Vlahakis},
  {Wiedner}, and {Xilouris}}]{vanderWerf2010}
{van der Werf} PP, {Isaak} KG, {Meijerink} R, {Spaans} M, {Rykala} A, {Fulton}
  T, {Loenen} AF, {Walter} F, {Wei{\ss}} A, {Armus} L, {Fischer} J, {Israel}
  FP, {Harris} AI, {Veilleux} S, {Henkel} C, {Savini} G, {Lord} S, {Smith} HA,
  {Gonz{\'a}lez-Alfonso} E, {Naylor} D, {Aalto} S, {Charmandaris} V, {Dasyra}
  KM, {Evans} A, {Gao} Y, {Greve} TR, {G{\"u}sten} R, {Kramer} C,
  {Mart{\'{\i}}n-Pintado} J, {Mazzarella} J, {Papadopoulos} PP, {Sanders} DB,
  {Spinoglio} L, {Stacey} G, {Vlahakis} C, {Wiedner} MC, {Xilouris} EM (2010)
  {Black hole accretion and star formation as drivers of gas excitation and
  chemistry in Markarian 231}. \aap 518:L42, \doi{10.1051/0004-6361/201014682},
  \eprint{1005.2877}

\bibitem[{{Venemans} et~al(2012){Venemans}, {McMahon}, {Walter}, {Decarli},
  {Cox}, {Neri}, {Hewett}, {Mortlock}, {Simpson}, and {Warren}}]{Venemans2012}
{Venemans} BP, {McMahon} RG, {Walter} F, {Decarli} R, {Cox} P, {Neri} R,
  {Hewett} P, {Mortlock} DJ, {Simpson} C, {Warren} SJ (2012) {Detection of
  Atomic Carbon [C II] 158 {$\mu$}m and Dust Emission from a z = 7.1 Quasar
  Host Galaxy}. \apjl 751:L25, \doi{10.1088/2041-8205/751/2/L25},
  \eprint{1203.5844}

\bibitem[{{Venemans} et~al(2016){Venemans}, {Walter}, {Zschaechner}, {Decarli},
  {De Rosa}, {Findlay}, {McMahon}, and {Sutherland}}]{Venemans2016}
{Venemans} BP, {Walter} F, {Zschaechner} L, {Decarli} R, {De Rosa} G, {Findlay}
  JR, {McMahon} RG, {Sutherland} WJ (2016) {Bright [C II] and Dust Emission in
  Three z $>$ 6.6 Quasar Host Galaxies Observed by ALMA}. \apj 816:37,
  \doi{10.3847/0004-637X/816/1/37}, \eprint{1511.07432}

\bibitem[{{Venemans} et~al(2017){Venemans}, {Walter}, {Decarli}, {Ferkinhoff},
  {Wei{\ss}}, {Findlay}, {McMahon}, {Sutherland}, and
  {Meijerink}}]{Venemans2017}
{Venemans} BP, {Walter} F, {Decarli} R, {Ferkinhoff} C, {Wei{\ss}} A, {Findlay}
  JR, {McMahon} RG, {Sutherland} WJ, {Meijerink} R (2017) {Molecular Gas in
  Three z$\sim$ 7 Quasar Host Galaxies}. \apj 845:154,
  \doi{10.3847/1538-4357/aa81cb}, \eprint{1707.05238}

\bibitem[{{Vieira} et~al(2013){Vieira}, {Marrone}, {Chapman}, {De Breuck},
  {Hezaveh}, {Wei{$\beta$}}, {Aguirre}, {Aird}, {Aravena}, {Ashby}, {Bayliss},
  {Benson}, {Biggs}, {Bleem}, {Bock}, {Bothwell}, {Bradford}, {Brodwin},
  {Carlstrom}, {Chang}, {Crawford}, {Crites}, {de Haan}, {Dobbs}, {Fomalont},
  {Fassnacht}, {George}, {Gladders}, {Gonzalez}, {Greve}, {Gullberg},
  {Halverson}, {High}, {Holder}, {Holzapfel}, {Hoover}, {Hrubes}, {Hunter},
  {Keisler}, {Lee}, {Leitch}, {Lueker}, {Luong-van}, {Malkan}, {McIntyre},
  {McMahon}, {Mehl}, {Menten}, {Meyer}, {Mocanu}, {Murphy}, {Natoli}, {Padin},
  {Plagge}, {Reichardt}, {Rest}, {Ruel}, {Ruhl}, {Sharon}, {Schaffer}, {Shaw},
  {Shirokoff}, {Spilker}, {Stalder}, {Staniszewski}, {Stark}, {Story},
  {Vanderlinde}, {Welikala}, and {Williamson}}]{Vieira2013}
{Vieira} JD, {Marrone} DP, {Chapman} SC, {De Breuck} C, {Hezaveh} YD,
  {Wei{$\beta$}} A, {Aguirre} JE, {Aird} KA, {Aravena} M, {Ashby} MLN,
  {Bayliss} M, {Benson} BA, {Biggs} AD, {Bleem} LE, {Bock} JJ, {Bothwell} M,
  {Bradford} CM, {Brodwin} M, {Carlstrom} JE, {Chang} CL, {Crawford} TM,
  {Crites} AT, {de Haan} T, {Dobbs} MA, {Fomalont} EB, {Fassnacht} CD, {George}
  EM, {Gladders} MD, {Gonzalez} AH, {Greve} TR, {Gullberg} B, {Halverson} NW,
  {High} FW, {Holder} GP, {Holzapfel} WL, {Hoover} S, {Hrubes} JD, {Hunter} TR,
  {Keisler} R, {Lee} AT, {Leitch} EM, {Lueker} M, {Luong-van} D, {Malkan} M,
  {McIntyre} V, {McMahon} JJ, {Mehl} J, {Menten} KM, {Meyer} SS, {Mocanu} LM,
  {Murphy} EJ, {Natoli} T, {Padin} S, {Plagge} T, {Reichardt} CL, {Rest} A,
  {Ruel} J, {Ruhl} JE, {Sharon} K, {Schaffer} KK, {Shaw} L, {Shirokoff} E,
  {Spilker} JS, {Stalder} B, {Staniszewski} Z, {Stark} AA, {Story} K,
  {Vanderlinde} K, {Welikala} N, {Williamson} R (2013) {Dusty starburst
  galaxies in the early Universe as revealed by gravitational lensing}. \nat
  495:344--347, \doi{10.1038/nature12001}, \eprint{1303.2723}

\bibitem[{{Walter} et~al(2014){Walter}, {Decarli}, {Sargent}, {Carilli},
  {Dickinson}, {Riechers}, {Ellis}, {Stark}, {Weiner}, {Aravena}, {Bell},
  {Bertoldi}, {Cox}, {Da Cunha}, {Daddi}, {Downes}, {Lentati}, {Maiolino},
  {Menten}, {Neri}, {Rix}, and {Weiss}}]{Walter2014}
{Walter} F, {Decarli} R, {Sargent} M, {Carilli} C, {Dickinson} M, {Riechers} D,
  {Ellis} R, {Stark} D, {Weiner} B, {Aravena} M, {Bell} E, {Bertoldi} F, {Cox}
  P, {Da Cunha} E, {Daddi} E, {Downes} D, {Lentati} L, {Maiolino} R, {Menten}
  KM, {Neri} R, {Rix} HW, {Weiss} A (2014) {A Molecular Line Scan in the Hubble
  Deep Field North: Constraints on the CO Luminosity Function and the Cosmic
  H$_{2}$ Density}. \apj 782:79, \doi{10.1088/0004-637X/782/2/79},
  \eprint{1312.6365}

\bibitem[{{Walter} et~al(2016){Walter}, {Decarli}, {Aravena}, {Carilli},
  {Bouwens}, {da Cunha}, {Daddi}, {Ivison}, {Riechers}, {Smail}, {Swinbank},
  {Weiss}, {Anguita}, {Assef}, {Bacon}, {Bauer}, {Bell}, {Bertoldi}, {Chapman},
  {Colina}, {Cortes}, {Cox}, {Dickinson}, {Elbaz}, {G{\'o}nzalez-L{\'o}pez},
  {Ibar}, {Inami}, {Infante}, {Hodge}, {Karim}, {Le Fevre}, {Magnelli}, {Neri},
  {Oesch}, {Ota}, {Popping}, {Rix}, {Sargent}, {Sheth}, {van der Wel}, {van der
  Werf}, and {Wagg}}]{Walter2016}
{Walter} F, {Decarli} R, {Aravena} M, {Carilli} C, {Bouwens} R, {da Cunha} E,
  {Daddi} E, {Ivison} RJ, {Riechers} D, {Smail} I, {Swinbank} M, {Weiss} A,
  {Anguita} T, {Assef} R, {Bacon} R, {Bauer} F, {Bell} EF, {Bertoldi} F,
  {Chapman} S, {Colina} L, {Cortes} PC, {Cox} P, {Dickinson} M, {Elbaz} D,
  {G{\'o}nzalez-L{\'o}pez} J, {Ibar} E, {Inami} H, {Infante} L, {Hodge} J,
  {Karim} A, {Le Fevre} O, {Magnelli} B, {Neri} R, {Oesch} P, {Ota} K,
  {Popping} G, {Rix} HW, {Sargent} M, {Sheth} K, {van der Wel} A, {van der
  Werf} P, {Wagg} J (2016) {ALMA Spectroscopic Survey in the Hubble Ultra Deep
  Field: Survey Description}. \apj 833:67, \doi{10.3847/1538-4357/833/1/67},
  \eprint{1607.06768}

\bibitem[{{Wang} et~al(2013){Wang}, {Wagg}, {Carilli}, {Walter}, {Lentati},
  {Fan}, {Riechers}, {Bertoldi}, {Narayanan}, {Strauss}, {Cox}, {Omont},
  {Menten}, {Knudsen}, {Neri}, and {Jiang}}]{Wang2013}
{Wang} R, {Wagg} J, {Carilli} CL, {Walter} F, {Lentati} L, {Fan} X, {Riechers}
  DA, {Bertoldi} F, {Narayanan} D, {Strauss} MA, {Cox} P, {Omont} A, {Menten}
  KM, {Knudsen} KK, {Neri} R, {Jiang} L (2013) {Star Formation and Gas
  Kinematics of Quasar Host Galaxies at z \~{} 6: New Insights from ALMA}. \apj
  773:44, \doi{10.1088/0004-637X/773/1/44}, \eprint{1302.4154}

\bibitem[{{Wei{\ss}} et~al(2007){Wei{\ss}}, {Downes}, {Neri}, {Walter},
  {Henkel}, {Wilner}, {Wagg}, and {Wiklind}}]{Weiss2007}
{Wei{\ss}} A, {Downes} D, {Neri} R, {Walter} F, {Henkel} C, {Wilner} DJ, {Wagg}
  J, {Wiklind} T (2007) {Highly-excited CO emission in APM 08279+5255 at z =
  3.9}. \aap 467:955--969, \doi{10.1051/0004-6361:20066117},
  \eprint{astro-ph/0702669}

\bibitem[{{Whitaker} et~al(2012){Whitaker}, {van Dokkum}, {Brammer}, and
  {Franx}}]{Whitaker2012}
{Whitaker} KE, {van Dokkum} PG, {Brammer} G, {Franx} M (2012) {The Star
  Formation Mass Sequence Out to z = 2.5}. \apjl 754:L29,
  \doi{10.1088/2041-8205/754/2/L29}, \eprint{1205.0547}

\bibitem[{{Whitaker} et~al(2014){Whitaker}, {Franx}, {Leja}, {van Dokkum},
  {Henry}, {Skelton}, {Fumagalli}, {Momcheva}, {Brammer}, {Labb{\'e}},
  {Nelson}, and {Rigby}}]{Whitaker2014}
{Whitaker} KE, {Franx} M, {Leja} J, {van Dokkum} PG, {Henry} A, {Skelton} RE,
  {Fumagalli} M, {Momcheva} IG, {Brammer} GB, {Labb{\'e}} I, {Nelson} EJ,
  {Rigby} JR (2014) {Constraining the Low-mass Slope of the Star Formation
  Sequence at 0.5 $<$ z $<$ 2.5}. \apj 795:104,
  \doi{10.1088/0004-637X/795/2/104}, \eprint{1407.1843}

\bibitem[{{Wiklind} et~al(2018){Wiklind}, {Combes}, and
  {Kanekar}}]{Wiklind2018}
{Wiklind} T, {Combes} F, {Kanekar} N (2018) {ALMA Observations of Molecular
  Absorption in the Gravitational Lens PMN 0134-0931}. ArXiv e-prints
  \eprint{1804.05377}

\bibitem[{{Willott} et~al(2015){Willott}, {Carilli}, {Wagg}, and
  {Wang}}]{Willott2015}
{Willott} CJ, {Carilli} CL, {Wagg} J, {Wang} R (2015) {Star Formation and the
  Interstellar Medium in z $>$ 6 UV-luminous Lyman-break Galaxies}. \apj
  807:180, \doi{10.1088/0004-637X/807/2/180}, \eprint{1504.05875}

\bibitem[{{Wuyts} et~al(2011){Wuyts}, {F{\"o}rster Schreiber}, {van der Wel},
  {Magnelli}, {Guo}, {Genzel}, {Lutz}, {Aussel}, {Barro}, {Berta}, {Cava},
  {Graci{\'a}-Carpio}, {Hathi}, {Huang}, {Kocevski}, {Koekemoer}, {Lee}, {Le
  Floc'h}, {McGrath}, {Nordon}, {Popesso}, {Pozzi}, {Riguccini}, {Rodighiero},
  {Saintonge}, and {Tacconi}}]{Wuyts2011}
{Wuyts} S, {F{\"o}rster Schreiber} NM, {van der Wel} A, {Magnelli} B, {Guo} Y,
  {Genzel} R, {Lutz} D, {Aussel} H, {Barro} G, {Berta} S, {Cava} A,
  {Graci{\'a}-Carpio} J, {Hathi} NP, {Huang} KH, {Kocevski} DD, {Koekemoer} AM,
  {Lee} KS, {Le Floc'h} E, {McGrath} EJ, {Nordon} R, {Popesso} P, {Pozzi} F,
  {Riguccini} L, {Rodighiero} G, {Saintonge} A, {Tacconi} L (2011) {Galaxy
  Structure and Mode of Star Formation in the SFR-Mass Plane from z \~{} 2.5 to
  z \~{} 0.1}. \apj 742:96, \doi{10.1088/0004-637X/742/2/96},
  \eprint{1107.0317}

\end{thebibliography}

%\begin{thebibliography}{}
%\bibitem{}
%Allam, S. S., Tucker, D. L., Lin, H. et al.: 2007 ApJ 662, L51
%\end{thebibliography}
\end{document}